\documentclass[10pt]{iopart}

\usepackage{braket,dsfont,color,amsfonts,bm,amstext,mathrsfs}
\usepackage[bottom=1in,left=1in,right=1in,top=1in]{geometry}
\usepackage{graphicx}       
    \graphicspath{ {Figs/} }
    
\newtheorem{definition}{Definition}[section]    
\newtheorem{theorem}{Theorem}[section]
\newtheorem{proposition}{Proposition}[section]

\newtheorem{lemma}{Lemma}[section]

\newcommand{\sgn}{\textup{sgn\,}}
\newcommand{\spn}{\text{span}\,}
\newcommand{\norm}[1]{\left\|#1\right\|}

\begin{document}

\title[Non-Hermitian dynamics in quadratic bosonic Hamiltonians]{Deconstructing effective non-Hermitian dynamics \\
in quadratic bosonic Hamiltonians}

\author{Vincent P. Flynn$^1$, Emilio Cobanera$^{2,1}$,
and Lorenza Viola$^1$}

\address{$^1$Department of Physics and Astronomy, Dartmouth
College, 6127 Wilder Laboratory, \\ Hanover, NH 03755, USA}
\address{$^2$Department of Mathematics and Physics, SUNY Polytechnic Institute, \\
100 Seymour Avenue, Utica, NY 13502, USA}

\begin{abstract}
Unlike their fermionic counterparts, the dynamics of Hermitian quadratic bosonic Hamiltonians are governed by a generally non-Hermitian Bogoliubov-de Gennes effective Hamiltonian. This underlying non-Hermiticity gives rise to a {\em dynamically stable} regime, whereby all observables undergo bounded evolution in time, and a {\em dynamically unstable} one, whereby evolution is unbounded for at least some observables. We show that stability-to-instability transitions may be classified in terms of a suitably {\em generalized $\mathcal{P}\mathcal{T}$ symmetry},  which can be broken when diagonalizability is lost at exceptional points in parameter space, but also when degenerate real eigenvalues split off the real axis while the system remains diagonalizable. By leveraging tools from Krein stability theory in indefinite inner-product spaces, we introduce an indicator of stability phase transitions, which naturally extends the notion of phase rigidity from non-Hermitian quantum mechanics to the bosonic setting. As a paradigmatic example, we fully characterize the stability phase diagram of a bosonic analogue to the Kitaev-Majorana chain under a wide class of boundary conditions. In particular, we establish a connection between phase-dependent transport properties and the onset of instability, and argue that stable regions in parameter space become of measure zero in the thermodynamic limit. Our analysis also reveals that boundary conditions that support Majorana zero modes in the fermionic Kitaev chain are precisely the same that support stability in the bosonic chain. 
\end{abstract}


\submitto{\NJP}
\maketitle

\section{Introduction}

Systems of non-interacting (``free'') fermions or bosons have long provided a simple yet paradigmatic setting for investigating the differences that their statistical properties elicit in both equilibrium and non-equilibrium many-body physics \cite{ripka}. The relevant model Hamiltonians, which are {\em quadratic} in the respective canonical fermionic or bosonic operators, either describe truly non-interacting degrees of freedom or, more commonly, they may arise from simplified treatments of interaction effects, for instance via random-phase or mean-field approximations \cite{ringschuck}. At equilibrium, a striking manifestation of the different underlying statistics stems from the ability of bosons to exhibit Bose-Einstein condensation into a macroscopically occupied quantum state \cite{pitaevskiiBook}, as opposed to the formation of a sharp Fermi surface in systems non-interacting fermions at zero temperature. Away from equilibrium, the \textit{dynamical} behavior of free fermions and bosons can also be dramatically different, however \cite{ripka,colpa,RK}. The Heisenberg equations of motion for the canonical creation and annihilation operators can be described in terms of an {\em effective single-particle Hamiltonian} (SPH), whose associated Schr\"odinger-like equation is precisely the Bogoliubov de-Gennes equation. Solving this equation yields the normal modes, which determine the elementary excitations of the model, as well as the dynamics of any physical observable of interest. 

For both fermions and bosons, the effective SPH inherits a complex structure, the charge conjugation operation, arising from the fact that canonical creation and annihilation operators are mutually adjoint. Similarly, effective SPHs inherit a type of Hermiticity structure from underlying statistics. However, while fermionic effective SPHs are Hermitian with respect to the canonical inner product, bosonic effective SPHs are Hermitian with respect to an indefinite inner product -- namely, they are  ``pseudo-Hermitian'' \cite{ali,aliphrep} (see also \cite{pauli} for an early physical application). As a consequence, fermionic effective SPHs are always diagonalizable by a unitary Bogoliubov transformation and possess a purely real spectrum, with the corresponding normal modes always exhibiting periodic (bounded) time evolution and obeying canonical anti-commutation relations. From a dynamical-system standpoint, quadratic fermionic Hamiltonians can only be {\em dynamically stable}. In contrast, a bosonic effective SPH can possess non-real eigenvalues and lose diagonalizability at {\em exceptional points} (EPs) in parameter space, where both eigenvalues and eigenvectors coalesce \cite{nimrod}; the corresponding normal modes can display both oscillatory and non-oscillatory dynamics, leading to unbounded growth (or decay) of physical observables in time, and can satisfy a wide range of algebraic relationships beyond the canonical commutation relations. In other words, both dynamically stable and {\em dynamically unstable} behavior is possible in general for a many-body system described by a quadratic bosonic Hamiltonian (QBH). 

Physically, the emergence of effective non-Hermitian dynamics at the single-particle level is a direct manifestation of bosonic statistics in systems where particle-number conservation is broken at the many-body level.
Such a scenario is thus typical for systems of massless bosons, conspicuous examples of which could include photons \cite{milonniBook}, 
phonons \cite{kantorovichBook}, and magnons \cite{safonovBook}. 
Even for systems of massive bosons such as cold atoms, the mean-field quasiparticles and the small fluctuations of Bose-Einstein condensates are well described by non-particle-conserving QBHs \cite{pitaevskiiBook}. As a result, the onset of dynamical instabilities has been linked to a wide range of phenomena, including certain Goldstone modes \cite{ripka}, parametric amplification by coherent driving in photonic systems \cite{schneeloch}, decay mechanisms for atoms in optical lattices \cite{martik}, robustness properties of topological edge modes \cite{barnett,baldespeano}, the formation of spin domains in spinor Bose-Einstein condensates \cite{kawaguchi}, and and chiral mode switching in cavity QED systems \cite{Wiersig}.

Interest in the dynamical behavior of QBHs has heightened in recent years due to several reasons. On the one hand, topology plays an extraordinarily important role for fermions. Even in the absence of strong interactions, topological features of fermionic SPHs explain remarkable, robust physical effects like the integer quantum Hall that informs the current metrological standard for resistance. Along with the many successes of the topological classification of mean-field fermionic matter \cite{ryuRMP}, this is prompting researchers to look for roles that topology may play for free bosons.  
Although no conclusive notion of a topological classification has been established as yet (see however \cite{kawabata2,qiaoru} for up-to-date discussions on this fast-evolving subject), topologically non-trivial bands are by now well-documented for QBHs and a rigorous bulk-boundary correspondence has been identified, relating these bands to surface bands away from zero-frequency \cite{baldespeano,qiaoru}. Likewise, topology has been conjectured to constrain dynamical rather than thermodynamical properties of free bosons \cite{clerkPRX}, bringing to the fore the possibility that topological features might be most apparent in systems that need not be thermodynamically stable. On the other hand, the emergence of effective non-Hermitian dynamics makes QBHs natural candidates for the wealth of distinctive phenomena that non-Hermitian open dynamical systems are known to exhibit \cite{clerkPT}, including asymmetric mode switching, enhanced EP sensing, and non-Hermitian skin effect \cite{nimrod,NonHermReview2018,YaoSkin}. Meanwhile, alongside theoretical progress, a variety of photonic setups are becoming experimentally available as powerful platforms for simulating topological states of light and matter, and probing their statistical properties away from equilibrium \cite{topophrev}.

\begin{figure}[b!]
\centering
\includegraphics[width=0.85\textwidth]{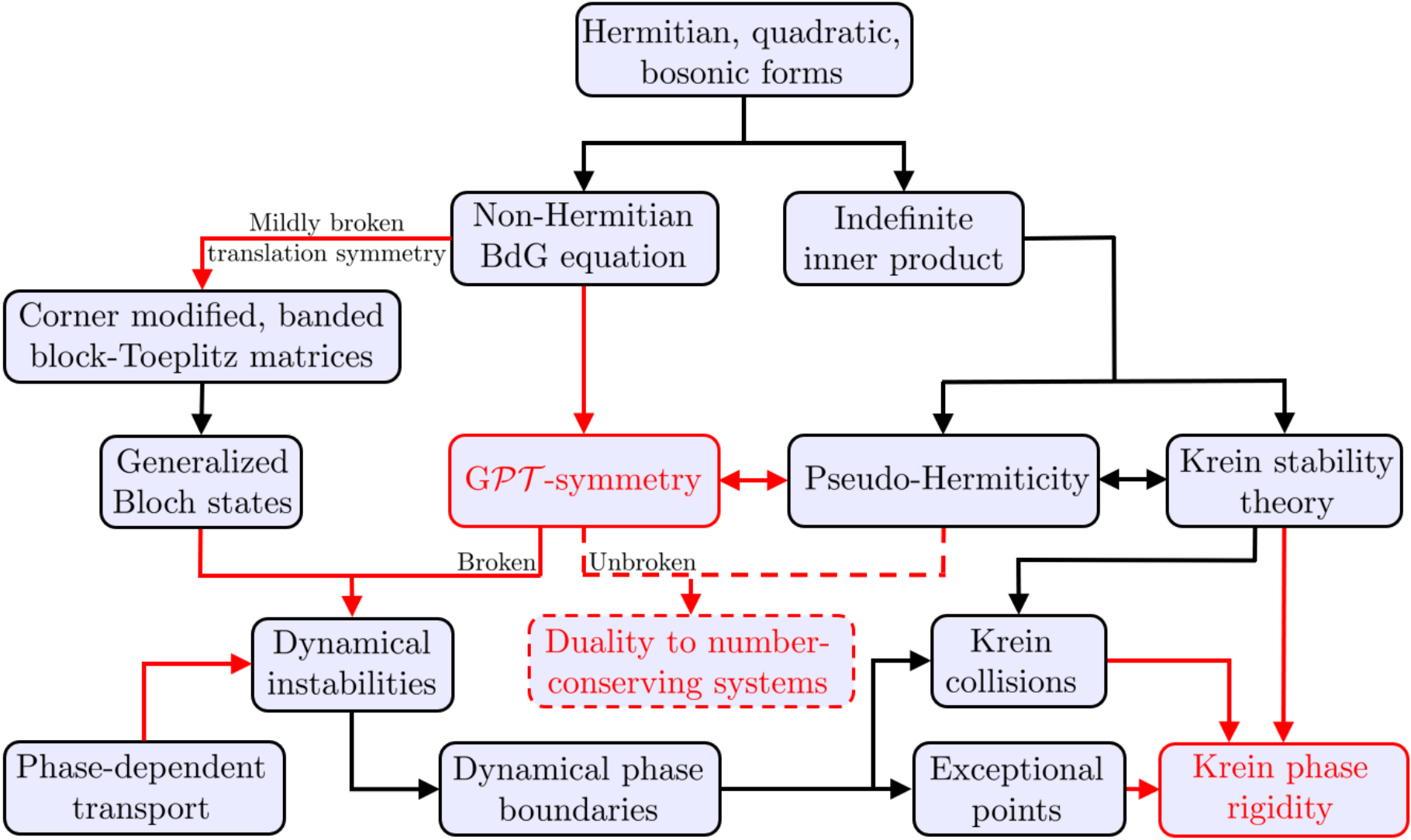}
\caption{Pictorial summary of the main concepts relevant to our analysis, along with their logical inter-connections. An arrow from box A to box B indicates that A leads naturally to B in some particular way. Black arrows (boxes) indicate connections (concepts) previously established in literature, while red arrows (boxes) indicate new connections (concepts) we established (introduced) in this paper. The red dashed arrows (and box) indicate connections and concepts we further elaborate on in a companion paper \cite{dualities}. }
\label{overview}
\end{figure}

Our goal in this paper is to obtain a deeper, unified understanding of the possible dynamical regimes that QBHs can support, and the ways in which transitions between different regimes may occur as parameters in the Hamiltonian and system size are changed. 
That is, for a given QBH, we aim to characterize its \textit{dynamical stability phase diagram}, 
for any given system size. We tackle this problem by combining tools from non-Hermitian quantum mechanics \cite{bender,aliphrep,nimrod} and fundamental results from linear algebra in indefinite inner-product spaces  \cite{gohberg} with a particular type of stability theory of linear time-invariant dynamical systems, known as ``Krein stability theory" \cite{yaku}. 
In essence, our work leads to the following general conclusions:
\begin{enumerate} 
\item All bosonic effective SPHs are {\em $\mathcal{P}\mathcal{T}$ symmetric} in a {\em suitable} 
sense;
\item Stability-to-instability transitions are associated with breaking of this generalized 
$\mathcal{P}\mathcal{T}$-symmetry;  
\item The transitions between dynamical phases can be detected by a new type of 
``phase rigidity" indicator that we call the {\em Krein phase rigidity} (KPR henceforth).
\end{enumerate}
With reference to Fig.\,\ref{overview} for a pictorial summary of the different elements that enter our analysis and their interconnections, let us put our main results above in context.

\smallskip

(i) \& (ii) {\it All bosonic effective SPHs possess a generalized $\mathcal{P}\mathcal{T}$ symmetry.---} There are many investigations into the connections between pseudo-Hermiticity, $\mathcal{P}\mathcal{T}$ symmetry, and more general antilinear symmetries in the mathematical physics literature. Motivated by the necessary conditions for a given diagonalizable non-Hermitian matrix to possess a real spectrum, it has been proposed that pseudo-Hermiticity should be regarded as the natural generalization of $\mathcal{P}\mathcal{T}$ symmetry \cite{ali,aliphrep}. As we emphasize in our work, the condition of pseudo-Hermiticity is geometric in origin: it is equivalent to self-adjointness with respect to a possibly indefinite inner product. A very recent investigation shows that $\mathcal{P}\mathcal{T}$ symmetry implies pseudo-Hermiticity, \textit{regardless of diagonalizability} \cite{Zhang}. In Sec. \ref{sub:PT}, we complete this line of reasoning, by proving that the converse is also true if one allows for a suitably modified, yet mathematically and physically reasonable, notion of $\mathcal{P}\mathcal{T}$ symmetry that we dub {\em generalized $\mathcal{P}\mathcal{T}$ (G$\mathcal{P}\mathcal{T}$) symmetry}. It follows that {\em all} bosonic effective SPHs, which are inherently pseudo-Hermitian as a reflection of bosonic canonical commutation relations, are G$\mathcal{P}\mathcal{T}$-symmetric. In particular, one can always classify bosonic dynamical phases according to whether this G$\mathcal{P}\mathcal{T}$ symmetry is broken or not. Physically, as we also show, unbroken G$\mathcal{P}\mathcal{T}$ symmetry is intimately connected to the fact that all normal modes may be chosen to obey canonical commutation rules and a normalizable Fock vacuum may be constructed, even for thermodynamically unstable systems. In fact, dynamical stability allows one to identify a dual, number-conserving Hamiltonian that is unitarily equivalent to the original, as we explicitly show in a separate study \cite{dualities}. 

\smallskip

(iii) \textit{Krein Phase Rigidity.---} One of the hallmarks of non-Hermitian quantum mechanics are EPs. The standard indicator of EPs for complex symmetric non-Hermitian Hamiltonians is a scalar quantity called the ``phase rigidity" \cite{PR1,PR2,rotter}. However, transitions between dynamical phases can occur without EPs. In Sec. \ref{Krein}, we introduce a KPR indicator which, in addition to detecting EPs, also leverages the indefinite inner-product geometry specific to bosons to detect transitions caused by the splitting of degenerate real eigenvalues into non-real ones, that entail no loss of diagonalizability -- a so-called ``Krein collision''. In this sense, our KPR is a powerful extension of the phase rigidity to bosonic many-body systems, reducing to it whenever the relevant effective SPH happens to be complex-symmetric. Since there exist protocols for simulating non-Hermitian Hamiltonians of varied physical origin using photonic arrays \cite{clerkPT}, this establishes one of many possible routes for extending the applicability of our techniques beyond systems of free bosons. The ability of the KPR to characterize stability phase diagrams and their phase boundaries (which are defined by G$\mathcal{P}\mathcal{T}$ symmetry breaking) is explicitly demonstrated through simple illustrative examples in both a single-mode and a two-mode cavity QED setting. 

\smallskip

To further exemplify the application of our general theory in a more complex scenario, in the second half of the paper (Sec. \ref{BKC}) we present a detailed investigation of the dynamical stability phase diagram of the {\em bosonic Kitaev-Majorana chain} (BKC) introduced in \cite{clerkPRX}. Aiming to a fuller comparison between the fermionic and bosonic models and the interplay between bulk and boundary, we explore a broader family of boundary conditions (BCs) that interpolates smoothly between open and periodic and allows for an arbitrary twisting angle. Most of our work on the BKC is analytical or supported by analytical results, which is important in several respects. For example, it can be difficult to distinguish numerically whether a stability transition point is associated to loss of diagonalizability of the effective SPH (an EP), or a Krein collision, or both. Another difficult but crucial problem is that of calculating the dynamical stability phase diagram in the limit of infinite system size. While dynamical stability phase diagrams and thermodynamic phase diagrams are \textit{a priori} very different objects, it is sensible to ask whether the limit of infinite system size can highlight gross regularities within the class of all possible dynamical phase diagrams. Our analysis for the BKC suggests that the answer is likely in the affirmative. For both finite and infinite size, the key to accessing spectral properties analytically is the exact diagonalization algorithm for {\em corner-modified block-banded Toeplitz matrices} of \cite{JPA}. In the context of fermions, this algorithm yields a generalized Bloch's theorem for clean systems under arbitrary boundary conditions (BCs) \cite{GBT}. Here, we explicitly apply this algorithm to bosonic SPHs for the first time, which in itself constitutes a new result of independent interest.

Finally, as we remarked above, in \cite{clerkPRX} the authors conjecture that certain topological properties of the BKC may be responsible for its peculiar dynamical features --  including the strong sensitivity to BCs and the ability to propagate excitations in a ``chiral'' fashion, depending on the phase of the excitation. The BKC model itself is constructed by applying a particular mapping from free fermions to free bosons to the fermionic Kitaev chain at zero chemical potential. In Sec. \ref{furtherimp}, we characterize this mapping in full generality and show that can it only preserve certain topological invariants at the cost of mapping fermions to either dynamically unstable bosons or bosons that are at the cusp of instability. For the BKC, this is precisely the difference between closed (unstable) and open (at the cusp of instability) BCs. Hence, Krein stability theory explains the fragility of the dynamically stable phase of the open BKC against bulk disorder observed in \cite{clerkPRX}. Altogether, our analytical solutions point to a remarkable correspondence between (open and $\pi/2$-twisted) BCs and parameter regimes that support Majorana zero modes in the fermionic Kitaev-Majorana chain versus, respectively, dynamical stability in its bosonic counterpart. Further to that, as we outline in Sec. \ref{furtherimp}, bosonic models may also be constructed, which host boundary-localized, Hermitian zero-frequency analogs to Majorana modes -- although, ascertaining a clear connection with topology remains a fascinating topic for 
further research.

\section{Background}
\label{Background}

QBHs have been extensively considered in the literature as integrable models for both thermodynamically stable systems \cite{colpa,ripka,ringschuck} and beyond \cite{RK,oriol,RG,namDiag2016,Sred2019}. In this section, after introducing the basic concepts, we provide a self-contained review of the mathematical 
techniques needed for casting these Hamiltonians in normal form, to a level of detail appropriate for subsequent analysis.

\subsection{From quadratic bosonic Hamiltonians to effective non-Hermitian dynamics}

Let $a_i^\dag$ and $a_i$ denote canonical bosonic creation and annihilation operators for a mode $i$, satisfying $[a_i,a_j]= 0= [a_i^\dag,a_j^\dag]$ and $[a_i,a_j^\dag]=\delta_{ij}1_F$, with $1_F$ denoting the identity operator on Fock space. A QBH is an operator on Fock space of the form  
\begin{equation}
\label{genham}
\widehat{H} = \sum_{i,j=1}^N K_{ij}a_i^\dag a_j + \frac{1}{2}\left( \Delta_{ij} a_i^\dag a_j^\dag +\Delta_{ij}^* a_i a_j \right) , 
\qquad K_{ij}, \Delta_{ij}\in {\mathbb C}.
\end{equation}
The requirement that $\widehat{H}$ be Hermitian imposes the condition $K_{ij} = K_{ji}^*$. Furthermore, the bosonic commutation relations allow us to take $\Delta_{ij}=\Delta_{ji}$. In terms of the Nambu (column) array $\hat{\Phi} \equiv [a_1,a_1^\dag \ldots, a_N,a_N^\dag]^T$, the above QBH can be re-written as $\widehat{H} = \frac{1}{2}\hat{\Phi}^\dag H \hat{\Phi}-\frac{1}{2} \tr K$, with
\begin{equation}
\label{spH}
H = \left[\matrix{
h_{11} & \cdots & h_{1N} \cr \vdots & \ddots & \vdots \cr h_{N1} & \cdots & h_{NN}
} \right] =H^\dagger, 
\qquad 
h_{ij} = \left[\matrix{
K_{ij} & \Delta_{ij} \cr \Delta_{ij}^* & K_{ij}^*}\right].
\end{equation}
Note that our definition of Nambu arrays differs from the standard ordering $\hat{\Phi}'\equiv [a_1,\ldots,a_N,a_1^\dag,\ldots,a_N^\dag]^T$. While this rearrangement clearly does not affect the physics, it will be instrumental to bring the role of translation invariance to the fore and enable a more straightforward application of the diagonalization techniques to be employed in Sec.\,\ref{GBTprimer}. Beside being Hermitian, the matrix $H$ in Eq.\,(\ref{spH}) obeys the condition 
\begin{equation}
\label{xsym}
H^* = \tau_1 H \tau_1 , \qquad \tau_1 \equiv \mathds{1}_N \otimes \left[\matrix{
0 & 1 \cr 1 & 0
}\right] \equiv \mathds{1}_N\otimes \sigma_1.
\end{equation}
We will also denote $\tau_2\equiv \mathds{1}_N\otimes \sigma_2$ and $\tau_3\equiv\mathds{1}_N\otimes \sigma_3$, 
with $\sigma_j, j=1,2,3$, being the usual Pauli matrices.  

While formally $H$ plays the role of a SPH, a key difference between bosonic and fermionic quadratic forms arises from the fact that diagonalizing $H$ does {\em not}, in general, imply diagonalization of $\widehat{H}$ \cite{ripka}, nor does it characterize the dynamics that $\widehat{H}$ generates. The latter is determined by the solution of the Heisenberg equations which, in units where $\hbar=1,$ may be compactly written as 
\begin{equation}
\label{Heis}
i\frac{d}{dt}\hat{\Phi}(t) = -[\widehat{H},\hat{\Phi}(t)] \equiv  G\hat{\Phi}(t), \qquad G =\tau_3 H .
\end{equation} 
A more transparent way to interpret this equation may be obtained by considering an arbitrary (column) vector 
$\ket{\alpha}\in {\mathbb C}^{2N}$ and define $\widehat{\alpha} \equiv \bra{\alpha} \tau_3 \hat{\Phi}$ (e.g., for a single mode, $\widehat{\alpha} = \alpha_1^* a_1 -\alpha_2^* a_1^\dagger$).  It then follows that (i) many-body commutators induce a geometric structure on single-particle space, in the sense that $[\widehat{\alpha},\widehat{\beta}^\dag] = \braket{\alpha|\tau_3|\beta}1_F$; and (ii) the many-body adjoint operation similarly induces a charge-conjugation operation, $\widehat{\alpha}^\dag = -\widehat{\mathcal{C} \alpha}$, with $\mathcal{C}\equiv \tau_1\mathcal{K}=\mathcal{C}^{-1}$ and $\mathcal{K}$ being complex conjugation. In order to exemplify the single-particle structure of the dynamics in Eq.\,(\ref{Heis}), we further define the convention $\widehat{\alpha}(t) \equiv  \bra{\alpha(t)}\tau_3\hat{\Phi}(0)$, that is, we absorb the time-dependence of the bosonic operators $a_j,a_j^\dag$ into the vector of coefficients $\ket{\alpha}$. Then, by taking advantage of the identity \(-[\widehat{H},\widehat{\alpha}(t)] = \widehat{G\alpha}\), one finds that $\widehat{\alpha}(t)$ satisfies the Heisenberg equation of motion if and only if
\begin{equation}
\label{Heis2}
\frac{d}{dt}\ket{\alpha(t)} = i G \ket{\alpha(t)}.
\end{equation}
Some symmetries of this single-particle equation are inherited from, and/or can be can be lifted back, to the original many-body problem. For example, \(\mathcal{C}G=-G\mathcal{C}\) in terms of the charge conjugation operation identified above. As for ordinary, commuting symmetries, suppose \(U\) is a ``\(\tau_3\)-unitary" (or para-unitary)
matrix, that is, $U^{-1}=\tau_3 U^\dag \tau_3$ such that \([G,U]=0\). Then, \(U \ket{\alpha(t)}\) 
solves equation Eq.\,\eref{Heis2} provided \( \ket{\alpha(t)}\) does and $\widehat{H}$ is invariant under the canonical 
transformation $\hat{\Phi}\mapsto U \hat{\Phi}$. Similarly, if $[G,\Theta]=0$ for some anti-$\tau_3$-unitary operator, 
that is, $\Theta=U\mathcal{K}$ with $U$ $\tau_3$-unitary and $\mathcal{K}$ complex conjugation, 
then \(\Theta \ket{\alpha(-t)}\) solves Eq.\,\eref{Heis2} provided \( \ket{\alpha(t)}\) does and
$\widehat{H}$ is invariant under the the anti-linear transformation $c\Phi\mapsto c^*U\Phi$. 

According to Eq.\,\eref{Heis2}, the dynamics of a free bosonic system are governed by the \textit{effective SPH} $G$ which need {\em not} be Hermitian or even normal (in fact, $G$ is Hermitian if and only if the ``pairing" contribution $\Delta_{ij}$ vanishes). As a consequence, $G$ can have non-real eigenfrequencies as well as non-trivial Jordan chains. Since, from a dynamical-system standpoint, Eq.\,(\ref{Heis2}) defines a linear time-invariant system with state matrix $-iG$, the normal modes of the dynamics generated by $\widehat{H}$ are built up from the (generalized) eigenvectors of $G$. Two distinct notions of stability are then relevant for the QBHs in question: 
\begin{enumerate}
\item The system $\widehat{H}$ is \textit{dynamically stable} if $G$ is diagonalizable and all of its eigenvalues are {\em real}. 
\item The system $\widehat{H}$ is \textit{thermodynamically stable} if there exists a {\em finite} lower bound on the expectations of $\widehat{H}$. 
\end{enumerate}
On the one hand, the normal modes of a dynamically stable QBH exhibit {\em strictly bounded motion} \cite{baldespeano}: the onset of dynamical instability is signaled by the appearance of a non-trivial Jordan chain or a complex eigenvalue in the eigensystem of $G$ \footnote{Note that the condition of all eigenvalues of $G$ being real is stronger than the standard (Hurwitz) stability condition for the linear system in Eq.\,\eref{Heis2}, which would only require every eigenvalue to have a strictly positive imaginary part. This stems from the symmetry structure that the bosonic nature of the many-body problem imposes on the eigenvalue spectrum, see Sec.\,\ref{normalform}.}. 
On the other hand, the notion of thermodynamic stability matters because a mean-field ground state need not exist in the bosonic Fock space: for example, if there is a bosonic excitation of negative energy, in the thermodynamic limit one may occupy it with an arbitrary number of bosons and lower the energy of the system without ever hitting a ground state. Such an instability is sometimes called a \textit{Landau instability} \cite{kawaguchi} and we will see an example in Sec.\,\ref{BKC}. 

Thermodynamic stability can be diagnosed in terms of the the Hermitian matrix \(H=\tau_3G\): If \(H\) is positive semi-definite, 
then $\widehat{H}$ is thermodynamically stable \cite{ripka,colpa}. The converse implication is more subtle
\cite{namDiag2016,qiaoru}.  In particular, if $H \geq 0$ and $G$ is 
diagonalizable, then the associated QBH is both thermodynamically and dynamically stable. Nonetheless,
the two notions of stability are \textit{independent}, as one can see from the following examples: (1) A system 
of $N$ decoupled harmonic oscillators is thermodynamically and dynamically stable. (2) A  quantum harmonic 
chain with nearest-neighbor (NN) couplings (one-dimensional acoustic phonons) is thermodynamically stable 
and displays a non-trivial Jordan chain associated to the conserved total momentum and center-of-mass operators. 
Hence, it is dynamically unstable. (3) As we will see in Sec. \ref{BKC}, there are parameters regimes of the bosonic 
analogue of the Kitaev chain introduced in \cite{clerkPRX}, in which the model is dynamically yet not thermodynamically stable.

\subsection{Normal form of a quadratic bosonic Hamiltonian}
\label{normalform}

In order to elucidate how the eigensystem of $G$ determines the normal modes of Eq.\,\eref{Heis} and a normal form of the corresponding QBH, we start from identifying some intrinsic symmetry properties that the spectrum of $G$ enjoys. Firstly, Hermiticity of $H$ implies that $G^\dag = \tau_3 G \tau_3$. As a consequence, if $\omega_n$ is an eigenvalue of $G$, then so is $\omega_n^*$. Secondly, the charge conjugation symmetry \(\mathcal{C}G=-G\mathcal{C}\) is equivalent to $G^*= -\tau_1 G \tau_1$. Hence, if $\omega_n$ is an eigenvalue of $G$, then so is $-\omega_n^*$. Accordingly, the eigenvalues of $G$ come in quartets $\{\omega_n,\omega_n^*,-\omega_n,-\omega_n^*\}$. These two properties also have consequences on the eigenvectors of $G$ and $G^\dag$. Specifically, let $\ket{\psi_n}$ be an eigenvector 
of $G$ associated to the eigenvalue $\omega_n$. Then,
(i) $\tau_3\ket{\psi_n}$ is an eigenvector of $G^\dag$ with eigenvalue $\omega_n$, and 
(ii) $\mathcal{C}\ket{\psi_n}$ is an eigenvector of $G$ with eigenvalue $-\omega_n^*$. Finally, note that if $\omega_n\not\in\mathbb{R}$, then $\braket{\psi_n|\tau_3|\psi_n}=0$.

\subsubsection{Diagonalizable case.} 
\label{sub:diag}
Suppose that \(G\) is diagonalizable, in which case the eigenvectors of \(G\) form a complete basis of $\mathbb{C}^{2N}$. Then, aside from minor notational differences, we may repeat the analysis in \cite{RK}.  Let $\ket{\psi_{n*}}$ denote the eigenvector of $G$ with eigenvalue $\omega_n^*$ if $\textup{Im}(\omega_n)\neq 0$, and $\ket{\psi_{n*}} = \text{sgn}\left(\braket{\psi_n|\tau_3|\psi_n}\right) \ket{\psi_n}$ otherwise. There are $2N$ eigenvectors $\ket{\psi_n}$ of $G$, which correspond to eigenvalues $\omega_{n},$ $n=1,\ldots, 2N$, and satisfy $\braket{\psi_{n*}|\tau_3|\psi_m}=\delta_{nm}$, that is, there exists a \(\tau_3\)-orthonormal basis. In terms of this basis, 
\begin{equation*}
G = \sum_{n=1}^{2N} \omega_n \! \ket{\psi_n}\bra{\psi_{n*}} \tau_3 .
\end{equation*}
This spectral decomposition of \(G=\tau_3H\) leads to the desired normal form of the QBH, namely: 
\numparts\begin{eqnarray}
\label{diagnf}
\widehat{H} &=& \frac{1}{2}\sum_{n=1}^{2N} \omega_n \widehat{\psi}_n^\dag \widehat{\psi}_{n*} - \frac{1}{2}\tr K, \\
\widehat{\psi}_n^\dag &=& \hat{\Phi}^\dag \tau_3 \ket{\psi_n} = -\bra{\widetilde{\psi}_n}\tau_3 \hat{\Phi} ,\quad \widehat{\psi}_{n*} = \bra{\psi_{n*}}\tau_3\hat{\Phi} = - \hat{\Phi}^\dag \tau_3 \ket{\widetilde{\psi}_{n*}},
\end{eqnarray}\endnumparts
where $\ket{\widetilde{\psi}_n}\equiv \mathcal{C}\ket{\psi_n}$, $[\widehat{\psi}_{n*},\widehat{\psi}_m^\dag]= \delta_{nm}1_F$, and $[\widehat{\psi}_{n*},\widehat{\psi}_{m*}]=0= [\widehat{\psi}_{n},\widehat{\psi}_{m}]$.  At this point the analysis 
splits depending on whether \(\omega_n\) is or is not real. If $\omega_n \not \in {\mathbb R}$, the pair $(\widehat{\psi}_{n},\widehat{\psi}_{n*}^\dag)$ is called a \emph{canonical pseudo-bosonic normal mode} \cite{RK} and there is no room for further simplification. By contrast, if $\omega_n \in {\mathbb R}$, then one can choose the eigenvectors so that  $\ket{\psi_{n*}}\propto \ket{\psi_n}$ and 
$\ket{\widetilde{\psi}_n}$ is an eigenvector with eigenvalue $-\omega_n$. Furthermore,  
$\braket{\psi_n|\tau_3|\psi_n}=-\braket{\widetilde{\psi}_n|\tau_3|\widetilde{\psi}_n}\neq 0$ \footnote{More precisely, if $\omega_n\neq 0$, we have $\braket{\psi_n|\tau_3|\psi_n} = \omega_n^{-1}\braket{\psi_n|\tau_3 G|\psi_n} = \omega_n^{-1}\braket{\psi_n|H|\psi_n} \neq 0$. The case $\omega_n=0$ requires separate consideration; see \cite{colpa,qiaoru} for a self-contained study of the zero subspace.}. 
Hence, one can renormalize these states so that, without loss of generality, $\braket{\psi_n|\tau_3|\psi_n}  =1$. This procedure
yields a term of the form $\omega_n \widehat{\psi}_n^\dag \widehat{\psi}_n$ in the normal form of 
$\widehat{H}$. The pair $(\widehat{\psi}_n,\widehat{\psi}_n^\dag)$ is a \textit{bosonic normal mode}, that is, it 
satisfies the canonical commutation relation $[\widehat{\psi}_n,\widehat{\psi}_n^\dag]=1_F$.  

The transformation  $\hat{\Phi} \mapsto \hat{\Psi} \equiv [\widehat{\psi}_{1*}, \widehat{\psi}_{1}^\dag,\ldots,\widehat{\psi}_{N*}, \widehat{\psi}_{N}^\dag]^T$ from the physical bosonic modes $\hat{\Phi}$ to the normal modes of the system
is furnished by the modal matrix $M$, whose columns are the eigenvectors of $G$, that is, $\hat{\Psi} = M^{-1} \hat{\Phi}$. Inverting this transformation allows us to find the desired time dependence of $\hat{\Phi}$. Explicitly, 
\begin{equation*}
\widehat{\psi}_{n*}(t) = e^{-i\omega_n^* t} \,\widehat{\psi}_{n*}(0), \qquad \widehat{\psi}_{n}(t) = 
e^{-i\omega_n t}\, \widehat{\psi}_{n}(0).
\end{equation*}
Hence, the system is dynamically stable only if all the eigenvalues are real (we have excluded non-trivial
Jordan chains by assumption). Otherwise, some normal modes are amplified exponentially with time, which 
can happen for example in parametrically driven optical systems. Each amplified mode is paired with an 
exponentially decaying (de-amplified) mode. Such a system is thermodynamically stable if it is dynamically 
stable and all the \textit{creation} operators are associated to \textit{positive} eigenfrequencies. 

\subsubsection{General case.} 
\label{sub:general}
If $G$ is not diagonalizable, at least one eigenvalue of $G$, say, $\omega_0$, must be associated to a non-trivial Jordan chain. Allowing for degeneracy, let $\ket{\chi_{j1}}$ for $j=1,\ldots,N_0$ denote a complete set of independent eigenvectors corresponding to \(\omega_0\).  Moreover, for each $j$, let $r_j >1$ be the length of the corresponding Jordan chain, satisfying
\begin{equation}
\mkern-58mu (G-\omega_0 \mathds{1}_{2N})\ket{\chi_{j1}} = 0, \qquad (G-\omega_0 \mathds{1}_{2N})\ket{\chi_{jk}} = \ket{\chi_{j(k-1)}},\quad 2\leq k \leq r_j . 
\end{equation}
We can then see that the normal modes defined by $\widehat{\chi}_{jk}\equiv \bra{\chi_{jk}}\!\tau_3\hat{\Phi}$ manifest as Jordan chains in the adjoint action of $\widehat{H}$, namely, 
\begin{equation}\label{jcs}
\!\!\!-[\widehat{H},\widehat{\chi}_{j1}] =\omega_0 \widehat{\chi}_{j1},\qquad -[\widehat{H},\widehat{\chi}_{jk}] = \omega_0 \widehat{\chi}_{jk}+ \widehat{\chi}_{j(k-1)},
\quad 2\leq k \leq r_j,
\end{equation}
and exhibit exponentially modulated polynomial time evolution,
\begin{equation}
\label{generictevo}
\widehat{\chi}_{jk}(t) = e^{-i\omega_0 t}\sum_{\ell=0}^{k-1}\frac{(-i t)^\ell}{\ell!}\widehat{\chi}_{j(k-\ell)}(0).
\end{equation}
Hence, dynamical stability is excluded by non-trivial Jordan chains but, as it turns out, thermodynamic stability is not.

Let us call the modes $\widehat{\chi}_{jk}$, $k>1$, generalized normal modes of rank $k$. As we saw, one can enforce bosonic or pseudo-bosonic commutation relations for the diagonalizable sectors \cite{RG} (i.e., $r_j=1$).
By contrast, the commutation relations of the generalized normal modes are more difficult to pin down. To gain physical insight into this problem, let us first focus on thermodynamically stable systems. When $H\geq 0$, it is known that there can only be Jordan chains at $\omega_0=0$ and their lengths are at most two (see Theorem 5.7.2 in \cite{gohberg}). The eigenvector and generalized eigenvector of a given Jordan chain can then be used to create pair of Hermitian normal modes satisfying Heisenberg-Weyl commutations relations. Physically, these Jordan chains often represent Goldstone modes \cite{ripka}. Mathematically, the situation is neatly exemplified by a single mode with Hamiltonian \(\widehat{H}=p^2/2m\): then, \(p\) is the zero mode and any combination \(c_1x+c_2p\), with \(c_1\neq 0\), can officiate as the generalized zero mode. Naturally, with hindsight, one chooses \(c_1=1, c_2=0\), but there is nothing in the problem that makes this choice canonical. At best, one can require 
invariance under the charge conjugation operation and restrict \(c_1,c_2 \in {\mathbb R}\). More generally (regardless of thermodynamic stability), since a generalized eigenvector can always be shifted by a constant multiple of an eigenvector, the commutation relations of the corresponding normal modes may obey a wide range of commutation relations. Notably, this ambiguity is employed in \cite{RK}  to construct bosonic generalized normal modes at a non-zero frequency. 

While the commutation relations of the normal modes are malleable, it is important to note that the evolution of observables is {\em independent} of the particular generalized eigenbasis used to determine their time dependence by way of Eq.\,\eref{generictevo}. For concreteness, one can fix the commutation relations of a given Jordan chain by utilizing a particular Jordan normal form 
available for matrices satisfying $G^\dag = \tau_3 G \tau_3$. By Theorem 5.1.1 in \cite{gohberg}, we then know there exists an invertible matrix $T$, such that $J=TGT^{-1}$ is a Jordan normal form for $G$ and $\tau_3 = T^\dag P T$, with $P$ a block-diagonal matrix whose $j$-th block is an $r_j\times r_j$ matrix,  
and either all $1$'s or $-1$'s on the anti-diagonal and $0$'s elsewhere. By applying $T^{-1}$ to the canonical basis of $\mathbb{C}^{2N}$, we can construct Jordan chains at eigenfrequency $\omega_j$ satisfying the orthonormality condition $\braket{\chi_{jk}|\tau_3|\chi_{\ell p}} = \varepsilon_j\delta_{j\ell_*}\delta_{k,r_j+1-p}$, with $\varepsilon_j\in\{-1,1\}$ and $\ell_*$ the index labeling a 
Jordan chain at $\omega_\ell^*$. 
The commutation relations of the corresponding normal modes then satisfy $[\widehat{\chi}_{jk},\widehat{\chi}_{\ell p}^\dag]=\varepsilon_j\delta_{j\ell_*}\delta_{k,r_j+1-p}1_F$. The commutators $[\widehat{\chi}_{jk},\widehat{\chi}_{\ell p}] = -
\braket{\chi_{jk}| \tau_2 \mathcal{K}    |\chi_{\ell p}}1_F$,
can then be determined on a case-by-case basis.

\subsection{Diagonalization of corner-modified, banded block-Toeplitz matrices}
\label{GBTprimer}

As mentioned in the Introduction, our main goal is to develop a general theory of the possible dynamical behaviors of QBHs, by shedding light, in particular, on the extreme sensitivity of their dynamical response to BCs.
A key tool for our analysis is an exact diagonalization procedure for \(G\) (loosely speaking, since \(G\) need not be diagonalizable) which, while originally developed with fermions in mind \cite{JPA,GBT,PRL}, works more generally for clean (disorder-free) quadratic Hamiltonians subject to arbitrary BCs. The key property is that \(G\) belongs to a class of structured matrices known as {\em corner-modified, banded block-Toeplitz} (BBT). 
In what follows, we summarize the essential steps of this diagonalization procedure both to explicitly show-case its first application to bosons and to make the presentation as self-contained as possible. 
For clarity, technical details are deferred to \ref{GBTappendix}. 

Corner-modified BBT matrices arise naturally from QBHs whose couplings have finite range and possess translation invariance ``up to a boundary''. While the framework developed in \cite{JPA,GBT,PRL} is more general (e.g., it allows for multiple internal modes), we specialize it here to a one-dimensional lattice of $N$ sites with bosonic modes $(a_j,a_j^\dag)$ attached to each site, and define $\hat{\phi}_j = [a_{j}, a_{j}^\dag]^T$, $1\leq j \leq N$. The Nambu array of Sec.\,\ref{Background} then reads $\hat{\Phi}= [\hat{\phi}_1,\ldots,\hat{\phi}_N]^T$, and the relevant class of QBHs has the form 
\begin{equation}
\label{GBTHams}
\widehat{H} = \frac{1}{2}\sum_{r=0}^R  \sum_{j=1}^{N-r} \hat{\phi}_j^\dag h_r \hat{\phi}_{j+r} + \sum_{b,b'} \hat{\phi}^\dag_b W_{bb'} \hat{\phi}_{b'}+\text{H.c.}, 
\end{equation}
where $b,b'\in\{1,\ldots,R,N-R+1,\ldots,N\}$, and $h_r$, $W_{bb'}$ are $2\times 2$ matrices that couple the bosonic modes at different sites in the ``bulk'' and ``boundary'' respectively, with range $R<N/2$. In practice, one often has $R \ll N$ (e.g., $R=1$ for NN couplings). As one can see from the structure of $\widehat{H}$ in Eq.\,\eref{GBTHams}, translation invariance is maintained up to a boundary slab of thickness $R$. Hence, we speak of \textit{mildly broken translation invariance} in these systems. The effective SPH associated to Eq.\,\eref{GBTHams}
can be split as $G=G_O + V$, with 
\begin{equation}
\fl G_O \equiv  \left[\matrix{
g_0  & \cdots & g_R & & & 0 & \cdots & 0
\cr
\vdots & \ddots & & \ddots & & & \ddots & \vdots
\cr
g_{-R} & & \ddots & & \ddots & & & 0
\cr
 & \ddots & & & & & & 
\cr
  & & & & & & \ddots&
\cr
0 & & & \ddots & & \ddots & & g_R
\cr
\vdots & \ddots & & & \ddots & & \ddots & \vdots
\cr
0 & \cdots & 0 & & & g_{-R} & \cdots & g_0
}\right], 
\hspace{.2em}
V \equiv 
\left[\matrix{
v_{11}^{(l)} & \cdots & v_{1R}^{(l)} & & 0 & & v_{11}  & \cdots & v_{1R} 
\cr
\vdots & \ddots & \vdots & & \vdots & & \vdots & \ddots & \vdots 
\cr
v_{R1}^{(l)} & \cdots & v_{RR}^{(l)} & & \vdots & & v_{R1} & \cdots & v_{RR}
\cr
 & & & & & & & &
\cr
0 & \cdots & \cdots & & 0 & & \cdots & \cdots & 0
\cr
 & & & & & & & &
 \cr
v_{11}^\dag & \cdots & v_{1R}^\dag & & 0 & & v_{11}^{(r)}  & \cdots & v_{1R} ^{(r)}
\cr
\vdots & \ddots & \vdots & & \vdots & & \vdots & \ddots & \vdots 
\cr
v_{R1}^\dag & \cdots & v_{RR}^\dag & & 0 & & v_{R1}^{(r)}  & \cdots & v_{RR} ^{(r)}
}\right],
\end{equation} 
and where we have 
\begin{equation*}
 \fl g_r = \sigma_3 h_r, \quad g_{-r} = \sigma_3 h_r^\dag = \sigma_3 g_r^\dag \sigma_3,
\quad
v_{bb'}^{(l)} = \sigma_3 W_{bb'}, \quad v_{bb'}^{(r)} = \sigma_3 W_{N-b+1,N-b'+1}, \quad v_{bb'} = \sigma_3 W_{b,N-b'+1}.
\end{equation*} 
Mathematically, $G_O$ is an example of a BBT matrix, that is, a block matrix whose entries are constant along diagonals. It acts naturally on the tensor-product space \(\mathds{C}^N\otimes\mathds{C}^2\equiv \mathcal{H}_L\otimes \mathcal{H}_I\), where the first (second) factor carries the lattice (internal) degrees of freedom. The bandwidth of $G_O$, that is, the number of non-zero diagonals, is $2R+1$. Physically, one recognizes $G_O$ as the effective SPH of the system subject to open BCs. 

The role of (mildly broken) translation invariance becomes more apparent if we define the \textit{left-shift operator}, $T \equiv \sum_{j=1}^{N-1} \ket{j}\bra{j+1}$, which acts only on $\mathcal{H}_L$ and is the finite-lattice truncation of the translation operator $\bm{T}\equiv \sum_{j\in\mathbb{Z}}\ket{j}\bra{j+1}$.  Importantly, $\bm{T}$ is taken to act on the infinite lattice space $\spn\{\ket{j}\}_{j\in\mathbb{Z}}$, without the corresponding $\ell^2$-inner product; thus, while remaining invertible, $\bm{T}$ is no longer unitary and may possess generalized eigenvectors in general. Using shift operators, the above \(G_O\) rewrites as
\begin{equation}
G_O = \mathds{1}_N \otimes h_0 + \sum_{r=1}^R \Big( T^r \otimes g_r + (T^\dag)^r \otimes g_{-r}\Big).
\label{leftshift}
\end{equation}
The effective SPH \(\bm{G}\) of the translation-invariant system is recovered by replacing \(T\) with $\bm{T}$. The matrix $V$ is, in turn, a corner modification that encodes BCs other than open. Periodic BCs, for example, can be recovered by a suitable choice of \(V\) \cite{GBT}. In the following, we will also need the matrix polynomial \( G(z)=\mathds{1}_N \otimes h_0 + \sum_{r=1}^R \left( z^r \otimes g_r + z^{-r} \otimes g_{-r}\right), \) which is the analytic continuation of the effective single-particle Bloch Hamiltonian off the Brillouin zone.

Next, let us define the \textit{bulk} and \textit{boundary projectors} as $P_B\equiv \sum_{j=R+1}^{N-R}\ket{j}\bra{j}\otimes \mathds{1}_2$ and $P_\partial = \mathds{1}_{2N} - P_B$, respectively. The goal is to solve the eigenvalue equation $G_O\ket{\psi} = \omega \ket{\psi}$. Since $P_B + P_\partial = \mathds{1}_{2N}$ and $P_B V = 0$, the eigenproblem is equivalent to the following ``bulk-boundary system of equations" \cite{PRL}:
\numparts
\begin{eqnarray}
&P_B G_O \ket{\psi}& = \omega P_B\ket{\psi} , \label{BBbulk} \\
P_\partial (G_O &+ V)\ket{\psi}& = \omega P_\partial \ket{\psi} . 
\label{BBbdry}
\end{eqnarray}\endnumparts
The diagonalization proceeds by first solving the \textit{bulk equation}, Eq.\,\eref{BBbulk}, parametrically in \(\omega\), and then employing the resulting solutions as an Ansatz for the \textit{boundary equation}, Eq.\,\eref{BBbdry}. One can show that, generically, such a strategy yields {\em all} of the eigenvectors of \(G_O+V\) and can also be applied for computing generalized eigenvectors \cite{JPA}.  

For fixed $\omega\in\mathbb{C}$, the complete set of solutions to the bulk equation \eref{BBbulk} breaks up into three different types of solutions (for a derivation, see \ref{GBTappendix}). Solutions of the first type are obtained by restricting to the finite-lattice solutions of the translation-invariant equation $(\bm{G}-\omega)^n\Psi=0$, for some suitable \(n\), and thus arise from eigenvectors and generalized eigenvectors of \(\bm{G}\). Specifically, these solutions take the form
\begin{equation*}
\ket{\psi_{\ell s}} = \sum_{\nu=1}^{s_\ell}\ket{z_\ell,\nu}\otimes \ket{u_{\ell s v}} ,
\end{equation*}
where the \(z_\ell\) are the roots of the equation \(\det(G(z)-\omega)=0\) with algebraic multiplicity \(s_\ell\), and the vectors $\ket{z_\ell,\nu}$ are as follows: for $\nu=1$, $\ket{z_\ell,1}=\sum_{j=1}^N z_\ell^j \ket{j}$ represents a {\em generalized Bloch wave}, with possibly complex momentum \(k_\ell=-i\log(z_\ell)\); for \(\nu>1\) the \(\ket{z_\ell,\nu}\)  are proportional to $\partial_z^{\nu-1} \ket{z_\ell,\nu}$, and hence contain amplitudes with a power-law pre-factor to the exponential weight $z^j$. The other two types of solutions that can arise are localized on the boundary of the system and are no longer controlled by $\bm{G}$ and the corresponding (non-unitary) translation symmetry. Rather, they emerge entirely due to the truncation from the bi-infinite lattice to a finite one. We will denote these left $(-)$ and right $(+)$ localized {\em emergent solutions} by $\ket{\psi_\ell^\pm}$, with $\ell=1,\ldots,s_0\equiv 2R-\frac{1}{2}\sum_{\ell=1}^n s_\ell$. Here, $s_0$ is the multiplicity of $z=0$ as a root of  \(\det(G(z)-\omega)=0\). Finally, we remark that there may exist exceptional, isolated values of \(\omega\), which physically correspond to dispersion-less ``flat bands'' and whose associated eigenvectors are not included among the previous three types of solutions. While we refer to \cite{JPA} for more discussion, flat bands will not be encountered in the models under consideration in this paper.

The complete set of solutions to the bulk equation may thus be parameterized as follows:
\begin{equation*}
\ket{\omega,\bm{\alpha}} = \sum_{\ell=1}^n\sum_{s=1}^{s_\ell}\alpha_{\ell s}\ket{\psi_{\ell s}} + \sum_{\ell=1}^{s_0} \alpha^-_\ell\ket{\psi^-_\ell} + \sum_{\ell=1}^{s_0} \alpha^+_\ell\ket{\psi^+_\ell} ,
\end{equation*}
where $\bm{\alpha}\equiv [\alpha_{11},\ldots,\alpha_{n s_n},\alpha^-_{1},\ldots,\alpha^-_{s_0},\alpha^+_{1},\ldots,\alpha^+_{s_0}]^T\in \mathbb{C}^{4R}$. 
Using $\ket{\omega,\bm{\alpha}}$ as an Ansatz for the boundary equation, Eq.\,\eref{BBbdry}, leads to the identity
\begin{equation}
\label{bdryeqn}
\mkern-70mu P_\partial (G-\omega \mathds{1}_{2N})\ket{\omega,\bm{\alpha}} = \sum_b \ket{b}(B(\omega) \bm{\alpha})_b, \quad b\in\{1,\ldots,R,N-R+1,\ldots,N\}. 
\end{equation}
Here, the \textit{boundary matrix} $B(\omega)$ has elements $B_{bs}(\omega)$ that, in our case, consist of $2\times 1$ blocks and are given by $B_{bs}(\omega) = \braket{b|\left( G-\omega \mathds{1}_{2N}\right)|\Psi}$, with $\ket{\Psi}\equiv [\ket{\psi_{11}},\ldots,\ket{\psi_{n s_n}},\ket{\psi^-_{1}},\ldots,\ket{\psi^-_{s_0}},\ket{\psi^+_{1}},\ldots,\ket{\psi^+_{s_0}}]^T$. 
Eq.\,\eref{bdryeqn} tells us that 
if $B(\omega)\bm{\alpha} =0$,  
then $\ket{\omega, \bm{\alpha}}$ solves {\em both} the bulk and boundary equations and hence is an eigenvector of $G_O$ with eigenvalue $\omega$, as desired. 

For diagonalizable matrices, the above procedure yields a Bloch-like diagonal basis. However, $G$ may fail to be diagonalizable, in which case the generalized eigenvectors of \(G\) are needed in addition to the eigenvectors to complete a basis. One can calculate some generalized eigenvectors in Bloch-like form by repeating the above procedure (see also \ref{GBTappendix}) to determine ker$\,(G-\omega \mathds{1}_{2N})^p$ for various powers $p$ and each eigenvalue $\omega$. However, there is a constraint $p<(N-1)/R\equiv p_{\text{max}}$ on how large \(p\) can be because, for \(p \geq p_{\text{max}}\), $(G-\omega \mathds{1}_{2N})^p$ need not be a corner-modified BBT matrix. If there are any, generalized eigenvectors of rank greater than $p_{\text{max}}-1$ may have to be determined by means other than the above bulk-boundary separation. As it turns out, the models we consider offer examples of this peculiar phenomenon.

\section{Manifestations of effective non-Hermitian dynamics}
\label{sec3}

As we have just seen, the effective single-particle description of a QBH is based on mapping the Heisenberg dynamics of Fock space operators to the the dynamics of finite-dimensional vectors by way of Eq.\,\eref{Heis}. The resulting dynamical system, Eq.\,\eref{Heis2}, features a structured matrix \(G\) that need not be Hermitian. The only two constraints on \(G\) are $G^\dag=\tau_3 G \tau_3$, and $G^* = -\tau_1 G \tau_1$. These constraints are the direct mathematical manifestation of the quantum statistics of bosons at the effective single-particle level. In this section, we develop a general framework for analyzing the dynamical system Eq.\,\eref{Heis2} subject to the two above constraints, with special emphasis on the consequences at the many-body level. 

\subsection{Pseudo-Hermiticity and generalized $\mathcal{P}\mathcal{T}$ symmetry} 
\label{sub:PT}

The condition $G^\dag=\tau_3 G \tau_3$ is an instance of a more general mathematical concept known as \textit{pseudo-Hermiticity}. A matrix $M$ is pseudo-Hermitian if there exists a Hermitian, invertible matrix $\eta$ such that $M^\dag = \eta M \eta^{-1}$ \cite{pauli,ali,aliphrep}.  A pseudo-Hermitian matrix need not be diagonalizable unless the spectrum of \(\eta\) is positive- or negative-definite, and that is precisely not the case for bosons since \(\eta=\tau_3\).  
As we are now going to show, the conditions of pseudo-Hermiticity can nonetheless be recast as a particular form of anti-linear symmetry.
Specifically, such an anti-linear symmetry 
turns out to be closely related to the $\mathcal{P}\mathcal{T}$ symmetry of some non-Hermitian quantum
systems \cite{bender,bendermann}. In the current usage of the term, an $n\times n$ complex matrix $M$ is said to be $ \mathcal{P}\mathcal{T}$-{\em symmetric} if it commutes with an anti-linear operator of the form $\mathcal{P}\mathcal{T}$, with $\mathcal{P}$ linear and \textit{involutory}, 
that is, $\mathcal{P}^2=\mathds{1}_n$, and $\mathcal{T}$ the operation of complex conjugation \textit{with respect to the canonical basis}. The physical origin or meaning of the $\mathcal{P}\mathcal{T}$-symmetry 
is not important from a mathematical perspective. There are a series of results in the literature establishing a web of 
relationships between pseudo-Hermiticity and $\mathcal{P}\mathcal{T}$-symmetry in the strict sense just described. In particular: 
\begin{itemize}
\item 
Every $\mathcal{P}\mathcal{T}$-symmetric matrix is pseudo-Hermitian \cite{Zhang}. 
\item 
A \textit{diagonalizable} matrix is pseudo-Hermitian if and only if it commutes with an anti-linear invertible mapping \cite{ali}. If, in addition, the spectrum the matrix is real and discrete, then there is a basis in which its anti-linear symmetry can be decomposed as a product of an involutory linear operator and complex conjugation with respect to this basis \cite{aliphrep}. 
\item 
{\em Weak pseudo-Hermiticity} (a condition that is more general that pseudo-Hermiticity but coincides with it if the spectrum is discrete) is equivalent to the existence of an involutory anti-linear symmetry \cite{solombrino2002,solombrino2003}.
\end{itemize}

In the following proposition, we show that a pseudo-Hermitian matrix is {\em always} $\mathcal{P}\mathcal{T}$-symmetric provided that one relaxes the notion of $\mathcal{P}\mathcal{T}$ symmetry slightly:
\begin{definition} {\bf (G$\mathcal{P}\mathcal{T}$ symmetry)}
A linear transformation $M$ 
is G$\mathcal{P}\mathcal{T}$-symmetric if there exist (i) an invertible anti-linear map $\Theta$ 
such that $[M,\Theta]=0$; and (ii) some basis relative to which $\Theta=\mathcal{P}\mathcal{T}$, with 
$\mathcal{T}$ the operation of complex conjugation with respect to this basis and $\mathcal{P}$ an involutory linear map.
\end{definition}
A $\mathcal{P}\mathcal{T}$-symmetric matrix in the usual sense is automatically G$\mathcal{P}\mathcal{T}$-symmetric with respect the canonical basis. We then have:

\begin{proposition}
\label{pseudoequivpt}
Let $M$ denote a linear transformation of a finite-dimensional Hilbert space $\mathcal{H}$. 
Then $M$ is pseudo-Hermitian if and only if $M$ is G$\mathcal{P}\mathcal{T}$-symmetric.
\end{proposition}
\smallskip
\noindent
{\em Proof:}
If $M$ commutes with an invertible anti-linear transformation, like a G$\mathcal{P}\mathcal{T}$ symmetry, then it is necessarily pseudo-Hermitian by Theorem 3 in \cite{solombrino2003}. To establish the opposite implication, suppose $M$ is pseudo-Hermitian. Let $\lambda_1,\ldots,\lambda_\alpha$ denote the real eigenvalues of $M$ (if any) and $\mu_1,\ldots,\mu_\beta$ denote the non-real eigenvalues of $M$ in the upper half-plane (i.e., $\text{Im}(\mu_j)>0$) for $j=1,\ldots,\beta$). Furthermore, let $r_j$ and $p_j$ denote the lengths of the Jordan chains corresponding to $\lambda_j$ and $\mu_j$, respectively, in the Jordan normal form of $M$. Pseudo-Hermiticity implies that for each Jordan chain of length $p_{j}$ corresponding to eigenvalue $\mu_j^*$, there is a Jordan chain of length $p_j$ corresponding to the eigenvalue $\mu_j^*$ (see e.g. Proposition 4.2.3 in \cite{gohberg}). Hence, we can construct a basis of $\mathcal{H}$ consisting of generalized eigenvectors of $M$, say, $\mathcal{B}\equiv \{\ket{v_{jk}},\ket{w_{jk}},\ket{\bar{w}_{jk}}\}$, where $\ket{v_{jk}}$, $\ket{w_{jk}}$, and $\ket{\bar{w}_{jk}}$ denote rank-$k$ generalized eigenvectors of $M$ at eigenvalues $\lambda_j$, $\mu_j$, and $\mu_j^*$, respectively, with $(M-\lambda_j \mathds{1}_n)\ket{v_{jk}}=\ket{v_{j(k-1)}}$  for $1<k\leq r_j$ and $(M-\mu_j \mathds{1}_n)\ket{w_{jk}}=\ket{w_{j(k-1)}}$,  $(M-\mu_j^* \mathds{1}_n)\ket{\bar{w}_{jk}}=\ket{\bar{w}_{j(k-1)}}$  for $1<k\leq p_j$. Next, define an involutory linear transformation $\mathcal{P}$ on $\mathcal{B}$ by $\mathcal{P}\ket{v_{jk}}\equiv \ket{v_{jk}}$, $\mathcal{P}\ket{w_{jk}}\equiv \ket{\bar{w}_{jk}}$, and $\mathcal{P}\ket{\bar{w}_{jk}} \equiv \ket{w_{jk}}$. Furthermore, let $\mathcal{T}$ denote complex conjugation with respect to the basis $\mathcal{B}$ and let 
$\Theta\equiv\mathcal{P}\mathcal{T}$. It follows immediately that $[M,\Theta]\ket{v_{jk}}=0$. Furthermore, we also have 
\begin{equation*}
M \Theta \ket{w_{jk}} = M\ket{\bar{w}_{jk}} = \mu_j^*\ket{\bar{w}_{jk}} + \ket{\bar{w}_{j(k-1)}} = \Theta M\ket{w_{jk}}, 
\end{equation*}
where we take $\ket{\bar{w}_{j(k-1)}}=0$ for $k=1$. Thus, $[M,\Theta]=0$ and \(\Theta\) is a 
G$\mathcal{P}\mathcal{T}$ symmetry of \(M\). \hfill$\Box$
\smallskip

\noindent
There is a shorter but non-constructive proof of necessity. Theorem 3 in \cite{solombrino2003} asserts that every pseudo-Hermitian operator possesses an anti-linear involutory symmetry. Furthermore, every anti-linear involutory operator coincides with complex conjugation in some basis \cite{solombrinobook}. Since conjugation is of the type $\mathcal{P}\mathcal{T}$, every pseudo-Hermitian operator possesses a G$\mathcal{P}\mathcal{T}$ symmetry.

The following theorem is simply the instantiation of Proposition \ref{pseudoequivpt} to effective SPHs 
$G$ of QBHs. Recall that these matrices are pseudo-Hermitian with respect to $\tau_3$, \(G^\dagger=\tau_3G\tau_3\). 
\begin{theorem}
\label{GPTbosons}
Bosonic effective single-particle Hamiltonians are necessarily G$\mathcal{P}\mathcal{T}$-symmetric.
\end{theorem}
The theorem is also true for fermions. The SPHs of fermionic systems are Hermitian, and hence pseudo-Hermitian with respect to $\eta=\mathds{1}_{2N}$. In this case, the G$\mathcal{P}\mathcal{T}$ symmetry of Proposition \ref{pseudoequivpt} is simply complex conjugation with respect to the eigenbasis of the SPH. The range of possibilities is richer for bosons because non-trivial Jordan chains are possible even for thermodynamically stable systems and so, even in the most elementary systems, this symmetry can be broken -- a situation with no counterpart in fermions.

We say that the G$\mathcal{P}\mathcal{T}$-symmetry \(\Theta\) of a bosonic matrix G is \textit{unbroken} if there exists a basis of simultaneous eigenvectors of \(G\) and \(\Theta\). In the G$\mathcal{P}\mathcal{T}$-unbroken phase, \(G\) is diagonalizable (by definition) and one can check that the spectrum of \(G\) is necessarily real. Hence, \(G\) is dynamically stable and, by combining these spectral features with results from Sec.\,\ref{normalform}, we see that the normal modes of the system may be chosen to satisfy canonical commutation relations.
In the G$\mathcal{P}\mathcal{T}$-broken phase, in contrast, either \(G\) fails to be diagonalizable or the spectrum of \(G\) includes complex eigenvalues, and so \(G\) is dynamically unstable and at least some of the normal modes need not satisfy canonical commutation relations. Therefore, for bosons, the emergence of dynamical instabilities and non-canonical normal modes is {\em rooted in symmetry breaking}. With this mechanism firmly established, we call a transition between a dynamically stable and an unstable regime a \textit{dynamical phase transition} (not to be confused with the notion of a phase transition occurring in the thermodynamic limit away from equilibrium, for instance under the action of a time-dependent Hamiltonian).

Physically, a many-body feature distinctive of the G$\mathcal{P}\mathcal{T}$-unbroken phase is the existence of a {\em convergent} quasi-particle vacuum. Since all the normal modes of the system can be chosen to satisfy bosonic commutation relations, it follows that $G$ can be diagonalized by a $\tau_3$-unitary matrix $L$ satisfying $L^{-1}=\tau_3 L^\dag \tau_3$ and $L^* = \tau_1 L \tau_1$. Explicitly, let 
\begin{equation*}
L \equiv \sum_{m,j=1}^N \ket{m}\bra{j}\otimes \left[\matrix{
X_{mj} & -Y_{mj} \cr -Y_{mj}^* & X_{mj}^* ,
}\right] , 
\end{equation*}
with $XX^\dag-YY^\dag  = \mathds{1}_{N}$, $XY^T-YX^T=0$, $X^\dag X - Y^T Y^* = \mathds{1}_{N}$, and $X^TY^* - Y^\dag X=0$. By contruction, this transformation maps the physical bosonic modes $\hat{\Phi}$ to the bosonic normal modes $\hat{\Psi}= [\widehat{\psi}_1,\widehat{\psi}_1^\dag,\ldots,\widehat{\psi}_N ,\widehat{\psi}_N^\dag]^T = L\hat{\Phi}$ of
\(\widehat{H}=\frac{1}{2}\hat{\Phi}^\dagger\tau_3G\hat{\Phi}   -\frac{1}{2} \tr K   \). The quasi-particle vacuum $\ket{\overline{0}}$, that is, the state such that $\widehat{\psi}_j\ket{\overline{0}}=0$ for all $j$, is formally given by \cite{ripka,RG, ringschuck}
\begin{equation*}
\ket{\overline{0}} = \det(XX^\dag)^{-1/4}\exp\bigg[ \frac{1}{2}\sum_{i,j=1}^N(X^{-1}Y)_{ij}a_i^\dag a_j^\dag\bigg] \ket{0} ,
\end{equation*}
where $a_j\ket{0}=0$ for all \(j\). One needs to check that $\ket{\overline{0}}$ is normalizable just as \(|0\rangle\) is. The normalizability of $\ket{\overline{0}}$ is equivalent to the condition that all the singular values of $X^{-1}Y$ be less than one. It is convenient to recast this condition as the requirement that all the eigenvalues of $Z^\dag Z$, with $Z\equiv X^{-1} Y$, be strictly less than one. Since \(L\) is a canonical transformation, we know that  $XX^\dag = \mathds{1}_N + YY^\dag$. It follows that  $|\det Y|^2< |\det X|^2$ and this inequality in turn implies that $\det Z^\dag Z <1$ (note that this also implies $X$ is always invertible). Hence, the quasi-particle vacuum $\ket{\overline{0}}$ is always normalizable, as claimed. 

Knowing that  $\ket{\overline{0}}$ is normalizable, one may compute a basis of eigenvectors of $\widehat{H}$ as
\begin{equation}
\ket{\overline{n}_1,\ldots,\overline{n}_N} = \prod_{m=1}^N \frac{ (\widehat{\psi}_m^\dag)^{n_m}}{\sqrt{n_m!}}\ket{\overline{0}},\quad \widehat{H}\ket{\overline{n}_1,\ldots,\overline{n}_N} = \Big( \sum_{m=1}^N \omega_m 
 - \frac{1}{2}\tr K 
\Big) \ket{\overline{n}_1,\ldots,\overline{n}_N}, 
\end{equation}
where the eigenfrequencies $\omega_m$ are precisely the eigenvalues of \(G\). This familiar diagonalization procedure for \(\widehat{H}\) is only possible in the G$\mathcal{P}\mathcal{T}$-unbroken phase. Note that the quasi-particle vacuum $\ket{\overline{0}}$ need not be the ground state unless the system also happens to be thermodynamically stable.

\subsection{Characterizing G$\mathcal{P}\mathcal{T}$-symmetry-breaking phase transitions} 
\label{Krein} 

The G$\mathcal{P}\mathcal{T}$ symmetry of QBHs breaks if the effective SPH matrix $G$ approaches a point in parameter space where it loses diagonalizablility -- that is, an EP -- or, when the spectrum splits off the real axis into the complex plane, regardless of diagonalizability. For the subclass of non-Hermitian matrices that are symmetric, so-called  \textit{phase rigidity} (PR) has been extensively used to detect EPs \cite{PR1,PR2,rotter}. Let $\ket{\psi}$ be an eigenvector of $M=M^T$ with eigenvalue $\lambda$. Then, $\ket{\psi^*}\equiv \left(\ket{\psi}\right)^*$ is an eigenvector of $M^\dag$ with eigenvalue $\lambda^*$. If one imposes on $\ket{\psi}$  the \textit{bi-orthonormalization condition} $\braket{\psi^*|\psi}=1$, then one can show that the PR, defined by  
\begin{equation}
\label{CSPR}
\rho \equiv \frac{\braket{\psi^*|\psi}}{\braket{\psi|\psi}}=\frac{1}{\norm{\psi}^2},\qquad \norm{\psi}^2 \equiv \braket{\psi|\psi},\quad \rho\in [0,1],
\end{equation}
vanishes smoothly as $M$ approaches an EP. While useful, such an indicator is not sufficient to our purpose of detecting 
dynamical stability-to-instability transitions in bosons,  
because bosonic matrices $G$ need not be symmetric to begin with and, in addition, such dynamical transitions can occur without loss of diagonalizability. In what follows, we introduce a new PR indicator and argue that it successfully characterizes transitions between G$\mathcal{P}\mathcal{T}$-broken and unbroken phases of QBHs. As it turns out, to do so it is necessary to leverage mathematical results from stability theory of linear dynamical systems governed by pseudo-Hermitian matrices, also known as {\em Krein stability theory} \cite{yaku}, as we recall next. 

\subsubsection{Tools from Krein stability theory.} 
Let \(\eta\) denote an invertible, Hermitian linear transformation of $\mathbb{C}^{2N}$ featuring both positive and negative eigenvalues. The space  $\mathbb{C}^{2N}$ endowed with the indefinite inner product $(\cdot|\cdot)=\left\langle\cdot|\eta|\cdot\right\rangle$ is called a {\em Krein space}.  An \(\eta\)-Hermitian linear transformation satisfies \((M\psi|\phi)=(\psi|M\phi)\), for all $\psi, \phi$. This condition can be seen to be equivalent to \(M\) being pseudo-Hermitian in the sense of Sec.\,\ref{sub:PT}, namely, $M^\dagger = \eta M \eta^{-1}$. A vector $\ket{\alpha}\in\mathbb{C}^{2N}$ is $\eta$\textit{-positive} if $\braket{\alpha|\eta|\alpha}>0$ and $\eta$\textit{-negative} if $\braket{\alpha|\eta|\alpha}<0$. If $\braket{\alpha|\eta|\alpha}=0$, then $\ket{\alpha}$ is $\eta$\textit{-null}. If $\ket{\alpha}$ is either $\eta$-positive or $\eta$-negative, the sign of $\braket{\alpha|\eta|\alpha}$ is the \textit{Krein signature} of $\ket{\alpha}$. Given an \(\eta\)-Hermitian matrix \(M\), the eigenspace $\mathcal{E}_\lambda$ for the eigenvalue $\lambda$ can be classified as follows: If all the eigenvectors of $M$ with eigenvalue $\lambda$ are $\eta$-positive (negative), we say that $\mathcal{E}_\lambda$, or $\lambda$ itself, is $\eta$\textit{-positive ($\eta$-negative) definite}. If we do not wish to specify the sign, we just call $\mathcal{E}_\lambda$, or \(\lambda\), $\eta$-definite. If $\mathcal{E}_\lambda$ is not definite, then we call it, or $\lambda$, $\eta$\textit{-indefinite}. The following definition and results (adapted from \cite{yaku}, Chapter III) are then relevant in our present context:

\begin{definition} {\bf (Krein collision)} Let \(\lambda\) denote an $\eta$-indefinite eigenvalue
of a pseudo-Hermitian matrix. There is a {\em Krein collision} at $\lambda$ if there exists both an 
eigenvector for \(\lambda\) with a Krein signature of $+1$ and an eigenvector for $\lambda$ with Krein signature $-1$.
\end{definition}

\begin{lemma}
\label{Kreinthms}
Let $M_0$ denote an $\eta$-pseudo-Hermitian matrix and $\lambda_0$ a real eigenvalue of \(M_0\). Then: \\ 
\hspace*{3mm}(i) If $\mathcal{E}_{\lambda_0}$ is $\eta$-definite, then all Jordan chains associated to $\lambda_0$ 
are of length one. In addition, there exist $\varepsilon,\delta>0$ such that if \(M\) is $\eta$-Hermitian and
$\norm{M-M_0}<\delta$, then all the eigenvalues \(\lambda\) of \(M\) such that $|\lambda-\lambda_0|<\varepsilon$ 
are real and the length of the Jordan chains associated to them is one. \\
\hspace*{2mm}(ii) If $\mathcal{E}_{\lambda_0}$ is $\eta$-indefinite and all Jordan chains associated to $\lambda_0$ 
are of length one, then for every $\varepsilon>0$ there exists an $\eta$-Hermitian matrix $M$ such that
$\norm{M-M_0}<\varepsilon$ and there are non-real eigenvalues of \(M\) 
in an open neighborhood of $\lambda_0$.  
\end{lemma}

This lemma has direct relevance to QBHs, as noticed in \cite{baldespeano,baldesflow} (see also \cite{nakamura} for related, albeit less general, results). When the bosonic matrix \(G\) is diagonalizable and all of its eigenvalues are definite, then small perturbations of the system cannot drive it away from dynamic stability. By contrast, if some eigenvalue of \(G\) hosts a Krein collision, then it can be de-stabilized 
by an arbitrarily small perturbation. These results have important consequences for the stability of topological edge modes of QBHs \cite{baldespeano}. 
Another interesting consequence of Lemma\,\ref{Kreinthms} is that the zero-frequency bosonic normal modes of dynamically stable QBHs are fragile. If \(G\) is dynamically stable, then these zero modes are constructed out of pairs of kernel vectors of $G$ of opposite Krein signature. Put differently, QBHs that host zero modes are either dynamically unstable or at the cusp of dynamical instability by Lemma \ref{Kreinthms}(ii). We investigate in detail bosonic zero-modes of thermodynamically stable systems in \cite{qiaoru}.

\subsubsection{Krein phase rigidity.} 
\label{sub:kpr}
Returning to the problem of detecting G$\mathcal{P}\mathcal{T}$ symmetry breaking in QBHs, we introduce our proposed indicator in the following:  

\begin{definition} {\bf (Krein phase rigidity)} 
Let $\ket{\psi}$ be an eigenvector of $G$ with eigenvalue $\omega$ and not in the range of 
\(G-\omega\mathds{1}_{2N}\).
Let $\ket{\phi}$ denote its bi-orthonormal partner, that is, the eigenvector of $G^\dag$ with eigenvalue $\omega^*$, such that
$\braket{\phi|\psi}=1$. The {\em Krein phase rigidity} (KPR) of $\ket{\psi}$ is the quantity 
\begin{equation}\label{pr}
r \equiv \frac{\braket{\phi|\psi}}{\norm{\psi}\norm{\phi}}=\frac{1}{\norm{\psi}}\frac{1}{\norm{\phi}}, \qquad r\in [0,1].
\end{equation} 
\end{definition}

While being intuitively appealing, this formula can be simplified. From Sec.\,\ref{normalform}, we can calculate $\ket{\phi}$ as follows. If $\omega \in {\mathbb R}$, then $\ket{\phi}=\kappa \tau_3\ket{\psi}$ with, $\kappa=\sgn\,\braket{\psi|\tau_3|\psi}$ (note that we can ensure that the overlap \(\braket{\psi|\tau_3|\psi}\) is non-vanishing even if \(\omega\) is degenerate). If $\omega \not\in {\mathbb R}$, then $\ket{\phi}=\tau_3\ket{\psi_*}$, with $\ket{\psi_*}$ the suitably normalized eigenvector of $G$ with eigenvalue $\omega^*$. Hence, if $\omega\in {\mathbb R}$, then the KPR immediately simplifies to $r=1/\norm{\psi}^2$, whereas if $\omega\not\in {\mathbb R}$, one can still always normalize $\ket{\psi}$ and $\ket{\phi}$ so that $\norm{\psi}=\norm{\phi}$, and then again $r=1/\norm{\psi}^2$. Also, observe that the condition $\norm{\psi} = \norm{\psi_*}$ can be enforced by renormalizing $\ket{\psi'}=z\ket{\psi}$ and $\ket{\psi'_*}= (1/z^*)\ket{\psi_*}$, with $|z|^2=\norm{\psi_*}/\norm{\psi}$ so that $\braket{\psi'|\tau_3|\psi'_*}=1$ and $\norm{\psi'}=\norm{\psi'_*}$. We will impose this normalization condition in all applications hereafter. 

Thus, the KPR in Eq.\,\eref{pr} may be calculated by a formula seemingly identical to the PR of Eq.\,\eref{CSPR}, once the input states are suitably normalized. In fact, the KPR coincides with the PR when $G$ happens to be symmetric. To see why this is the case, consider for simplicity a non-degenerate, real eigenvalue $\omega$ of \(G=G^T\), with $\ket{\psi}$, $\ket{\psi'}$ the corresponding eigenvectors satisfying $\braket{\psi|\tau_3|\psi}=\kappa\in\{1,-1\}$ and $\braket{\psi'^*|\psi'}=1$, respectively. Since \(\omega\) is not degenerate, $\ket{\psi}=\mu \ket{\psi'}$, for some $\mu\in {\mathbb C}$. Comparing Eqs. \eref{CSPR} and \eref{pr}, we see that KPR and PR coincide if we can set $|\mu|=1$. Since \(G\) is symmetric, both $\tau_3\ket{\psi}$ and $\ket{\psi'}$ are eigenvectors of $G^\dag$ and pseudo-Hermiticity further implies that $\omega$ is also non-degenerate as an eigenvalue of $G^\dag$. It follows that $\tau_3\ket{\psi}=\nu\ket{\psi'^*}$ for some \(\nu\). Equating norms and noting that $\tau_3$ is unitary, we concludes that $|\mu|=|\nu|$. Then $\kappa=\braket{\psi|\tau_3|\psi}=\mu\nu^*$ and taking the modulus of this equation one finally obtains that $|\mu|=1$, as desired. While the corresponding arguments for a non-degenerate, complex eigenvalue are more involved, the equality still holds.

Despite the above formal similarities, and the fact that the KPR consistently reduces to the PR for matrices that are pseudo-Hermitian {\em and} symmetric, the KPR is {\em not} the same quantity as the PR. 
Crucially, for the PR, the mapping from the eigenvector $\ket{\psi}$ to its bi-orthogonal partner $\ket{\psi^*}$ in Eq.\,\eref{pr} is anti-linear (complex conjugation), whereas for the KPR of bosonic matrices, the corresponding mapping is {\em linear} and given by $\tau_3$. Thus, the bi-orthonormalization condition associated to the PR translates into a $\tau_3$-normalization condition for the KPR. This difference is ultimately responsible for the KPR to serve as a useful indicator of G$\mathcal{P}\mathcal{T}$ symmetry breaking for {\em arbitrary} pseudo-Hermitian matrices -- in particular, bosonic ones, which are our main focus in this paper. Specifically, our claim is as follows:

\begin{figure}[t]
\centering
\includegraphics[width=.7\textwidth]{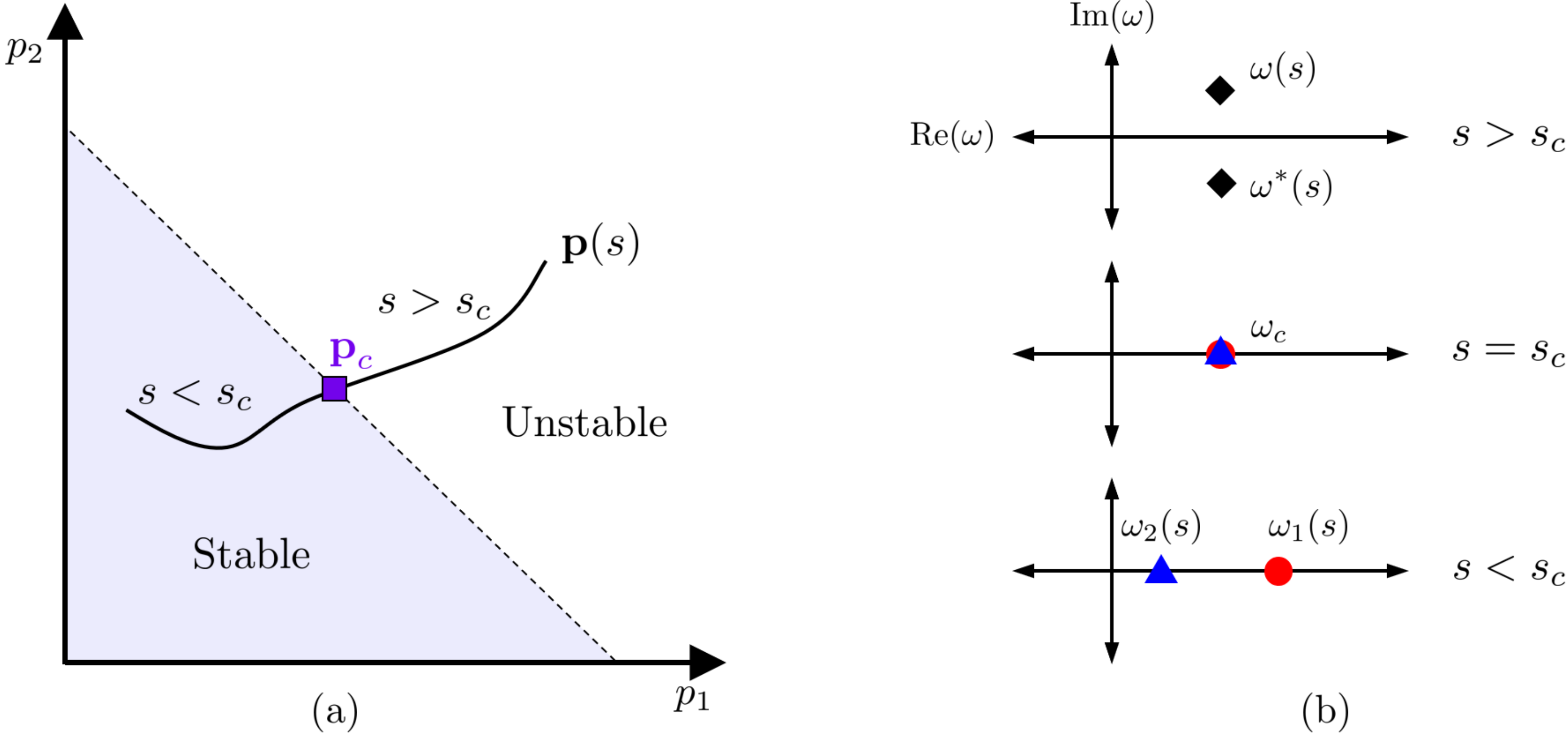}
\caption{(a) Dynamical stability phase diagram for an effective SPH that depends on two real parameters. (b) 
Spectral flow around a phase boundary that hosts Krein collisions. The red circles, blue triangles, and black diamonds indicate eigenvectors with Krein signature $1, -1$, and $0$, respectively. }
\label{KreinExample}
\end{figure}

\vspace{.18cm}

\noindent \textbf{Claim 3.1} {\em 
If a bosonic matrix $G$ undergoes a dynamical phase transition as a function of some
parameter or parameters, then there exists at least one eigenvector of $G$ such that its
KPR vanishes at the phase boundary.}

\vspace{.18cm}
\noindent
\noindent \textit{Heuristic justification}: We first argue that the KPR vanishes at a dynamical phase boundary that hosts a Krein collision. Consider a bosonic matrix $G(\mathbf{p})$ that depends on some parameters $\mathbf{p}$. Suppose that at some point $\mathbf{p}_c$,  
there is an eigenvalue $\omega_c$ of  \(G(\mathbf{p}_c)\) that hosts a Krein collision. Then there exist two eigenvectors $\ket{\psi^c_j}$, $j=1,2$, such that  $\braket{\psi_i^c|\tau_3|\psi_j^c}=\delta_{ij}\kappa_j$, with $\kappa_j$ the Krein signature of  $\ket{\psi_j^c}$ and $\kappa_1=-\kappa_2$. Consider a smooth path $\mathbf{p}(s)$, with $\mathbf{p}(s_c)=\mathbf{p}_c$ and, at this point in the argument, let us add the requirement that $\mathbf{p}_c$ lies on a phase boundary: that is, the system is dynamically stable for, say, $s\leq s_c$ and unstable for \(s>s_c\) (see also Fig. \ref{KreinExample}). From Lemma \ref{Kreinthms}(ii), one expects generically that $\omega_c$ should split into a pair of eigenvalues $\omega(s),\omega(s)^*$ related by complex conjugation, for some $s>s_c$, and that $\omega_c$ should split into a pair of real eigenvalues $\omega_1(s), \omega_2(s)$, for some $s\leq s_c$. Now, let $\ket{\psi_1(s)}$ ($\ket{\psi_2(s)}$) be the eigenvector of $G(\mathbf{p}(s))$ corresponding to the eigenvalue $\omega(s)$ ($\omega(s)^*$) for $s>s_c$ and $\omega_1(s)$ ($\omega_2(s)$) for $s\leq s_c$. Furthermore, suppose these eigenvectors obey the pseudo-bosonic normalization condition $\braket{\psi_2(s)|\tau_3|\psi_1(s)} = 1$ for $s>s_c$ and the bosonic normalization condition $\braket{\psi_i(s)|\tau_3|\psi_j(s)} = \delta_{ij}\kappa_j$ for $s<s_c$, and evolve smoothly as a function of $s$. Letting $\ket{\psi_1'(s)}\equiv \kappa_1\ket{\psi_1(s)}$ for $s< s_c$ and $ \ket{\psi_1'(s)}\equiv \ket{\psi_2(s)}$ for $s>s_c$ (that is, $\tau_3\ket{\psi_1'(s)}$ is the bi-orthonormal partner of $\ket{\psi_1(s)}$), the KPR along the path is then 
\begin{equation*}
r(s) = \frac{\braket{\psi_1'(s)|\tau_3|\psi_1(s)}}{\norm{\psi_1'(s)}\norm{\psi_1(s)}} = \frac{1}{\norm{\psi_1(s)}^2} ,
\end{equation*}
where we again enforce the normalization $\norm{\psi_1(s)}=\norm{\psi_2(s)}$ for $s>s_c$. Equivalently, $r(s) = \braket{\overline{\psi'}_1(s)|\tau_3|\overline{\psi}_1(s)}$, where the overline indicates the vectors are normalized in the usual sense. When $s\to s_c$ from below, the KPR is given by $\kappa_1\braket{\overline{\psi}_1(s)|\tau_3|\overline{\psi}_1(s)}$. Since $\ket{\psi_1(s)}$ evolves smoothly to a $\tau_3$-null eigenvector at $s> s_c$, we must have that $r(s)\to 0$ smoothly approaches zero from the left. When $s\to s_c$ from above, the KPR is given by $\braket{\overline{\psi}_2(s)|\tau_3|\overline{\psi}_1(s)}$. Since $\ket{\psi_1(s)}$ and $\ket{\psi_2(s)}$ evolve smoothly to mutually $\tau_3$-orthogonal vectors for $s<s_c$, we must have that $r(s)$ smoothly approaches $0$ from the right. We conclude that $\lim_{s\to s_c}r(s)=0$ and hence the KPR should vanish at a phase boundary that hosts a Krein collision.

Next, we argue that the KPR should also detect a dynamical phase boundary that hosts an EP, even if \(G\neq G^T\) (recall that we already know the claim to be true for \(G=G^T\), since $r=\rho$). The setup is the same as above, but instead of having two linearly independent eigenvectors $\ket{\psi_j^c}$ at $s=s_c$, we now have one normalized eigenvector $\ket{\chi_a}$ and one normalized generalized eigenvector $\ket{\chi_b}$. Imposing a smooth evolution in parameter space implies that $\ket{\overline{\psi}_1(s)}$ and $\ket{\overline{\psi}_2(s)}$ will both approach the eigenvector $\ket{\chi_a}$ as $s\to s_c$ from above and below. Since $\ket{\chi_a}$ splits into two $\tau_3$-orthogonal eigenvectors for $s>s_c$, we must have that $\braket{\chi_a|\tau_3|\chi_a}=0$ and so $r(s) = \braket{\overline{\psi}_1(s)|\tau_3|\overline{\psi}_2(s)}$ will vanish as $s\to s_c$ from above. \hfill$\Box$
\smallskip

\begin{figure}[t]
\centering
\includegraphics[width=0.9\textwidth]{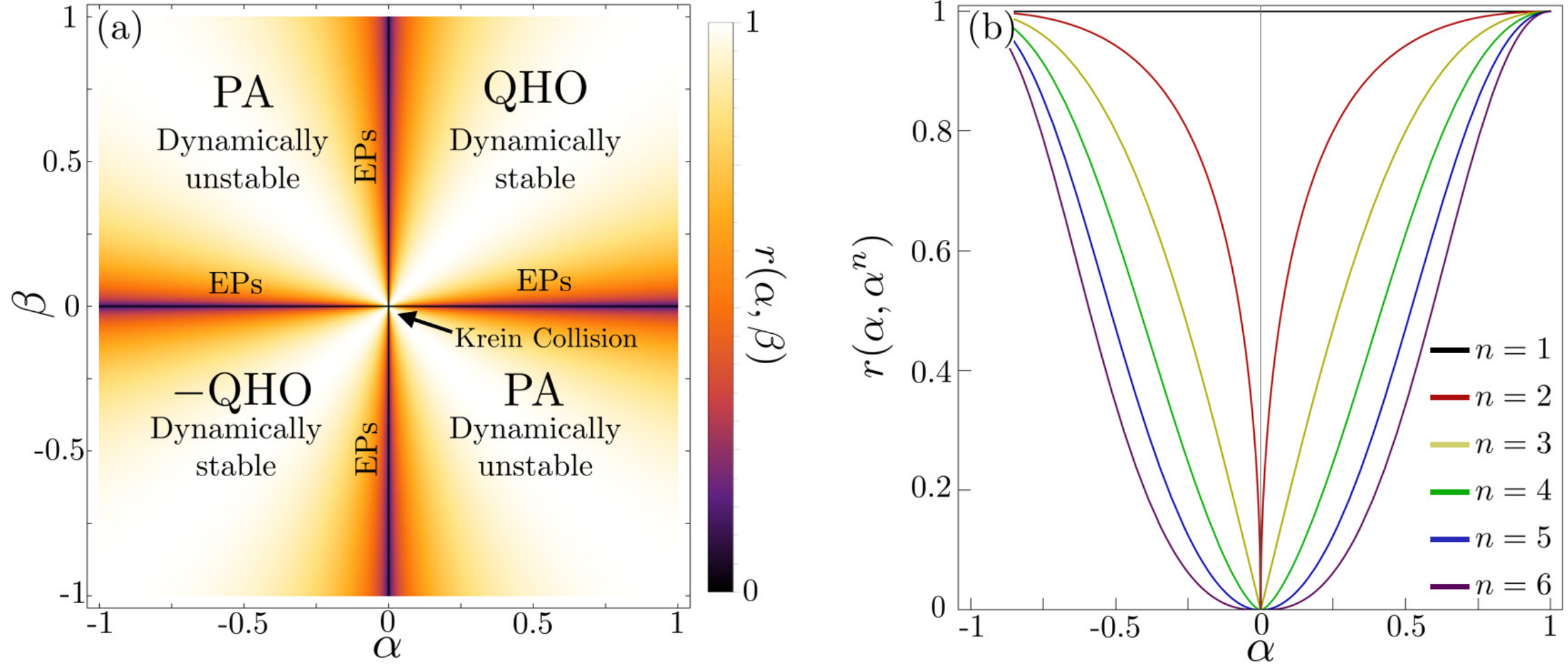}
\vspace*{-2mm}
\caption{(a) A plot of the KPR $r(\alpha,\beta)$ for the Hamiltonian Eq.\,\eref{QHOHam}, with the dynamical phase diagram overlayed. Notice that $r(\alpha,\beta)$ vanishes between the phase boundaries separating the stable quantum harmonic oscillator (QHO) phase and the unstable parametric amplifier (PA) phases. (b) A plot of the KPR $r(\alpha,\beta)$ evaluated on the contours $\beta=\alpha^n$ for $n=1,\ldots,6$. Notice that the KPR vanishes at $\alpha=\beta=0$ even though the system is diagonalizable along these contours.}
\label{QHOfig}
\end{figure}

\subsubsection{Example 1: A single-mode model.} 
\label{sub:example}
It is instructive to check our claim for the simplest model that may support non-trivial dynamical features.
Consider the quadratic single-mode QBH given by 
\begin{equation}
\label{QHOHam}
\widehat{H} = \alpha p^2 + \beta x^2 = \frac{\alpha+\beta}{2}\left( a^\dag a + a a^\dag\right) - \frac{\alpha-\beta}{2}\left( a^{\dag 2} + a^2\right) ,\quad \alpha,\beta\in \mathbb{R}.
\end{equation}
The corresponding effective SPH reads  
\begin{equation}
\label{QHOG}
G(\alpha,\beta) = \left[\matrix{
\beta+\alpha & \beta-\alpha \cr \alpha-\beta & -\alpha-\beta
}\right] ,
\end{equation}
with eigenvalues $\omega_{\pm}=\pm 2\sqrt{\alpha\beta}$.
If $\sgn(\alpha)=\sgn(\beta)$, the system can be described a a dynamically but not necessarily thermodynamically 
stable quantum harmonic oscillator (QHO). For $\sgn(\alpha)\neq \sgn(\beta)$, the system is dynamically
unstable and equivalent to a degenerate parametric amplifier (PA) (see \cite{barton} for an in-depth analysis).
The phase boundaries correspond to \(\alpha=0\) or \(\beta=0\). When only one of the parameters is zero, 
\(G\) fails to be diagonalizable and the system can be described as a free particle. By contrast,
$G$ is trivially diagonalizable for $\alpha=0=\beta$ because it is the zero matrix. The system
can be interpreted as a zero-frequency QHO. This dynamical phase diagram is summarized in Fig. \ref{QHOfig}.
The isolated point \(\alpha=0=\beta\) is an example of a dynamical transition point at which diagonalizability is retained. In Sec.\,\ref{BKC} we will investigate a model where these types of phase boundaries 
extend beyond isolated points (albeit for a fixed BC).

Let us now investigate the phase boundaries in terms of the KPR. For $G\neq 0$, the KPR is
\begin{equation*}
r(\alpha,\beta) 
= \frac{2\sqrt{|\alpha||\beta|}}{|\alpha|+|\beta|} ,
\end{equation*}
for both eigenvectors (see again Fig. \ref{QHOfig}(a)). The KPR vanishes at $(\alpha=0,\beta\neq 0)$ and $(\alpha\neq 0,\beta=0)$ as expected, since these points separate a dynamically stable phase from an unstable one and correspond to EPs of \(G\).  The situation at  \(\alpha=0=\beta\) is more delicate because four dynamical phases meet at the origin and a path through the origin could connect two dynamically (un)stable phases. Our heuristic argument for the behavior of the KPR suggests that it may not be defined at such a point and indeed, the limit \(\lim_{(\alpha,\beta)\rightarrow 0}r\) does not exist in the sense that it is contour-dependent. However, we do expect the KPR to vanish on any path through the origin that connects a dynamically stable phase to an unstable one. For concreteness, let $\beta = f(\alpha)$, with 
$f(\alpha)$ real analytic at $\alpha=0$ and \(f(0)=0\). Since $f(\alpha) = c_1 \alpha + c_2\alpha^2 + \cdots$, the KPR for $|\alpha|\ll 1$ is well approximated by 
\begin{equation}
\label{rc1}
r(\alpha,f(\alpha)) \simeq \frac{2\sqrt{|c_1|}}{1+|c_1|}, 
\end{equation}
which can take any value in $[0,1]$ \textit{as long as \(c_1\neq 0\)}. This behavior is exactly as expected according to our heuristic arguments because paths that behave linearly to lowest order near the origin cross from one (un)stable phase to another (un)stable phase. By contrast, paths with \(c_1=0\) and even leading order cross from a stable to an unstable phase or, for odd leading order, flatten along the boundary of EPs of \(G\). Either way, the KPR vanishes as it should, signaling that the point  $\alpha=0=\beta$ is indeed contained in a dynamical phase boundary. In Fig.\,\ref{QHOfig}(b) we demonstrate this fact by evaluating $r(\alpha,\beta)$ along the contours $\beta=\alpha^n$ for $n=1,\ldots,6$. For $n=1$ the KPR is constant while for $n>1$ the KPR vanishes, as expected.We will encounter this contour-dependent behavior of the KPR around Krein collisions again in the multi-mode model studied in Sec.\,\ref{BKC}. 

It is also instructive to investigate the behavior of the KPR at \(\alpha,\beta=0\) from the point of view of the eigenvectors of \(G(\alpha,\beta)\). Since \(G(0,0)=0\), any choice of $\tau_3$-normalized, charge-conjugate vectors can act as the bosonic normal modes and so there is a Krein collision at \(0\) frequency. The coefficient $c_1$ in Eq.\,\eref{rc1} can be traced back to the normalized eigenvectors of \(G(\alpha,\beta=c_1\alpha)\), that read 
\begin{equation*}
\ket{\overline{\psi}_\pm} = \frac{1}{\sqrt{2(1+c_1)}}\left[\matrix{
1\pm\sqrt{c_1}\cr 1\mp\sqrt{c_1}
}\right]. 
\end{equation*}
Clearly, different choices of $c_1\neq 0$ yield different limiting eigenvectors at $\alpha=0$ and, for \(c_1\rightarrow 0\), 
the eigenvectors converge to the $\tau_3$-null vector associated with the momentum operator $p$.

\subsubsection{Example 2: A cavity QED model.} 
\label{sub:cQED}
In the previous example, the one Krein collision occurred only in the limit where the QBH itself vanished. This being the only way to obtain a Krein collision for a single-mode, we must add an additional mode in order to have a ``minimal model'' of a QBH that is non-vanishing and possesses a Krein collision. The example we consider arises from the cavity QED Hamiltonian studied in Ref.\,\cite{Wiersig}, which describes $N$ identical neutral spin-$1/2$ particles interacting with a single cavity mode:
\begin{equation}\label{cQEDexact}
\widehat{H} = \omega_c a^\dag a+\omega_s S_z + g(a^\dag+a)(S_++S_-) ,
\end{equation} 
with $a^\dag$ ($a$) the creation (annihilation) operator associated with the optical cavity mode, $S_z$ the collective $z$-direction spin operator, and $S_+$ ($S_-$) the collective spin raising (lowering) operator. The (positive) frequencies $\omega_c$ and $\omega_s$ correspond to the resonant frequency of the cavity and the transition frequency of the atoms, respectively, whereas the atom-cavity  coupling strength is given by $g\in\mathbb{R}$. 

Following Ref.\,\cite{Wiersig}, we take $N$ to be sufficiently large and the atoms to be (approximately) in a strongly polarized state, whereby we can make the large-spin (Holstein-Primakoff) approximation
$S_z = {N}/{2}-b^\dag b$, $S_+ \simeq \sqrt{N}b$, $S_-\simeq \sqrt{N}b^\dag$,
with $b^\dag$ ($b$) the bosonic creation (annihilation) operator associated with lowering (raising) the collective spin by $1/2$.
This gives the linearized Hamiltonian
\begin{equation}
\label{H0}
\widehat{H} \simeq \widehat{H}_0 \equiv \omega_c a^\dag a - \omega_s b^\dag b + \chi \left( a^\dag + a\right) \left( b^\dag +b\right),
\qquad \chi\equiv g\sqrt{N} .
\end{equation}
Right away, we can see that $\widehat{H}_0$ is not bounded below, thus it is not thermodynamically stable (consider the expectation values of states of the form $\ket{0,n_s}$, corresponding to no photons and $n_s$ $b$-bosons). This is consistent with the fact that our approximation is only valid for $\braket{b^\dag b}\ll N/2$ and so the unbounded series of states $\ket{0,n_s}$ become less and less physical as $n_s\to\infty$. The relevant Nambu array in this case is simply $\hat{\Phi}\equiv [a\,\, a^\dag\,\, b\,\,b^\dag]^T$. Then, with $\delta\equiv \omega_c-\omega_s$ being the detuning parameter, me may rewrite 
\begin{equation*}
\widehat{H}_0 =\frac{1}{2}\hat{\Phi}^\dag \tau_3 G_0 \hat{\Phi} - \frac{\delta}{2}, \quad  G_0\equiv \left[\matrix{
\omega_c & 0 & \chi & \chi
\cr
0 & -\omega_c & -\chi & -\chi
\cr
\chi & \chi & -\omega_s & 0
\cr
-\chi & -\chi & 0 & \omega_s
}\right] .
\end{equation*}

\begin{figure}[t!]
\centering
\includegraphics[width=\textwidth]{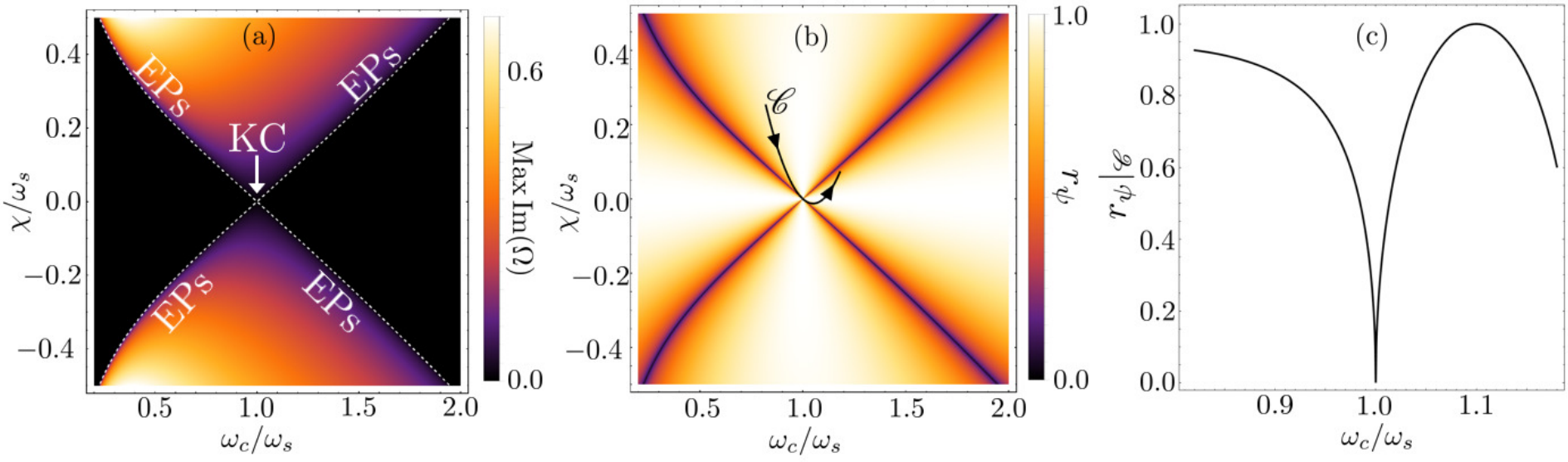}
\caption{(a) Numerical assessment of dynamical stability as a function of $\omega_c/\omega_s$ and $\chi/\omega_s$ sampled on a grid of spacing 0.001 with $\omega_s=1$. Since pseudo-Hermitian matrices with non-real eigenvalues must always have an eigenvalue with a positive imaginary part, the regions in black indicate the dynamically stable phase, with the complement being the dynamically unstable one. The stability phase boundaries (white dashed lines) are labelled based on whether they host EPs or KCs. (b) The KPR $r_\psi$ of a representative eigenvector $\ket{\psi}$ computed numerically in the same parameter space as (a). (c) The KPR $r_\psi$ computed numerically along the contour $\mathscr{C}$ in (b) defined by $y=5x^2-x/2$, with $x=\delta/\omega_s$ and $y=\chi/\omega_s$. Note that the KPR vanishes despite no loss of diagonalizability.}\label{cqfig}
\end{figure}

For simplicity, let us define two dimensionless parameters $x\equiv \delta/\omega_s \in(-1,\infty)$ and $y\equiv \chi/\omega_s \in (-\infty,\infty)$ and let $f(x,y)\equiv x^2(x+2)^2-16 y^2(x+1)$. The eigenvalues of $G$ then take the form 
\begin{equation*}
\Omega_{1\pm} = \pm\frac{\omega_s}{\sqrt{2}}\sqrt{x^2+2x+2+\sqrt{f(x,y)}}, 
\qquad 
\Omega_{2\pm} = \pm\frac{\omega_s}{\sqrt{2}}\sqrt{x^2+2x+2-\sqrt{f(x,y)}}. 
\end{equation*}
One can easily check that these normal mode frequencies become non-real when $f(x,y)<0$, which yields phase boundaries ($f(x,y)=0$) defined by $y = y_\pm(x) \equiv \pm (x^2+2x) /(4\sqrt{x+1}) .$ 
For $x\neq 0$, these boundaries host EPs (as confirmed in Ref.\,\cite{Wiersig}). Surprisingly, when the two boundaries meet, $y_+=y_-$, we have $(x,y)=(0,0)$, which corresponds to the decoupled system $(\chi=0)$ on resonance ($\omega_c=\omega_s$). Since the system remains diagonalizable $(\widehat{H}_0 = \omega_c(a^\dag a-b^\dag b))$, this point hosts a Krein collision.  In Fig.\,\ref{cqfig}(a), we present the dynamical phase diagram along with the classification of the various phase boundaries. Interestingly, the Krein collision lies \emph{precisely} at the locus of EP boundaries, just as in the single-mode example. Moreover, the fact that Krein collision is associated with a decoupling of modes is a theme that will return in Sec.\,\ref{sub:pd}. 

In Ref.\,\cite{Wiersig}, it was pointed out that the stability phase transitions of this model are a consequence of time-reversal symmetry breaking. Time-reversal, whose single-particle manifestation is simply complex conjugation with respect to the canonical basis of $\mathbb{C}^4$ in this case, is a particular instance of the G$\mathcal{P}\mathcal{T}$ symmetry notion we introduced in Sec.\,\ref{sub:PT}, with $\mathcal{P}=\mathds{1}_4$. As we know from Sec.\,\ref{sub:PT}, \emph{all} stability phase transitions in QBHs are associated with a G$\mathcal{P}\mathcal{T}$ symmetry breaking. The time-reversal symmetry breaking in this model is a manifestation of this general feature. Moreover, we may detect this symmetry breaking via the KPR. As a concrete demonstration, we plot the KPR of a representative eigenvector in Fig.\,\ref{cqfig}(b). As expected, it vanishes precisely at the stability phase boundaries. Interestingly, it takes on every value in $[0,1]$ in any open neighborhood about the Krein collision, just as it did for the single-mode model. The stability phase diagrams for these two models are, in a sense, topologically equivalent. To make clear that the KPR vanishes through a Krein-collided phase boundary, in Fig.\,\ref{cqfig}(c) we additionally plot it along a diagonalizable contour between two distinct stability phases, whereby it vanishes precisely at the Krein collision.

\section{Case study: The bosonic Kitaev-Majorana chain}
\label{BKC}

The foundation of our theory of dynamical stability for QBHs lies on the two results from the previous section: (i) Regimes of dynamical stability and instability can be, respectively, understood as unbroken or broken phases of an anti-linear G$\mathcal{P}\mathcal{T}$ symmetry; and (ii)
transitions between two distinct G$\mathcal{P}\mathcal{T}$ phases can be detected by the KPR introduced in Eq. \eref{pr}.  
In this section, we put our theory to work on an interesting model, the BKC of \cite{clerkPRX}. While this one-dimensional QBH was introduced as a bosonic analogue of the fermionic Kitaev chain, the conventional notion of a quantum phase diagram is not applicable because the system is thermodynamically unstable. The expectation in \cite{clerkPRX} was that the topology of the effective SPH would control, or impact significantly, the dynamical properties of the model. Our work was partly motivated by a desire to ground this conjecture.
We will come back to this issue in Sec.\,\ref{furtherimp}.   

\subsection{The model}

\begin{figure}[b!]
\centering
\includegraphics[width=.54\textwidth]{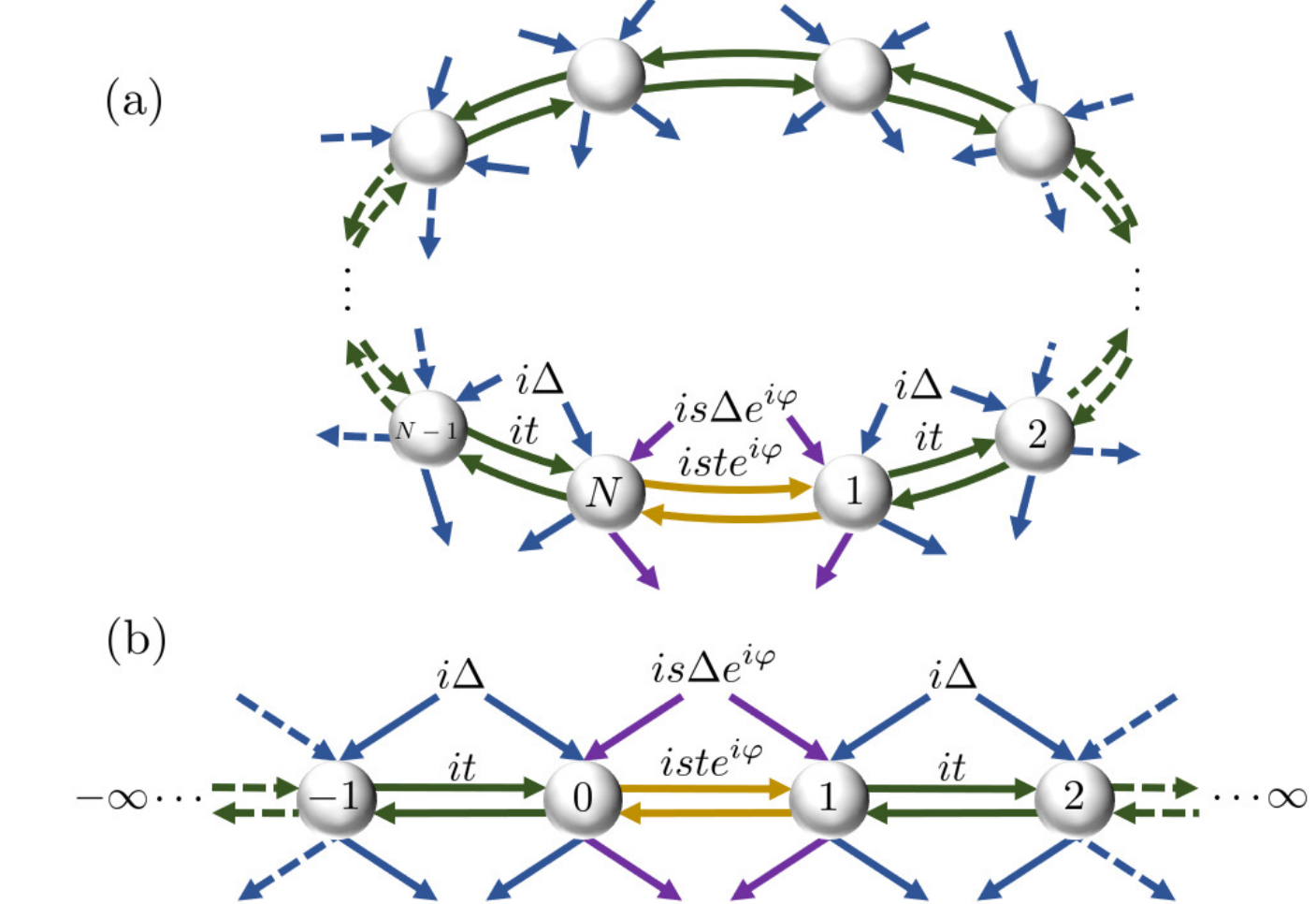}
\caption{Pictorial representation of the BKC model of Eqs.\,\eref{BKCHam}-\eref{BKCB}, along with the relevant parameters for (a) a finite, $N$-site lattice and (b) an infinite lattice.}
\label{BKCfig}
\end{figure}

The QBH we investigate may be written in the form $\widehat{H}(s,\varphi) \equiv \widehat{H}_O + s\widehat{W}(\varphi)$, with $s\in[0,1]$ and 
\begin{eqnarray}
\widehat{H}_O &\equiv& \frac{1}{2}\sum_{j=1}^{N-1}\Big( it a_{j+1}^\dag a_j + i\Delta a_{j+1}^\dag a_j^\dag + \text{H.c.}\Big), \qquad 
t,\Delta>0 , \label{BKCHam} \\
\widehat{W}(\varphi)& \equiv& \frac{1}{2}\Big( it e^{i\varphi} a_1^\dag a_N + i\Delta e^{i\varphi} a_1^\dag a_N^\dag +\text{H.c.}\Big), \qquad 
\: \varphi\in[0,\pi]. \label{BKCB}
\end{eqnarray} 
Specifically, the QBH \(\widehat{H}_O\) is the bosonic Kitaev-Majorana chain of \cite{clerkPRX} subject to open BCs. With respect to the general form in Eq. \eref{genham}, we thus have $K_{ij}= \frac{i t}{2} (\delta_{i,j+1}- \delta_{i+1,j})$, $\Delta_{ij}= \frac{i \Delta}{2} (\delta_{i,j+1}+ \delta_{i+1,j})$ in the bulk, and $K_{1N}=\frac{ist}{2}e^{i\varphi}=K_{N1}^*$, $\Delta_{1N}=\frac{is\Delta}{2}e^{i\varphi}=\Delta_{N1}$ on the boundary. The boundary modification \(\widehat{W}(\varphi)\) imposes twisted BCs with a twisting angle $\varphi$, and the parameter \(s\) allows us to smoothly interpolate between open \(s=0\) and twisted \(s=1\) BCs (see Fig.\,\ref{BKCfig}).
The relevant bosonic matrix is $G(s,\varphi) \equiv G_O + V(s,\varphi)$, with
\begin{equation}
\label{BKCGO}
G_O =T\otimes g_{1}+T^\dag \otimes g_{-1}, \quad V(s,\varphi)= \ket{N}\bra{1}\otimes v_1(s,\varphi)+\ket{1}\bra{N} \otimes v_{-1}(s,\varphi) ,
\end{equation}
with the matrices 
\begin{equation*}
g_1 \equiv -\frac{i}{2}\left[\matrix{
t & -\Delta \cr -\Delta & t
}\right], \qquad v_{1}(s,\varphi)\equiv -\frac{is}{2}\left[\matrix{
t e^{-i\varphi} & -\Delta e^{i\varphi}
\cr
-\Delta e^{-i\varphi} & te^{i\varphi}
}\right], 
\end{equation*}
and, in addition, $g_{-1}=\sigma_3 g_1^\dag \sigma_3$, $v_{-1}(s,\varphi)=\sigma_3 v_1^\dag(s,\varphi) \sigma_3$. 
It is easy to see that \(\widehat{H}_O\) is not thermodynamically stable. A quick check shows that 
odd under time reversal, that is, 
$\widehat{\mathcal{T}} \widehat{H}_O\widehat{\mathcal{T}}^{-1}=-\widehat{H}_O, $
where $\widehat{\mathcal{T}}$ is the usual anti-unitary time-reversal operator satisfying $\widehat{\mathcal{T}}x_j\widehat{\mathcal{T}}^{-1} = x_j$, $\widehat{\mathcal{T}}p_j\widehat{\mathcal{T}}^{-1} = -p_j$, hence 
\(\widehat{\mathcal{T}} \hat{\Phi}\widehat{\mathcal{T}}^{-1}=\hat{\Phi}\). Thus, the spectrum of $\widehat{H}_O$ is symmetric about zero (chiral).

\begin{figure}[t!]
\centering
\includegraphics[width=.83\textwidth]{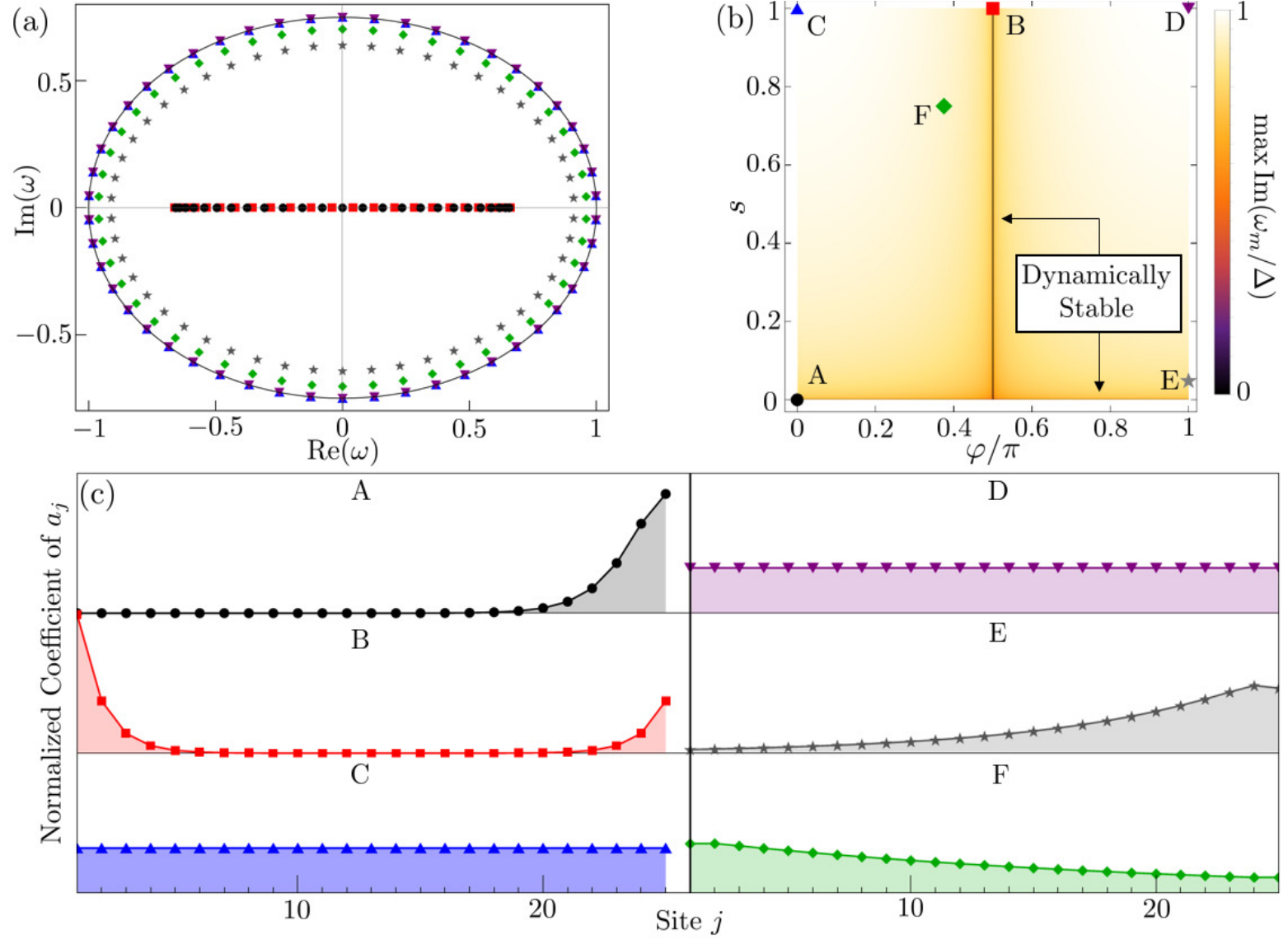}
\vspace*{-1mm}
\caption{Spectral properties of the SPH matrix $G(s,\varphi)$ for $t=1,$ $\Delta=0.75$ and $N=25$. (a) The eigenvalues of $G(s,\varphi)$ for various choices of boundary parameters $s$ and $\varphi$. (b) The largest imaginary part, in absolute value, for the eigenvalues of $G$ numerically calculated on a grid of spacing $0.002$. Points indicate values of $s$ and $\varphi$ where the eigenvalues and eigenvectors are sourced in (a) and (c), respectively. Points  A, B, C, and D correspond, in particular, to parameter values for which we obtained exact analytical solutions [see text]. (c) The normalized coefficients of $a_j$ for a representative normal mode of the chain at various choices for $s$ and $\varphi$. These normal modes are representative in the sense that their gross localization properties are independent of the particular eigenvector chosen.}
\label{spectra}
\end{figure}

In \cite{clerkPRX}, diagonalization of $\widehat{H}(s,\varphi)$ for $t\ne \Delta$ was achieved for open BCs ($s=0$) thanks to a suitably devised position-dependent local squeezing transformation and for periodic BCs ($s=1, \varphi=0$) by standard momentum-space techniques. Here, we employ the general techniques from Sec.\,\ref{GBTprimer} to both recover these solutions and analyze different exactly solvable parameter regimes (see also Fig.\,\ref{spectra}). While we include in \ref{diagdetails} full detail about the diagonalization of $G(s,\varphi)$ by our approach, we highlight in the following section the salient features of the resulting solutions, as relevant to the subsequent stability analysis.

\subsection{Exact characterization of spectral properties}

\subsubsection{Open boundary conditions.}
\label{obcs}
In this regime, the spectrum is doubly degenerate and given by $\omega_m = \sqrt{t^2-\Delta^2}\,\cos(m\pi/(N+1))$, with $m=1,\ldots, N$. For $t\neq \Delta$, the eigenvectors have the generalized Bloch form 
\begin{equation}
\label{obcevecs}
\ket{\psi_{m,\sigma}^\pm} = \mathcal{N}_m\sum_{j=1}^N (-\sigma)^{j/2}\sin\left( \frac{m\pi j}{N+1}\right) \ket{j}\ket{\xi_\sigma^\pm(j)}, \qquad \sigma \equiv  \sgn(t-\Delta), 
\end{equation}
where  $\mathcal{N}_m$ is a normalization constant and, in terms of the parameter $2r \equiv \ln [(t+\Delta)/|t-\Delta|]$, 
\begin{equation*}
\fl\ket{\xi_+^+(j)} = [\cosh(jr), \sinh(jr)]^T, \quad \ket{\xi_-^+(j)} = \left\{\matrix{
[\cosh(jr),\sinh(jr)]^T,& j\text{ even},
\cr
[\sinh(jr), \cosh(jr)]^T,& j\text{ odd},
}\right., \quad \ket{\xi_\sigma^-(j)} = \sigma_1 \ket{\xi_\sigma^+(j)}.
\end{equation*}
As also noted in \cite{clerkPRX}, {\em all} eigenvectors are exponentially localized near the boundary, consistent with the so-called ``non-Hermitian skin effect'' \cite{YaoSkin}. 
Thus, the system is dynamically stable in the hopping-dominated regime, \(t>\Delta\), and unstable when pairing dominates, \(t<\Delta\). 
For \(t>\Delta\), $\sigma=+1$ and the eigenvectors satisfy the bosonic normalization conditions $\braket{\psi_{m,+}^\pm|\tau_3|\psi_{\ell,+}^\pm} = \pm\delta_{m\ell}$ and $\braket{\psi_{m,+}^\pm|\tau_3|\psi_{\ell,+}^\mp} =0$. For \(t<\Delta\), $\sigma=-1$ and the eigenvectors can be chosen to satisfy the pseudo-bosonic normalization conditions $\braket{\psi_{N+1-m,-}^\pm|\tau_3|\psi_{\ell,-}^\pm} = \delta_{m\ell}$ and $\braket{\psi_{m,+}^\pm|\tau_3|\psi_{\ell,+}^\mp} =0$.

A transition from dynamical stability to instability occurs at \(t=\Delta\). On this line, we have 
$G_O (t=\Delta) = it\left( T^\dag\otimes\ket{+}\bra{+} -  T\otimes\ket{-}\bra{-}\right),$
where $\ket{\pm}$ are the normalized eigenstates of $\sigma_1$ with eigenvalues $\pm 1$ and 
$T$ the left-shift operator of Eq. \eref{leftshift}. The spectrum of \(G_O (t=\Delta) \) contains the zero 
eigenvalue only, with two associated Jordan chains of length \(N\).  

The normal modes of \(\widehat{H}_O\) can be chosen to be bosonic for $t>\Delta$ and pseudo-bosonic for $t<\Delta$ (recall Sec.\,\ref{sub:diag}). The bosonic modes are given by \(\widehat{\psi}_m = \bra{\psi_{m,+}^+}\tau_3\hat{\Phi}\)
in terms of the eigenstates from Eq.\,\eref{obcevecs}. Keeping in mind the chiral symmetry of the spectrum, the normal form of the QBH is
\begin{equation*}
\widehat{H}_O = \sum_{m=1}^{N_+} \omega_m (\widehat{\psi}_m^\dag \widehat{\psi}_m -\widehat{\psi}_{\overline{m}}^\dag \widehat{\psi}_{\overline{m}}^{} ), \qquad \overline{m}= N+1-m, 
\end{equation*}
with $N_+ =N/2$ for $N$ even and $N_+=(N-1)/2$ for $N$ odd, respectively. For $N$ odd there is an additional zero mode, at $m=(N+1)/2$, which commutes with the Hamiltonian. The quasi-particle vacuum is  
\begin{equation*}
\ket{\overline{0}} = \mathcal{M} \exp\bigg[ \frac{1}{2}\sum_{j=1}^M \tanh(jr)( a_j^\dag)^2\bigg]\ket{0}.
\end{equation*}
The normal form of \(\widehat{H}_O\) reveals a symmetry of the model embodied in the Bogoliubov transformation, 
\begin{eqnarray}
\label{degfreeBT}
\widehat{\psi}_m &\mapsto& \widehat{\psi}_m(s_m) \equiv \cosh(s_m) \widehat{\psi}_m + \sinh(s_m)\widehat{\psi}_{\overline{m}}^\dag,
\cr
\widehat{\psi}_{\overline{m}}^\dag &\mapsto& \widehat{\psi}_{\overline{m}}^\dag(s_m) \equiv \cosh(s_m) \widehat{\psi}_{\overline{m}}^\dag + \sinh(s_m)\widehat{\psi}_{m},
\end{eqnarray}
with $s_m\in\mathbb{R}$ arbitrary.
The new normal modes are 
\begin{equation}\label{degfreedom}
\widehat{\psi}_m(s_m) = \sqrt{\frac{2}{N+1}}\sum_{j=1}^N i^{-j}\sin\left(\frac{m\pi j}{N+1}\right)\left( \cosh(s_m+jr) a_j - \sinh(s_m+jr) a_j^\dag\right).
\end{equation} 
Hence, $s_m$ is a free parameter that determines the localization properties of each mode. For the particular choice $s_m=-j_0 r$ for all $m$, we recover the same parametric freedom identified in \cite{clerkPRX}.

At the transition point $t=\Delta$, the normal modes consist of two Jordan chains of the adjoint action of length \(N\) and, from the discussion in Sec.\,\ref{sub:general}, we know that their algebra is fairly arbitrary. The corresponding Jordan chains of \(G_O (t=\Delta)\) can be chosen to be $\ket{\chi_{1k}}=(it)^{-k}\ket{N+1-k}\ket{+}$ and $\ket{\chi_{2k}}=(-it)^{-ik}\ket{k}\ket{-}$, with $k=1,\ldots, N$, and map to (multiples of) the Hermitian quadratures
\begin{equation*}
\widehat{\chi}_{1k} = \bra{\chi_{1k}}\tau_3 \hat{\Phi} = (-it)^{-k} p_{N+1-k}, \quad \widehat{\chi}_{2k} = \bra{\chi_{2k}}\tau_3 \hat{\Phi} = (it)^{-k} x_{k}.
\end{equation*} 
These normal modes cannot be combined to yield bosonic normal modes for $k\neq (N+1)/2$, as $[x_k,p_{N+1-k}]=0$. 
For \(N\) odd and $k=(N+1)/2$, one can construct a single bosonic normal mode at zero frequency.

\subsubsection{Periodic and anti-periodic boundary conditions.}
Periodic and anti-periodic BCs (PBCs and APBCs) correspond to ($s=1,\varphi=0$) and ($s=1,\varphi=\pi$), respectively.  
Owing to translational invariance, the Hamiltonian can be block-diagonalized the Fourier transform
\( b_k \equiv \frac{1}{\sqrt{N}}\sum_{j=1}^N e^{-ijk} a_j ,
\)
with $\mathcal{K}^{(x)}_N$, $x=A,P$, labeling the set of wave vectors appropriate for each type of BC, that is, explicitly, 
\begin{eqnarray*}
\mathcal{K}^{(P)}_N = \left \{ \begin{array}{ll}
\{ 0,\pm 2\pi/N,\pm 4\pi/N,\ldots,\pm \pi(1-1/N) \}, & N\text{ odd},
\\
\{0, \pm 2\pi/N,\pm 4\pi/N,\ldots,\pm \pi(1-2/N),-\pi\}, & N\text{ even}, 
\end{array} \right. \\
\mathcal{K}^{(A)}_N = \left \{ \begin{array}{ll}
\{ \pm \pi/N,\pm 3\pi/N,\ldots,\pm\pi(1-1/N),\pi\},& N\text{ odd},
\\
\{ \pm \pi/N,\pm 3\pi/N,\ldots,\pm \pi(1-1/N)\},& N\text{ even}.
\end{array} \right.
\end{eqnarray*}
One finds then that  
\( \widehat{H}(1,0) = \sum_{k\in\mathcal{K}_N^{(P)}} \widehat{H}_k\) and
\( \widehat{H}(1,\pi) = \sum_{k\in\mathcal{K}_N^{(A)}}\widehat{H}_k\), 
with 
\begin{equation*}
\widehat{H}_k = \left\{\matrix{
\frac{t}{2}\sin(k) \left( b_k^\dag b_k - b_{-k} ^\dag b_{-k}\right) + \frac{i\Delta}{2}\cos(k)
\left( b_k^\dag b_{-k}^\dag - b_k b_{-k}\right),& k\neq \pm\pi ,
\cr
-\frac{i\Delta}{2}\left( (b_k^\dag)^2 - ( b_k)^2\right), & k =\pm \pi .
}\right.
\end{equation*}
Note that the blocks $\widehat{H}_{k=\pm\pi/2}$, when $\pm \pi/2$ is in the Brillouin zone (which happens only if $N$ is  a multiple of $4$ for PBCs), are already in normal form in terms of the bosonic Fourier modes $b_{\pm \pi/2}$. The associated eigenfrequencies $\omega_{\pm \pi/2} = \pm t$ are the only real ones. Also, recall that the values $k\in \{ 0,\pm \pi\}$ (when in the spectrum) are precisely the momentum modes that are unpaired in the fermionic Kitaev-Majorana chain. Interestingly, since $\widehat{H}_k \propto x_kp_k+p_k x_k$ for $k\in \{0,\pm \pi \}$, these blocks can be put in a pseudo-bosonic form by letting $\widehat{\psi}_k^\dag = -ip_k$ and $\widehat{\psi}_{k*} = x_k$. 
For $k\not \in \{0,\pm \pi/2, \pm \pi\}$, the normal form of the blocks is 
\begin{equation*}
\fl\widehat{H}_k = \omega_k \widehat{\psi}_k^\dag \widehat{\psi}_{k*} + \omega_k^* \widehat{\psi}_{k*}^\dag \widehat{\psi}_k,  \quad  \omega_k = t\sin(k) + i\Delta \cos(k),
\quad
\widehat{\psi}_k = \frac{1}{\sqrt{2}}\left( b_k-b_{-k}^\dag\right), \quad \widehat{\psi}_{k*} = \frac{1}{\sqrt{2}}\left( b_k+b_{-k}^\dag\right),
\end{equation*}
in terms of pseudo-bosons satisfying $[\widehat{\psi}_k,\widehat{\psi}_{q*}^\dag] = \delta_{kq}$ and $[\widehat{\psi}_k,\widehat{\psi}_q] = [\widehat{\psi}_{k*},\widehat{\psi}_{q*}]=0$.

\subsubsection{Twisted boundary conditions: $\varphi=\pi/2$.}
\label{tbcs}

This case is of particular interest because these BCs are known to pin localized Majorana zero modes for the 
fermionic Kitaev-Majorana chain \cite{katsura}.  Under this BC, our model Hamiltonian corresponds to a system with purely imaginary hopping and pairing in the bulk, and real hopping and paring between sites $1$ and $N$.
The spectrum  is given by $\omega_m=\sqrt{t^2-\Delta^2}\,\sin((m+1/2)\pi/N)$, with $m=0,\ldots,2N-1$. 
The eigenvectors again have a generalized Bloch form given by
\begin{equation}
\ket{\psi_{m,\sigma}} = \mathcal{N}_m\sum_{j=1}^N  e^{ijk_m}\ket{j} 
\ket{\zeta_m(j)}, \qquad \sigma =  \sgn(t-\Delta), 
\end{equation}
where $\mathcal{N}_m$ is an appropriate normalization constant and
\[
\ket{\zeta_m(j)} = \sigma_1^{m} \left[\matrix{
\sinh\left[\left( j - \frac{N+2}{2}\right) r'\right]
\cr
\cosh\left[\left( j - \frac{N+2}{2}\right) r'\right]
}\right] , \qquad r'=\left \{ \begin{array}{ll} r , & t > \Delta, \\  
r+i\pi/2 ,& t < \Delta .\end{array} \right.
\]
Interestingly, as for open BCs, a non-Hermitian skin effect still occurs, despite the lack of an explicit boundary. Likewise, 
the system is dynamically stable for \(t>\Delta\) and unstable for \(t<\Delta\), as for the open chain. At the transition line \(t=\Delta\), the effective SPH reads
$$G(1,\pi/2; t=\Delta) = it\left( T^\dag\otimes \ket{+}\bra{+}- T\otimes\ket{-}\bra{-}\right) -  t\left( \ket{1}\bra{N}+\ket{N}\bra{1}\right)\otimes\ket{-}\bra{+}$$ and, again similar to the open BC case, the spectrum consists only of the zero eigenvalue, with associated pair of Jordan chains of length \(N\) [see Eqs.\,\eref{TBCgevecsNeven}-\eref{TBCgevecsNodd} for explicit expressions]. 

\subsubsection{Twisted boundary conditions: $t=\Delta$ and $\varphi\in(0,\pi)$.}
We have established that, for $t=\Delta$, the bosonic matrix \(G(1,\varphi)\) is diagonalizable 
for $\varphi=0$ (PBC) and $\varphi=\pi$ (APBC) but not for $\varphi=\pi/2$. In addition, the system is dynamically unstable 
for $\varphi=0,\pi$. To interpolate between these three points, we diagonalize $G_T(\varphi) \equiv G(1,\varphi)$. The spectrum is
\begin{equation}
\label{tzDarbphi}
\omega_m = it\,(\cos(\varphi))^{1/N}\left\{\matrix{
e^{-2\pi i m/N}, & N \text{ even,}
\cr
e^{-\pi i m/N}, & N\text{ odd,}}
\right. 
\qquad m=1,\ldots,2N , 
\end{equation} 
where we take $(\cos(\varphi))^{1/N} \equiv |\cos(\varphi)|^{1/N} e^{i\pi/N}$, for $\varphi\in(\pi/2,\pi]$. In fact, we have that the spectrum of $G_T(\varphi; t=\Delta)$ is precisely $|\cos(\varphi)|^{1/N}$ times the PBC spectrum, for $\varphi\in[0,\pi/2]$, or the APBC spectrum, for $\varphi\in(\pi/2,\pi]$. Thus, as $\varphi$ increases away from $0$, the ellipses in the complex plane that corresponds to the PBC spectrum shrinks isotropically until collapsing to zero at $\varphi=\pi/2$. The spectrum then emerges from zero and approaches the APBC spectrum uniformly as $\varphi\to \pi$.

For $N$ even, each of the above eigenvalues is doubly degenerate: $\omega_m = \omega_{N+m}$, for $m=1,\ldots,N$. Thus, we can take $1\leq m \leq N$. Up to a normalization constant, the two eigenvectors for each $m$ are given by
\begin{eqnarray*}
\ket{\psi_{m,1}} &=& \mathcal{N}_{m,1}\left( z_m^{-1} \ket{z_m,1}\ket{+} + i \tan(\varphi)\ket{1}\ket{-}\right) , 
\qquad z_m \equiv it/\omega_m , \\
\ket{\psi_{m,2}} &=& \mathcal{N}_{m,2} \ket{-z_m^{-1},1}\ket{-}, 
\end{eqnarray*} 
where $\ket{z_m,1}, \ket{-z_m^{-1},1}$ are generalized Bloch waves (see also Eq.\,\eref{z1}), and $\ket{1}$ denoted the first canonical basis vector of $\mathbb{C}^{N}$.
For $N$ odd, the spectrum is non-degenerate and the eigenvectors are 
\begin{equation}
\label{tzDarbphiWF}
\ket{\psi_m} = \mathcal{N}_m \left\{\matrix{
z_m^{-1} \ket{z_m,1}\ket{+} + i\tan(\varphi)\ket{1}\ket{-},& m \text{ even},
\cr
\ket{-z_m^{-1},1}\ket{-},& m\text{ odd} ,
}\right.
\end{equation}
where again $\mathcal{N}_m$ is a normalization constant that is chosen so that
\( \braket{\psi_m|\tau_3|\psi_{\ell}} = \delta_{m*,\ell}. \)
Given these normalization conditions, one can construct the pseudo-bosonic normal 
modes just as for periodic and antiperiodic BCs. 

This exact solution is remarkable because it makes it possible to investigate analytically the 
flow of eigenvectors as the system approaches the EP that now exists at $\varphi=\pi/2$. Let us consider
the way in which the pseudo-bosonic modes parametrically evolve into the Jordan chains 
of generalized normal modes at $\varphi=\pi/2$. Focusing for simplicity on $N$ odd, let $\ket{\chi_{11}}\equiv 
\ket{1}\ket{-}$ be one of the two linearly dependent eigenvectors of $G_T(\pi/2; t=\Delta)$, 
and let \(\mathcal{O}_m \equiv \frac{|\braket{\chi_{11}|\psi_m}|}{\norm{\chi_{11}}\norm{\psi_m}} \) be the corresponding fidelity overlap with the eigenvector in Eq. \eref{tzDarbphiWF}. After determining the proper normalization constants $\mathcal{N}_m$ \eref{sub:jordan}, 
we find the following closed-form expression:
\begin{equation*}
\mathcal{O}_m = (1- |\cos(\varphi)|^{2/N})^{1/2}\left\{\matrix{
(1-|\cos(\varphi)|^2)^{-1/2},& m\text{ odd},
\cr
1,& m\text{ even},
}\right.
\end{equation*}
In both cases, we have that $\mathcal{O}_m\to 1$ as $\varphi\to \pi/2$. Thus, all the eigenvectors coalesce to a {\em single} eigenvector at the EP.  In particular, they all become perfectly localized at site $j=1$ as the system approaches the EP.
 
Before moving to an in-depth analysis of the stability properties of the chain, we briefly comment on the induced many-body action of the G$\mathcal{P}\mathcal{T}$-symmetry as a function of BCs. The stable BCs (e.g., open and $\pi/2$-twisted) support an unbroken G$\mathcal{P}\mathcal{T}$ symmetry that (anti-linearly) maps each of the bosonic quasi-particle creation and annihilation operator to itself. For the unstable cases (e.g., periodic and anti-periodic), the G$\mathcal{P}\mathcal{T}$ symmetry (anti-linearly) maps the normal modes to their pseudo-bosonic partners (e.g., $z \widehat{\psi}_k\mapsto z^* \widehat{\psi}_{k*}$, for any $z\in\mathbb{C}$). In particular, we observe that this symmetry is a \textit{function} of the BCs.

\subsection{Stability analysis} 

\subsubsection{Stability phase diagram as a function of BCs.}
As we saw in terms of closed-form solutions, the bosonic Kitaev-Majorana chain of Eq.\,\eref{BKCHam} displays both stable and unstable dynamical phases.  There is a G$\mathcal{P}\mathcal{T}$ symmetry that breaks at any stability-to-instability transition in one of two non-exclusive ways: either because the effective SPH loses diagonalizability, or because it develops Krein collisions that will generically cause a real eigenvalue to split into a pair of complex-conjugate eigenvalues. For for both open or $\pi/2$-twisted BCs, the onset of instability at $t=\Delta$, is of the first type, due to loss of diagonalizability. There are also, however, dynamical phase transitions induced by changes in the the boundary parameters $s$ and $\varphi$ for $t>\Delta$. In this section, we investigate both analytically and numerically these transitions. 

\begin{figure}[t!]
\centering
\includegraphics[width=1\textwidth]{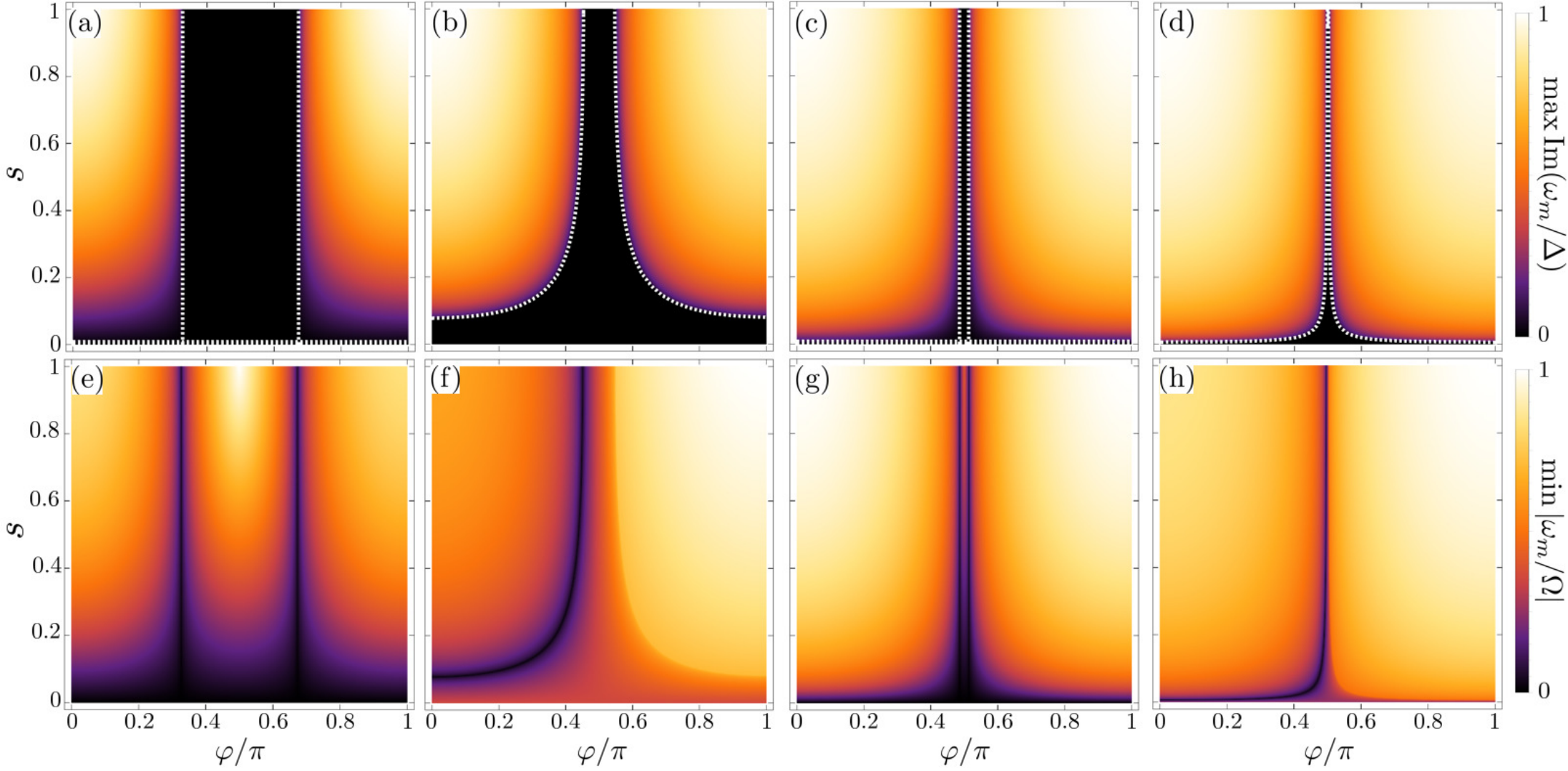}
\caption{(a)-(d): Numerical assessment of dynamical stability as a function of BCs, with $t=1$ and $\Delta=0.25$, and 
the boundary parameters sampled on a grid of spacing $0.002$. The systems size is (a) $N=5$, (b) $N=10$, (c) $N=15$, and (d) $N=20$. Phase boundaries are indicated by white, dashed lines [see Eq.\,\eref{bdrycurve}]. The $s\neq 0$ phase boundaries host $N$ (1) length-2 Jordan chains for $N$ odd (even), while the $s=0$ boundary hosts $N$ Krein collisions. (e)-(h): Minimum-modulus eigenvalue of $G(s,\varphi)$ sampled on the same grid as (a)-(d). The parameter values for (e), (f), (g), and (h) match those of (a), (b), (c), and (d), respectively. Here, $\Omega$ is the largest value of $\min|\omega_m/\Delta|$ over the whole sample grid.  The lines of zero modes (along with their mirror-symmetric partners in (f) and (h)) appear to define the dynamical phase boundaries.}
 \label{bdrynum}
\end{figure}

Figures\,\ref{spectra}(b) and \ref{bdrynum}(a)-(d) show the dynamical phase diagrams for various choices of parameters, obtained by numerically determining the eigenvalue spectrum for different system size $N$. From Lemma \ref{Kreinthms}, we know that the open chain can undergo a dynamical phase transition for arbitrarily small perturbations, due to Krein collisions. While Fig.\,\ref{spectra}(b) suggests that the region of dynamical stability which connects  OBCs to $\pi/2$-twisted BCs may be very nearly a line (hence, of zero measure), additional analysis for a smaller $\Delta/t$ ratio
reveals that the thickness of the region surrounding the line $\varphi=\pi/2$ does {\em not} vanish for finite lattice size. That is, for each $s$, the chain is stable for $\varphi\in[\pi/2-{\delta\varphi}_N(s)/2,\pi/2+{\delta\varphi}_N(s)/2]\equiv I_{{\varphi}_N} (s)$. For finite $N$, ${\delta \varphi}_N (s)>0$ strictly, and the \textit{minimum width} of the stability region is given by $\delta\varphi_N\equiv {\delta\varphi}_N(s=1)$. Analogously, for each $\varphi,N$, we let ${\delta s}_N(\varphi)$ be such that the system is dynamically stable for $s\in[0,{\delta s}_N(\varphi)]\equiv I_{s_N}(\varphi)$ and define the \textit{minimum height} of the stability region by ${\delta s}_N\equiv {\delta s}_N(\varphi=0)$. 

\begin{figure}[t!]
\centering
\includegraphics[width=0.95\textwidth]{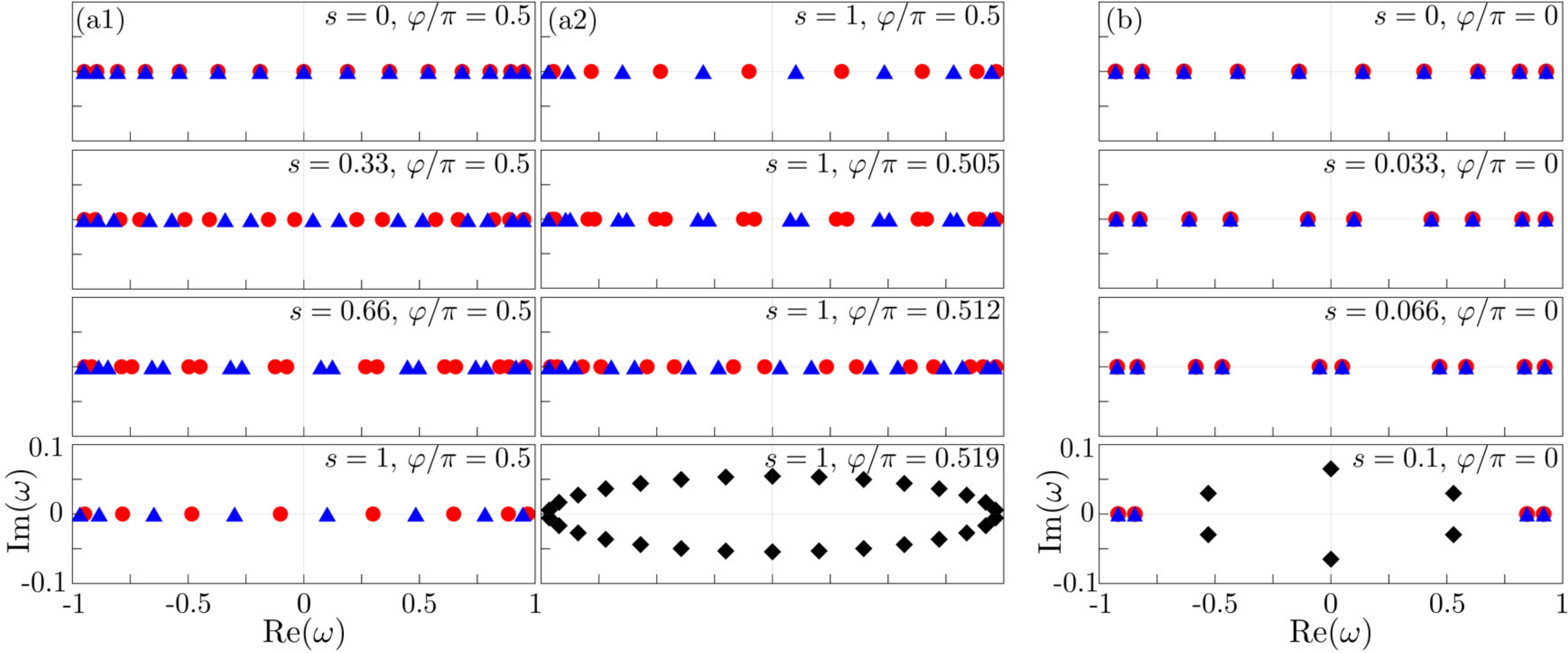}
\caption{The spectral flow of the chain as boundary parameters are varied with $t=1$, $\Delta=0.25$. The Krein signatures for the corresponding eigenvectors are indicated by red circles ($+1$), blue triangles ($-1$), and black diamonds (0), respectively. (a) $N=15$. Going down the left column (a1), we show how the eigenvalues evolve within the stable phase, as $s$ goes from $0$ to $1$ at $\varphi=\pi/2$. Recall that, for $N$ odd, $\varphi=\pi/2$, and $s=1$, each eigenvalue except the extremal ones are twofold degenerate. Stability is seen to be preserved along this flow. Going down the right column (a2), we show how that eigenvalues evolves around the transition between $\pi/2$-twisted and periodic, as $\varphi$ goes from $\pi/2$ to $1.03\cdot(\pi/2)$ at $s=1$. Note how eigenvalues are each split as $\varphi$ increases, and eventually move symmetrically off the real axis. (b) $N=10$. The spectral flow around the transition between open and periodic, as $s$ goes from $0$ to $0.1$ with $\varphi=0$. In this case, stability is retained for sufficiently small strength of the boundary perturbation due to non-zero $s$.}
\label{specflowFULL}
\end{figure}

We can characterize both the way in which phase boundaries depend upon parameters and their nature in terms of Krein collisions versus EPs by combining analytical and numerical techniques. To this end, it is useful to analyze the spectral flow through the transitions (shown in Fig.~\ref{specflowFULL} for the same $t,\Delta$ used in Fig.\,\ref{bdrynum}) as well as the presence of the zero eigenvalue in the spectrum, which we track by computing  minimum-modulus eigenvalue of $G(s,\varphi)$ (see Figs. \ref{bdrynum}(e)-(h)). Four main points are worth noticing:
\begin{enumerate}
\item For $N$ odd, Figs.\,\ref{bdrynum}(a) and \ref{bdrynum}(c) indicate that the open chain at $s=0$, where we know the system hosts a Krein collision at each eigenvalue, becomes dynamically unstable for arbitrarily small $s>0$, whenever $\varphi\not\in I_{\varphi_N}(s)$. Figs.\,\ref{specflowFULL}(a1) and \ref{specflowFULL}(b) also provide examples where the open chain remains stable under (boundary) perturbations corresponding to an increase of $s$. 

\item For $N$ odd, Fig.\,\ref{specflowFULL}(a2) further indicates that the dynamical phase transition between the $\pi/2$-twisted chain and the periodic chain at $s=1$  occurs when eigenvalues of opposite Krein signature become arbitrarily close. 
\item For $N$ even, Fig.\,\ref{specflowFULL}(b) reveals that the transition away from stability happens when pairs of eigenvalues with opposite Krein signature become arbitrarily close at finite $s>0$.

\item The value zero enters the spectrum of $G(s,\varphi)$ at the left dynamical phase boundary for arbitrary $N$,  
(in fact, at both boundaries if $N$ is odd), and the phase diagram is symmetric about $\varphi=\pi/2$. Interestingly, further numerics (data not shown) indicate that, for $N$ a multiple of 4, the right phase boundary is {\em still} defined as the locus in parameter space where the largest (in modulus) eigenvalue of $G(s,\varphi)$ is closest to zero as a function of \(s,\varphi\).
\end{enumerate}

Observation (i) demonstrates explicitly the behavior predicted by Lemma\,\ref{Kreinthms}, namely, stability-to-instability transitions are {\em generically} mediated by the coalescence of eigenvalues with opposite Krein signature. However, while this behavior is generic, it is {\em not} universal. Observations (ii) and (iii) together highlight an interesting {\em even-odd} effect, namely, the minimum stability height is only non-zero when $N$ is
even. Finally, observation (iv) provides evidence supporting the following conjecture:

\smallskip
\noindent
\textbf{Conjecture 1:} \textit{Generically, bosonic zero modes indicate dynamical phase boundaries}. 
\smallskip

While, as Fig.\,\ref{bdrynum} illustrates, the right phase boundary for even $N$ provides an example where {\em no} zero modes 
are hosted, we also see that if the two boundaries coalesce as $N$ grows, then, in the thermodynamic limit, {\em every dynamical phase transition of the BKC is defined by zero modes}. We now show that this does, in fact, happen. 
We do so by utilizing the techniques of Sec. \ref{GBTprimer} to determine the values of $s$ and $\varphi$ that support zero modes. Combining calculations that are detailed in \ref{GBTzero} with the symmetry of the dynamical phase diagram about $\varphi=\pi/2$, we find that the phase boundaries are parameterized by the curves 
\begin{equation}
\label{bdrycurve}
\cos(\varphi) =  \pm\frac{1}{2}\left\{\matrix{
(s+s^{-1})\text{sech}(Nr),& N \text{ even},
\cr
2\,\text{sech}(Nr),& N\text{ odd}, 
}\right. 
\end{equation}
which are shown in Fig.\,\ref{bdrynum}(a)-(d) are white, dashed lines. In particular, in term of the parameter $2r=\ln[(t+\Delta)/(t-\Delta)]$ previously introduced, the minimum height and width, which we introduced earlier on for characterizing the stable phase, are respectively determined by the following expressions:
\begin{equation}
{\delta s}_N = \left\{\matrix{
0& N \text{ odd} 
\cr
e^{-Nr}& N\text{ even}
}\right. ,\quad 
\sin({{\delta\varphi}_N}/{2}) = \text{sech}\left(N r\right). 
\end{equation}
Both quantities decrease rapidly with system size, confirming that that stable regions become of measure zero in parameter space in the thermodynamic limit. Thus, the transition lines in this case are precisely defined by the occurrence of zero modes, suggesting that Conjecture 1 may hold for \textit{all} QBHs in the thermodynamic limit. Similarly, for fixed $N$, both ${\delta s}_N$ and ${\delta\varphi}_N$ decrease as $\Delta$ approaches $t$, signaling the transition to the unstable \(\Delta>t\) phase. We note that in a recent paper \cite{hnchain}, a quantity analogous to ${\delta s}_N$ was calculated for the non-Hermitian fermionic Hatano-Nelson chain. As shown in \cite{clerkPRX}, the effective SPH of the BKC (without twisting) is unitarily equivalent to two copies of the Hatano-Nelson chain.  Since twisting the BCs manifestly destroys this unitary equivalence, it is remarkable that such strong similarities still exist.

As we explicitly show in \ref{GBTzero}, the zero eigenvalue hosts one (two) Jordan chain(s) of length two on the (left) phase boundaries for $N$ odd (even). This information allows us to better assess whether the rest of the spectrum hosts non-trivial Jordan chains or simple Krein collisions. For $N$ odd, we can do this numerically, by calculating the distance between the two eigenvectors (after correcting for arbitrary phases) that coalesce at each eigenvalue. We find that for odd system sizes between $N=5$ up to $N=55$, various choices of $t/\Delta\in(0,1)$, and at various points along the phase boundary, these distances vanish, indicating a loss of linear independence along phase boundaries at $s>0$. Thus, each eigenvalue hosts a Jordan chain of length two. For $N$ even, things are more complicated. As seen from Fig.\,\ref{specflowFULL}(b), the spectrum is always at least doubly degenerate, and splittings occur at different points in parameter space (the splitting at zero defining the phase boundary). In contrast, each eigenvalue splits simultaneously, at the same values of $s$ and $\varphi$, for $N$ odd. Due to these complications, we do not further assess the nature of these splittings;  we nonetheless conjecture that, like for zero frequency, they are induced by the formation of two length-two Jordan chains. 

Aside from the specific BKC example and the single-mode model considered in Sec.\,\ref{sec3}, Conjecture 1 is further motivated from a physical point of view. To see this, consider perturbing a QBH of the form Eq. \eref{genham} by adding a small term linear in the creation and annihilation operators (e.g., arising from a constant force). Intuition suggests that such a term may only elicit dynamical instability from a dynamically stable system if the latter possesses zero modes (e.g., a harmonic oscillator exhibits bounded motion even in a constant gravitational field, whereas a free particle or zero-frequency oscillator does not). Further, it can be shown that a QBH can ``absorb'' any such linear perturbation, meaning that the total Hamiltonian can be shown to be  unitarily equivalent to a purely QBH, {\em unless} it possesses zero modes. In this sense, zero modes are generically expected to define dynamical phase boundaries.

There is an intriguing point of contact between the BKC for $t>\Delta$ and the single-mode model of Sec.\,\ref{sec3}. It appears that the only phase boundary of the BKC that hosts (a macroscopic number of) Krein collisions is the line $s=0$, which physically represents a single (open) BC, hence a {\em single point} in parameter space. Similarly, there is only one point in the phase boundary of both the single-mode model, and the two-mode cavity QED model, that hosts a Krein collision, the origin. In contrast, the one-dimensional phase boundaries in both models are dominated by EPs. These observations suggest a second conjecture:

\smallskip
\noindent
\textbf{Conjecture 2:} \textit{Generically, the $(d-1)$-dimensional phase boundaries of the $d$-dimensional dynamical phase diagram of a QBH are characterized completely by EPs, whereas the $(d-2)$-dimensional boundaries are characterized completely by Krein collisions.  }
\smallskip

\begin{figure}[tb!]
\centering
\includegraphics[width=.8\textwidth]{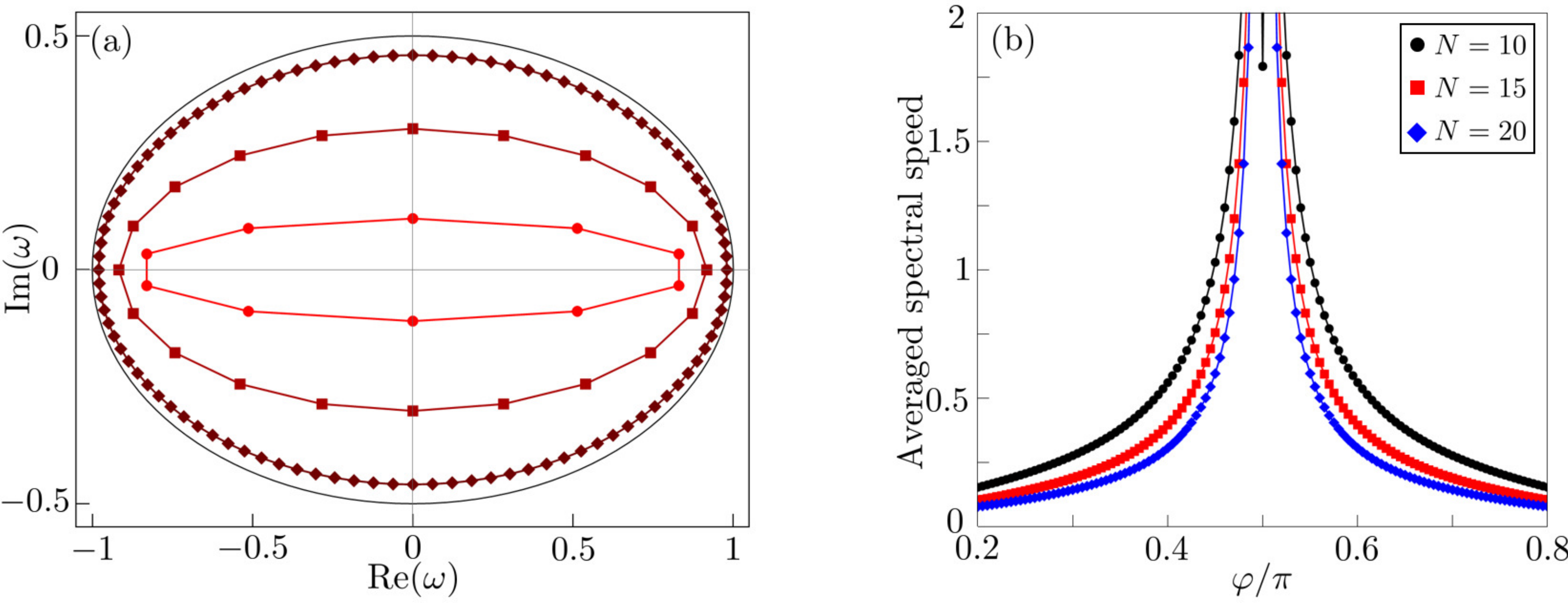}
\caption{(a) The spectrum with $t=1$, $\Delta=0.5$, $s=1$, and $\varphi=0.99\cdot (\pi/2)$ for system sizes, $N=10, 20,$ and $100$, from inner to outer. The solid outermost ellipse traces out the periodic spectrum for $N\to\infty$. (b) The spectral speed $d|\omega_m|/d\varphi$ for 
same $t$ and $\Delta$ for various $N$, averaged over all eigenvalues.}
\label{incrN}
\end{figure}

We conclude this section by investigating the spectral behavior near the dynamical phase transition as a function of system size. In Fig.\,\ref{incrN}(a), we show the spectrum of $G(1,\varphi)$ in the vicinity of the phase boundary around $\varphi=\pi/2$. We see that as $N$ increases, the spectrum clings more strongly to the ellipse defined by the periodic/anti-periodic spectrum. Thus, the ``speed" at which the spectrum splits from the real axis increases dramatically as $N$ increases. This may be quantified by examining the \textit{spectral speed}, $d|\omega_m|/d\varphi$, 
for which we conjecture that that $\lim_{N\to\infty} d|\omega_m|/d\varphi\propto\delta(\varphi-\pi/2)$. Although we cannot evaluate this quantity analytically for $t\neq \Delta$, we can use the exact analytical solution at $t=\Delta$, $s=1$ and $\varphi\in[0,\pi]$ as a point of comparison. From Eq.\,\eref{tzDarbphi}, it follows that 
\begin{equation*}
\frac{d|\omega_m|}{d\varphi} = \frac{1}{N} |\cos\varphi|^{1/N-1}.
\end{equation*}
In particular, we see in this case that $\lim_{N\to\infty} d|\omega_m|/d\varphi\propto\delta(\varphi-\pi/2)$, as conjectured. Thus, in the thermodynamic limit, this quantity contains an extreme non-analyticity, as we also illustrate in Fig.\,\ref{incrN}(b).

\subsubsection{Krein phase rigidity.} 
\begin{figure}[b!]
\centering
\includegraphics[width=.9\textwidth]{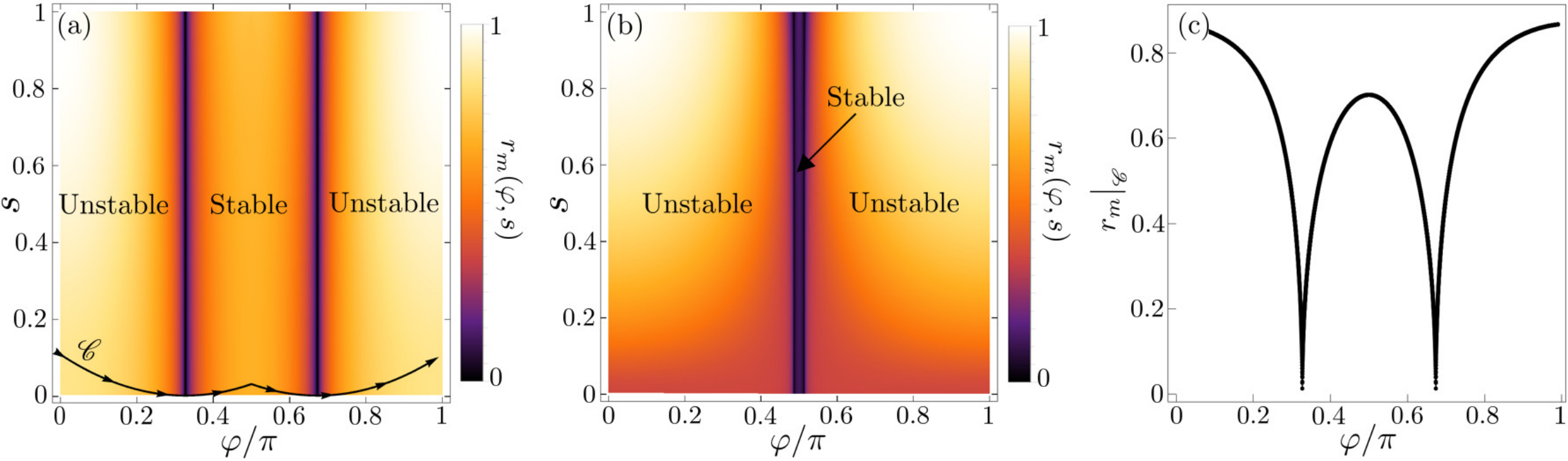}
\vspace*{-1mm}
\caption{The KPR evaluated numerically as a function of boundary parameters \(s, \varphi\), on a grid of spacing $0.002$, for $t=1$ and $\Delta=0.25$. (a) $N=5$ (b) $N=15$ (we restrict to $N$ odd to avoid difficulties in maintaining continuous eigenvector-tracking in the presence of the doubly degenerate spectrum for $N$ even). (c) The KPR evaluated along the contour $\mathscr{C}$ in (a), defined by $s(\varphi)\equiv (\varphi-\varphi^-)^2$, for $\varphi<\pi/2$, and $(\varphi-(\pi-\varphi^-))^2$ otherwise, with $\varphi^-$ being the angle defining the left dynamical phase boundary. }
\label{PR1}
\end{figure}

In Sec.\,\ref{Krein}, we argued that the Krein phase rigidity of Eq.\,\eref{pr} should be able to detect both EPs and Krein collisions at a boundary between a stable and an unstable dynamical phase. Here we demonstrate this capability in the context of the BKC by evaluating numerically the KPR of a representative eigenvector as a function of \(s\) and \(\varphi\), see Fig.\,\ref{PR1}.  We confirm that the KPR does, in fact, vanish at the phase boundaries for $s>0$. In order to understand the behavior of the KPR at $s=0$, we again point out that this corresponds to precisely one BC (open). Thus, the limiting value of the KPR at $s=0$ is contour-dependent, as it was in both the single-mode and the cavity QED models of Secs.\,\ref{sub:example} and \,\ref{sub:cQED}, respectively. In particular, we emphasize that the KPR evaluated along any contour in parameter space whose transition from instability $(\varphi\not\in I_{\varphi_N})$ to stability $(\varphi\in I_{\varphi_N})$ is mediated by the point $s=0$ will vanish at this point. This can be seen in Fig.\,\ref{PR1}(c), where the KPR is evaluated along a parabolic contour which passes through the point $s=0$ at precisely the twisting angles defining the dynamical phase boundaries and along which the effective SPH $G(s, \varphi)$ remains diagonalizable. As predicted in Sec.\,\ref{sec3}, the KPR detects these Krein-collision-dominated dynamical phase transitions, despite the lack of EPs.

\begin{figure}[t!]
\centering
\includegraphics[width=0.85\textwidth]{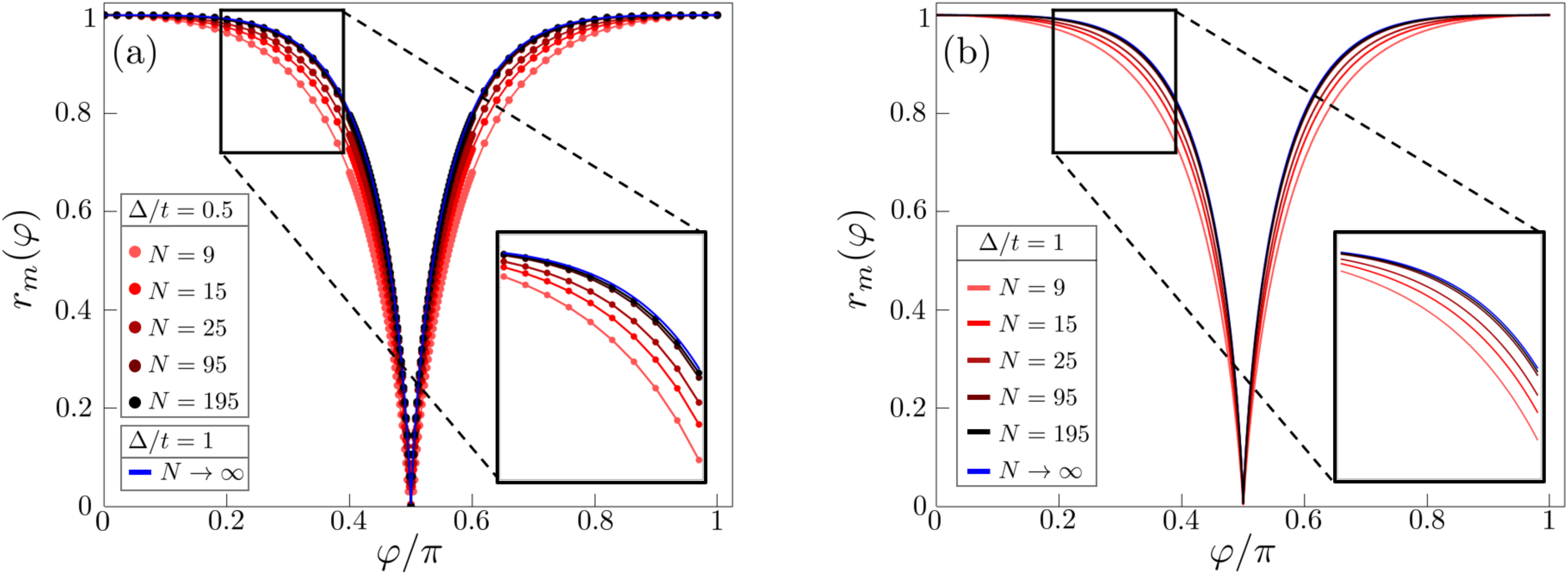}
\caption{The response of the KPR to system size (a) around the phase boundary near $\varphi=\pi/2$ for $t=1$, $\Delta=0.5$ and (b) around the EP near $\varphi=\pi/2$ for $t=\Delta=1$. In (a), the blue line follows from Eq.\,\eref{PREPTL}.}
\label{PR2}
\end{figure}

To illustrate the response of the KPR to system size, let us focus on twisted BCs, 
$s=1$ and $\varphi$ arbitrary, see 
Fig.\,\ref{PR2}(a). As $N$ increases, the KPR $r(\varphi)$ approaches a continuous curve, which however sharply detects the transition point $\varphi=\pi/2$, akin to a ``stability order parameter''. While this relatively tame behavior contrasts with the extreme sensitivity of the spectrum to system size, it is interesting to note that 
a similar, ``less extreme'' response to increase in system size has been reported for the eigenvectors of non-Hermitian asymmetric hopping models, through studies of fidelity decay and Loschmidt echo \cite{longhi}. 
We can compare the numerical results of Fig.\,\ref{PR2}(a) for \(t=1\) and \(\Delta=0.5\) with analytical results available
for \(t=\Delta\). In particular, our exact analytical solution for $G(1,\varphi; t=\Delta)$ allows us to investigate the 
KPR in the vicinity of the EP at $\varphi=\pi/2$. By taking $N$ to be odd and using the eigenvectors from Eq.\,\eref{tzDarbphiWF}, we find 
\begin{equation*}
r_m(\varphi) = \frac{1}{\braket{\psi_m|\psi_m}} = \frac{N |\cos\varphi |}{ |\cos\varphi|^2-1} \Big( |\cos\varphi|^{2/N}-1\Big), \qquad 
\varphi\in[0,\pi] .
\end{equation*}
A plot of $r_m(\varphi)$ for various system sizes $N$ is given in Fig.\,\ref{PR2}(b). 
In the limit as $N\to\infty$, 
\begin{equation}\label{PREPTL}
\lim_{N\to\infty}r_m(\varphi) = \frac{ |\cos\varphi|  \ln( |\cos\varphi| )}{ |\cos\varphi|^2-1}.
\end{equation} 
Since $r_m(\varphi)$ vanishes as $ |\cos\varphi| \to 0$, the EP at \(s=1, \varphi=\pi/2\) is indeed detected by the KPR. 

\section{Further implications}
\label{furtherimp}

\subsection{Phase-dependent transport precludes dynamical stability}
\label{sub:pd}

One of the interesting features of the BKC Hamiltonian of Eq.\,\eref{BKCHam} is that the real ($x$) and imaginary ($p$) parts of any coherent-state preparation of the chain propagate independently of one another. Such a {\em phase-dependent propagation} arises due to the decoupling of the Heisenberg equations of motion for the Hermitian quadratures $x_j$ from those governing the Hermitian quadratures \(p_j\) under open and periodic BCs: 
\begin{equation*}
\frac{d}{dt}\left[\matrix{
x_j\cr p_j
}\right] = \frac{1}{2}\left[\matrix{
J x_{j-1} - f x_{j+1} \cr
f p_{j-1} - J p_{j+1}
}\right] , \qquad J \equiv t+\Delta,\; f \equiv t -\Delta. 
\end{equation*}
This feature is referred to as ``phase-dependent chiral transport" in \cite{clerkPRX}. In this context, the word ``chiral"  is meant to further highlight the asymmetric way in which each quadrature is influenced by couplings between adjacent lattice sites, with maximal asymmetry and uni-directional transport being approached as $t\rightarrow \Delta$. 

Considering more general BCs, our analysis reveals that although the features characteristic of phase-dependent chiral transport survive for arbitrary $s\in[0,1]$ and $\varphi=0$, they are fragile from a dynamical perspective, since the system is either dynamically stable with Krein collisions in the spectrum or dynamically unstable. Is it possible to have phase-dependent chiral transport in chains with more favorable stability properties? Unfortunately, the answer is in the negative. The decoupling of the equations of motion is the key prerequisite for both phase-dependent and chiral transport, and this prerequisite condition forces the spectrum of $G$ to host Krein collisions, assuming it is not already unstable. More formally:

\begin{proposition}
\label{chiraltransport}
Let \(\widehat{H}\) be a dynamically stable QBH that supports phase-dependent transport. Then, the spectrum of the effective SPH $G$ necessarily hosts Krein collisions. In other words, the system is at the cusp of instability.
\end{proposition}
\smallskip
\noindent
{\em Proof:} 
With reference to Eq. \eref{genham} and the Heisenberg equations of motion in Eq. \eref{Heis}, the resulting equations for the quadrature modes are found as
\begin{equation*}
\frac{d}{dt}\left[\matrix{
x_j \cr p_j
}\right] = \sum_{k=1}^N\left[\matrix{C_{kj}x_k + T_{jk}p_k
\cr
-V_{jk}x_k - C_{jk}p_k
}\right], \quad C\equiv\textup{Im}(\Delta-K), \,V\equiv\textup{Re}(K+\Delta), \,T\equiv\textup{Re}(K-\Delta).
\end{equation*}  
Hence, the dynamics of the quadrature modes decouple if and only if both $K$ and $\Delta$ are purely imaginary and if this condition is fulfilled, then $[G,\tau_1]=0$. 
Since, by assumption, $G$ is dynamically stable, for each of the $N$ real eigenvalue pairs $(\omega,-\omega)$ there are eigenvectors $\ket{\psi^+}$ and $\ket{\psi^-}=\mathcal{C}\ket{\psi^+}$, satisfying $G\ket{\psi^\pm}=\pm\omega\ket{\psi^\pm}$, $\braket{\psi^\pm|\tau_3|\psi^\pm}=\pm 1$ and $\braket{\psi^\pm|\tau_3|\psi^\mp}=0$. As a consequence of $[G,\tau_1]=0$ we have that $\ket{\phi^+}=\tau_1\ket{\psi^+} = \left(\ket{\psi^-}\right)^*$ is an eigenvector of $G$ corresponding to eigenvalue $\omega$ and has Krein signature $\braket{\phi^+|\tau_3|\phi^+}=-1$. Thus, $\omega$ hosts a Krein collision and so the system sits on the cusp of instability in the sense of Lemma \ref{Kreinthms}(ii).  \hfill$\Box$
\smallskip

From a many-body perspective,  if the matrices $K$ and $\Delta$ have purely imaginary entries (the ``decoupling condition"), then the Hamiltonian $\widehat{H}$ is, like the GKC $\widehat{H}_O$, necessarily odd under time reversal. Thus, $\widehat{H}$  cannot be thermodynamically stable and even if it is dynamically stable, this feature is necessarily fragile.

\subsection{Interplay with topology}
\label{topo}
Looking for hints of topological physics in systems of free bosons, several researchers have 
developed mappings from free fermions to free bosons that preserve specific features of 
interest. For example, in our work in \cite{qiaoru} we explore a mapping from fermions to bosons
that preserve zero-energy modes. The BKC of the previous section is the result of another 
such mappings; specifically, by design, one that preserves a topological invariant, the winding number. To put things in context, we state here the mapping of \cite{clerkPRX} in greater generality. Let 
\begin{eqnarray*}
\widehat{H}_f=\sum_{i,j}[K_{ij}c^\dagger_ic_j+\frac{1}{2} (\Delta_{ij}c_i^\dagger c_j +\text{H.c.})], \qquad 
K^\dagger=K, \; \Delta^T=-\Delta, 
\end{eqnarray*}
denote a general quadratic fermionic Hamiltonian \cite{ripka}. The mapping of interest is restricted to the subclass of Hamiltonians obeying  \(K^T=K\) and \(\Delta^\dagger =\Delta\). Altogether, 
these additional conditions imply that \(K\) is purely real\ and \(\Delta\) is purely imaginary. The output of the mapping is the QBH
\begin{eqnarray}
\label{fgob}
\widehat{H}_b=\sum_{i,j}[\Delta_{ij}a^\dagger_ia_j+\frac{1}{2} (iK_{ij}a_i^\dagger a_j^\dag +\text{H.c.})] , 
\end{eqnarray}
that is,  with respect to Eq. \eref{genham}, we get $K_{ij}^b =\Delta_{ij}^f$, $\Delta_{ij}^b= i K_{ij}^f$ (recall that, for the BKC specifically, we have $K_{ij}= \frac{i t}{2} (\delta_{i,j+1}- \delta_{i+1,j})$, $\Delta_{ij}= \frac{i \Delta}{2} (\delta_{i,j+1}+ \delta_{i+1,j})$ in the bulk, and $K_{1N}=\frac{ist}{2}e^{i\varphi}=K_{N1}^*$, $\Delta_{1N}=\frac{is\Delta}{2}e^{i\varphi}=\Delta_{N1}$). Hence, according to Proposition 5.1, these bosonic counterparts of QFHs are either dynamically unstable 
or at the cusp of instability. This particular mapping from free fermions to free bosons 
may preserve topological invariants at the expense of sacrificing robust dynamical stability. 

This conclusion is tightly linked the factor of \(i\) that appears in front of \(K\) in Eq.\,\eref{fgob}. Interestingly, the deleterious factor of \(i\) is introduced precisely so that the winding number of the fermionic Kitaev chain survives the passage to bosons. 
For periodic BCs and zero chemical potential, the Hamiltonian of the fermionic Kitaev chain may be written in momentum space as
\begin{equation*}
\widehat{H}_{f} =\sum_{k\in\text{BZ}}\left[ t\cos(k)c_k^\dag c_k + i\frac{\Delta}{2}\sin(k)\left( c_k^\dag c_{-k}^\dag - \text{H.c.}\right)\right] \equiv \frac{1}{2}\sum_{k\in\text{BZ}}\hat{\Psi}^\dag_k (\mathbf{d}_f(k)\cdot \bm{\sigma}) \hat{\Psi}_k ,
\end{equation*}
where BZ denotes the appropriate Brillouin zone, 
$\hat{\Psi}_k=[c_k,c_{-k}^\dag]^T$, and $\mathbf{d}_f(k) = [0,-\Delta\sin(k),t\cos(k)]^T$. The SPHs $H=\mathbf{d}_f(k)\cdot\bm{\sigma}$ has a well-defined topological invariant -- namely, the number of times $\mathbf{d}_f(k)$ wraps around the origin in the $yz$-plane -- which is protected by a chiral symmetry. Implementing the above mapping, the bulk Hamiltonian for the corresponding QBH is given by $\mathbf{d}_b(k)\cdot \bm{\sigma}$ where, $\mathbf{d}_b(k) =[0,-\Delta \cos(k),t\sin(k)]^T$ 
In real space, this is precisely the Hamiltonian of Eq.\,\eref{BKCHam} (for PBCs). The vector $\mathbf{d}_b(k)\cdot \bm{\sigma}$ winds around the origin just like the original fermionic one does, 
but only because of the factor of \(i\) in front of \(K\) in Eq.\,\eref{fgob}.

Taking all of this into consideration, it seems both peculiar and suggestive that the BKC is 
dynamically unstable for periodic BCs and stable, but at the cusp of instability, for OBCs.
Is this feature somehow related to the winding number? The following
analysis suggest the the answer is in the negative. Consider again the Kitaev chain, but now 
including a non-zero chemical potential, 
\begin{equation*}
\widehat{H}_{f} =\sum_{k\in\text{BZ}}\left[ (\mu+t\cos(k))c_k^\dag c_k + i\frac{\Delta}{2}\sin(k)\left( c_k^\dag c_{-k}^\dag - \text{H.c.}\right)\right] \equiv \frac{1}{2}\sum_{k\in\text{BZ}}\hat{\Psi}^\dag_k (\mathbf{d'}_f(k)\cdot \bm{\sigma}) \hat{\Psi}_k ,
\end{equation*}
where $\mu,t,\Delta>0$ and now $\mathbf{d'}_f(k)=[0,-\Delta \sin(k),\mu+t\cos(k)]^T$. The winding number is 
non-zero for $\mu<t$ in the topologically non-trivial phase of the superconductor. In real space and
for open BCs the associated bosonic Hamiltonian is now 
\begin{equation}\label{zms}
\widehat{H}_b = \widehat{H}_O + \frac{i\mu}{2}\sum_{j=1}^N \left( (a_j^\dag)^2 - a_j^2\right) ,
\end{equation}
where $\widehat{H}_O$ is the BKC of Eq.\,\eref{BKCHam}. Therefore, the chemical potential term in the fermionic 
chain maps to a sum of degenerate parametric amplifier terms in the bosonic chain. These additional terms modify the bosonic effective SPH as $G_O \mapsto  G_O + i\mu \tau_1$, causing the originally doubly degenerate eigenvalues $\omega_m$ of \(G_O\) to split into $\omega_m \pm i \mu$, with $\omega_m=\sqrt{t^2-\Delta^2}\cos(m\pi/(N+1))$, $m=1,\ldots,N$. Thus, for $\mu\neq 0$, the system is \textit{always} dynamically unstable. And, this conclusion holds irrespective of the condition $\mu<t$ required for a non-zero winding invariant.  

At this point it would seem that we have managed to discount a direct topological origin to stability properties of QBH. However, as we noted in the Introduction, the role of topology for QBHs remains, at best, only partially understood as yet. The highly non-trivial nature of this interplay is nicely illustrated by two further observations pertaining to the comparison between the Kitaev chain and its bosonic counterpart. First, our analysis establishes that the BCs that render the BKC dynamically stable for all $N$ and $t>\Delta$ (open and $\pi/2$-twisted BCs) are {\em precisely the same} that host Majorana zero modes in the corresponding fermionic chain \cite{katsura}. Second, the Hamiltonian \(\widehat{H}_b\) of Eq.\,\eref{zms}
displays localized approximate zero modes for $t=\Delta$ and every finite $N$. Specifically, one can verify that the following left- and right-localized modes
\begin{equation*}
\widehat{\gamma}_L \equiv \sum_{j=1}^N\delta^{j-1} x_j = \widehat{\gamma}_L^\dag,\qquad 
\widehat{\gamma}_R \equiv \sum_{j=1}^N\delta^{N-j} p_j = \widehat{\gamma}_R^\dag , \qquad \delta=-\mu/t, 
\end{equation*}
satisfy $[\widehat{H}_b,\widehat{\gamma}_L] = it \delta^N x_N$, $[\widehat{H}_b,\widehat{\gamma}_R] = it \delta^N p_1$ and that, furthermore $[\widehat{\gamma}_L,\widehat{\gamma}_R] = i N \delta^{N-1}$. 
These Hermitian ``Majorana bosons" at zero frequency are normalizable, that is, exponentially localized
precisely if \(\mu<t\) -- {\em in perfect correspondence} with the topologically-sourced Majorana zero modes of
the fermionic chain. Captivating as they are, we do not know as yet whether these ``bosonic shadows'' of Majorana 
physics are topological in any suitable sense of the word.

\section{Conclusion and outlook}

We have systematically investigated the landscape of free-boson dynamical phase diagrams from a general dynamical stability perspective and by way of paradigmatic examples of increasing physical complexity: a single bosonic mode, a two-mode system modeling a realistic cavity QED setting, and a bosonic version of the Kitaev-Majorana chain. Our general framework for QBHs combines tools from pseudo-Hermitian, $\mathcal{P}\mathcal{T}$-symmetric, and non-Hermitian quantum mechanics with the Krein stability theory of dynamical systems in indefinite inner-product spaces.
Two key new results emerge from this analysis: First, all free-boson systems are $\mathcal{P}\mathcal{T}$-symmetric in a suitable sense and their dynamical phase diagrams are controlled by the fate of this symmetry. From a many-body standpoint, this symmetry is broken precisely when the QBH can no longer be diagonalized in terms of canonically bosonic Bogoliubov quasi-particles. Second, we argued 
that dynamical phase boundaries can be detected by a KPR indicator, which naturally extends the notion of phase rigidity widely employed within semiclassical (non-Hermitian) treatments of open quantum systems. Bosonic dynamical phase boundaries can consist of loci of exceptional points, where diagonalizability of the effective SPH is lost, but also, remarkably, or Krein collisions where degenerate real eigenvalues split into the complex plane without loss of diagonalizability. 

To illustrate and validate our framework, 
we obtained a complete characterization of the dynamical stability phase diagram of the BKC for a two-parameter family of BCs that interpolates between open and periodic BCs and includes twisted BCs as notable case. In particular, for both open and $\pi/2$-twisted BCs we were able to diagonalize in closed form the BKC, by employing for the first time an exact diagonalization procedure  developed in the context of free fermions as a generalization of Bloch's theorem to systems where translational symmetry is broken by BCs. In particular, the use of this procedure proved instrumental to access to study the stability phase diagram as a function of system size, supporting the emergence of extreme non-analyticity of the spectral response in the thermodynamic limit. We confirmed explicitly that the KPR vanishes at all the phase boundaries, as expected from our general arguments, and sharply detects G${\cal P}{\cal T}$-symmetry-breaking phase transitions in the thermodynamic limit. Remarkably, our analytical solutions prove that the BCs that host Majorana zero modes in the fermionic Kitaev-Majorana chain are precisely the same that allow for dynamical stability in its bosonic counterpart. 

One of the most interesting physical properties of the BKC is that it can support phase-dependent chiral transport, stemming from  the decoupling of the evolution of the real and imaginary parts of coherent excitations. Using tools from Krein stability theory, we showed that any QBH exhibiting a similar ``decoupling condition'' 
describes a thermodynamically unstable system, which is either dynamically unstable or sitting at the cusp of dynamical instability. As a consequence, unless some additional  protection mechanism is in place, stable phase-dependent transport is fragile against perturbations. Finally, the BKC is the result of applying a certain mapping to the fermionic chain. By exploring in more generality the idea of mappings between fermionic and bosonic systems, so that a specified set of topological invariants is preserved, we have shown that while topology may not directly influence the dynamical phase of a QBH, bosonic analogues to Majorana zero modes exist in a further generalized BKC model, for the same parameters that correspond to a topologically non-trivial fermionic phase. Achieving a clearer picture of what role topology may play (if any) in informing the dynamical properties of bosonic systems, along with an explanation of whether the seemingly special status of certain BCs is simply a coincidence or rather has a deeper significance, are natural next questions we leave to future research.

Our analysis also points to a number of additional directions for investigation.  On the one hand, for closed systems of bosons, it is worth to explore the role of KPR and G$\mathcal{P}\mathcal{T}$ symmetry beyond the mean-field approximation. For example, a question of fundamental relevance would be to determine whether the KPR may be used to assess the validity of the underlying quadratic approximations for interacting systems, both at equilibrium or possibly under a time-dependent driving. 
Furthermore, is there a role for Krein stability theory or G$\mathcal{P}\mathcal{T}$ symmetry-breaking to play in the full Fock space of systems of interacting bosons? 
On the other hand, our framework and tools can be extended to a large class of quadratic (fermionic or bosonic) open systems -- either described, semi-classically, in terms of non-Hermitian many-body effective Hamiltonians or, within a fully quantum formalism, by Lindblad (Markovian) master equations \cite{ViolaTAC}. In particular, we expect that the KPR may find natural applications in the context of exploring topological phenomena related to $\mathcal{P}\mathcal{T}$-symmetric quantum quenches \cite{Qiu2019} or  dynamically encircling EPs \cite{Du2020}, or in the context of $\mathcal{P}\mathcal{T}$-symmetry-breaking enhanced quantum metrology \cite{Nori2016}. Likewise, while for quadratic Lindbladians there exist (generically) non-Hermitian matrix analogues to the effective SPH \cite{prosen,poletti}, important differences are also to be expected and, indeed, have been recently pointed out for instance in the nature of the underlying EPs \cite{NoriEP2019}.
One of our next steps will thus be to understand the extent to which the KPR 
(or some suitable modification of it) may still provide a useful diagnostic tool in open quantum dynamical settings and, if so, be employed to characterize steady-state stability phase diagram in driven-dissipative many-body quantum systems, including the possible emergence of topological features or exotic transport phenomena \cite{Rudner2019,Song2019}.

\section*{Acknowledgments}

It is a pleasure to thank Abhijeet Alase and Gerardo Ortiz for many stimulating discussions, and Vittorio Peano, Hermann Schulz-Baldes, and Phan Th\`{a}nh Nam for valuable correspondence. V.~P.~F. is grateful to Schr\"odinger, the cat, for being a continuing source of inspiration. 
Work at Dartmouth was partially supported by the US NSF through Grant No. PHY-1620541 and the Constance and Walter Burke Special Projects Fund in Quantum Information Science. E.C. acknowledges partial support from a 2019 seed grant from Suny Polytechnic Research Office.

\appendix

\section{Diagonalization of corner-modified, banded block-Toeplitz matrices}
\label{GBTappendix}

In Sec.\,\ref{GBTprimer} we described the solutions for the bulk equation, Eq. \eref{BBbulk}, and how they can be used to construct a complete basis of generalized eigenvectors (with minor restrictions on the lengths of the Jordan chains) of a clean, 
finite-range bosonic Hamiltonian. In this appendix, we present the derivation of this Ansatz.

Recall that to each $G_O$, we can associate a translation-invariant auxiliary effective Hamiltonian $\bm{G}$. Then if $\bm{G}\Psi = \omega \Psi$, it follows that $\ket{\psi} \equiv \bm{P}_{1,N}\Psi$, with $\bm{P}_{1,N} = \sum_{j=1}^N \ket{j}\bra{j}\otimes \mathds{1}_2$, is a solution of the bulk equation. In the generic case where $\det g_{\pm R}\neq 0$, this method yields the complete set of solutions to the bulk equations, i.e., $\ker P_B(G_O-\omega \mathds{1}_{2N}) = \bm{P}_{1,N}\ker(\bm{G}-\omega \bm{1})$. For the time being, we restrict ourselves to this case.

If we were interested in diagonalizing $\bm{G}$ on its own, we would restrict to only the eigenvectors that are normalizable on the corresponding Hilbert space. Crucially, this does {\em not} capture the full kernel of $\bm{G}-\omega \bm{1}$, however: since we only consider the finite-lattice projections, the non-normalizable elements of $\ker(\bm{G}-\omega \bm{1})$ also provide solutions to the bulk equation. Furthermore, in the space of all bi-infinite sequences, the left and right translation operators $\bm{T}$ and $\bm{T}^{-1}$ are no-longer unitary and so these operators need not have spectra restricted to the unit circle. 

As noted in the main text, the translation invariance of $\bm{G}$ manifests as the vanishing commutators $[\bm{G},\bm{T}] = [\bm{G},\bm{T}^{-1}]=0$.  Hence, it is possible to construct simultaneous eigenvectors of $\bm{G}$, $\bm{T}$, and $\bm{T}^{-1}$. The simultaneous eigenvectors of $\bm{T}$ and $\bm{T}^{-1}$ are given by $\Phi_{z,1} \equiv \sum_{j\in \mathbb{Z}} z^j \ket{j}$, where $z$ is an arbitrary, non-zero complex number. Explicitly, $\bm{T} \Phi_{z,1} = z\Phi_{z,1}$ and  $\bm{T}^{-1} \Phi_{z,1} = z^{-1} \Phi_{z,1},$ which immediately lead to the identity
\begin{equation}
\bm{G} \Phi_{z,1} \ket{u} = \Phi_{z,1} G(z)\ket{u}, \qquad G(z)  \equiv h_0 + \sum_{r=1}^R\left( z^r g_r + z^{-r} g_{-r}\right), 
\label{eq:reduced}
\end{equation}
where $\ket{u}\in\mathbb{C}^2$ is arbitrary. We call $G(z)$  the \textit{reduced bulk effective Hamiltonian} and note that $G(z=e^{ik})$ is the usual Bloch Hamiltonian that arises in 1D systems under Born-von-Karman (periodic) BCs. Thus, $G(z)$ is the analytic continuation of $G(e^{ik})$ off the unit circle. Furthermore, we see that for any $z\neq 0$ such that $G(z)\ket{u}=\omega \ket{u}$, $\Phi_{z,1}\ket{u}$ is an eigenvector of $\bm{G}$ with eigenvalue $\omega$. 

To continue, we define the complex characteristic polynomial $P(\omega,z) \equiv z^{4R} \det ( H(z) - \omega \mathds{1}_2).$ We call an eigenvalue $\omega$ {\em regular} if $P(\omega,z)$ is not the zero polynomial. Otherwise, we say $\omega$ is {\em singular}. For the applications in this paper, it suffices to restrict to the eigenvalues $\omega$ that are regular. For a fixed $\omega$, let $\{z_\ell\}_{\ell=1}^n$ denote the $n$ distinct roots of $P(\omega,z)$ and $\{s_\ell\}_{\ell=1}^n$ denote their corresponding multiplicities. Generically, $G(z_\ell)$ will have $s_\ell$ eigenvectors $\{\ket{u_{\ell s}}\}_{s=1}^{s_\ell}$ satisfying $G(z_\ell)\ket{u_{\ell s}} = \omega \ket{u_{\ell s}}$, in which case, the vectors
\begin{equation}\label{z1}
\ket{z_\ell,1} \otimes \ket{u_{\ell s}}  \equiv \sum_{j=1}^N z^j_\ell \ket{j} \otimes \ket{u_{\ell s}} = \bm{P}_{1,N} \Phi_{z_\ell,1}\ket{u_{\ell s}}, 
\end{equation}
are solutions to the bulk equation, and akin to Bloch waves with complex momentum.

When the reduced bulk Hamiltonian $G(z_\ell)$ has less than $s_\ell$ eigenvectors, the remaining solutions are constructed from the generalized eigenvectors of the left and right translation operators. The sequences 
\begin{equation*}
\Phi_{z,\nu} \equiv \frac{1}{(\nu -1)!}\partial_z^{\nu-1} \Phi_{z,1}
\end{equation*}
span the kernel of $(\bm{T}-z)^s$ for $\nu=1,\ldots, s$. Furthermore,
\begin{equation*}
\bm{G} \Phi_{z,n} \ket{u} = \frac{1}{(n-1)!} \partial_z^{n-1}\Phi_{z,1} H(z) \ket{u}.
\end{equation*}
One can then show that the sequence $\Psi \equiv \sum_{n=1}^\nu \Phi_{z,n} \ket{u_n}$ satisfy
\begin{equation*}
\bm{G}\Psi = \sum_{n=1}^\nu \sum_{m'=1}^\nu \Phi_{z,m} [G_\nu(z)]_{mm'} \ket{u_{m'}},
\end{equation*}
where $G_\nu(z)$ is an upper-triangular block-Toeplitz matrix with non-zero blocks
\begin{equation}\label{Gnuz}
[G_\nu(z)]_{mm'} = \frac{1}{(m'-m)!} \partial_z^{m'-m} G(z),\quad 1\leq m \leq m' \leq \nu.
\end{equation}
It can then be shown that the eigenspace of $\bm{G}$ corresponding to eigenvalue $\omega$ is a direct sum of $n$ vector spaces spanned by generalized eigenvectors of $\bm{T}^{\pm 1}$ of the form
\begin{equation*}
\Psi_{\ell s} = \sum_{\nu=1}^{s_\ell} \Phi_{z_\ell,\nu} \ket{u_{\ell s \nu}},
\end{equation*}
where the linearly independent vectors $\{ u_{\ell s \nu}\}$ are chosen in such a way that $G_{s_\ell}(z_\ell)\ket{ u_{\ell s}} = \omega \ket{u_\ell s}$ with $\ket{u_{\ell s}} = [\ket{u_{\ell s 1}},\ldots, \ket{u_{\ell s s_\ell}}]^T$. With these, we obtain $\sum_{\ell =1}^n s_\ell$ solutions to the bulk equation given by
\begin{equation*}
\ket{\psi_{\ell s}} = \sum_{\nu =1}^{s_\ell} \ket{z_\ell,\nu}\ket{u_{\ell s \nu}}, \quad \ket{z_\ell,\nu} = \bm{P}_{1,N}\Phi_{z,\nu}.
\end{equation*}

If $g_{\pm R}$ are not invertible, then there exists $2s_0 \equiv 4R-\sum_{\ell=1}^n s_\ell$ additional boundary localized solutions to the bulk equation, where $s_0$ is the multiplicity of $z=0$ as a root of the characteristic polynomial $P(\omega,z)$ for a given regular eigenvalue $\omega$. We will now demonstrate how to construct the left $(j=1)$ localized solutions. Since these solutions emerge due to the truncation of the bi-infinite lattice to a finite one, we consider the half-infinite auxiliary effective Hamiltonian and unilateral shift operators 
 \begin{equation*}
\bm{G}_- \equiv \bm{1}_-\otimes g_0 + \sum_{r=1}^R\left(\bm{T}^r_- \otimes h_r + \bm{T}^{* r}_-\otimes g_{-r}\right), \quad
\bm{T}_- \equiv \sum_{j=1}^\infty \ket{j}\bra{j+1},  \quad
\bm T_-^* \equiv \sum_{j=1}^\infty \ket{j+1}\bra{j}.
 \end{equation*}
The corresponding half-infinite bulk projector is 
\begin{equation*}
\bm{P}_B^- \equiv \sum_{j=R+1}^\infty  \ket{j}\bra{j}\otimes \mathds{1}_2 = \bm{T}^{*R}_- \bm{T}_-^R\otimes \mathds{1}_2.
\end{equation*}
Now, suppose there is a vector $\bm{\Upsilon}^-$, that solves the half-infinite bulk equation $\bm{P}_B^-\left( \bm{G}_--\omega \bm{1}_-\right) \bm{\Upsilon}^-=0$. Then one can verify that $\ket{\psi}=\bm{P}_{1,N}\bm{\Upsilon}^-$ is a solution to the bulk equation. The emergent solutions are precisely those derived from the half-infinite bulk equation and not the bi-infinite eigenvalue problem. 
Since $\bm{T}_-\bm{T}^{*}_-=\bm{1}_-$, we may write $\bm{P}_B^-\left( \bm{G}_--\omega \bm{1}_-\right) =\bm{T}^{*R}_- K^{-}(\omega,\bm{T}_-)$, where $K^{-}(\omega,z)$ is the matrix polynomial 
$
 K^-(\omega,z) \equiv z^R \left( G(z) -\omega \mathds{1}_2\right). $
Thus, the $s_0$ left-localized emergent solutions to the bulk equation are determined by the kernel of the matrix $K_{s_0}^-(\omega,z_0=0)\equiv K^-(\omega)$, with $K^-_\nu(\omega,z)$ constructed exactly as in Eq.\,\eref{Gnuz}. Given a basis $\{\ket{u_s^-}\}_{s=1}^{s_0}$ for $\ker K^-(\omega)$, with $\ket{u_s^-} = [ \ket{u_{s1}^-}, \ket{u_{s2}^-} \ldots \ket{u_{s s_0}^-}]^T$, we can construct $s_0$ left localized solutions to the bulk equation given by
\begin{equation*}
\ket{\psi_s^-} = \sum_{j=1}^{s_0} \ket{j} \ket{u_{sj}^-}.
\end{equation*}

The remaining $s_0$ right-localized solutions, with support on $j=N$, can be found in an analogous way. Explicitly, they can be constructed using the kernel vectors $\{\ket{u_s^+}\}_{s=1}^{s_0}$ of the matrix $K^+(\omega) = \tau_3[K^-(\omega)]^\dag\tau_3$. That is, if $\ket{u_s^+} = [ \ket{u_{s1}^+}, \ket{u_{s2}^+}, \ldots ,\ket{u_{s s_0}^+}]^T$, then the vectors
\begin{equation*}
\ket{\psi_s^+} = \sum_{j=1}^{s_0}\ket{N-s_0+j}\ket{u_{sj}^+} , \qquad s=1,\ldots,s_0,
\end{equation*}
provide right-localized solutions to the bulk equation.

\section{Diagonalization of the bosonic Kitaev-Majorana chain}
\label{diagdetails}

\subsection{Open boundary conditions}
\label{appobc}

First, note that the the internal matrices $g_{\pm 1}$ commute. A basis of simultaneous eigenvectors is thus given by $\ket{\pm} \equiv (1/\sqrt{2})[1,\pm 1]^T$. This allows us to write
\begin{equation*}
-i G_O = \frac{1}{2}\left( J T^\dag - f T\right)\otimes \ket{+}\bra{+} + \frac{1}{2}\left( f T^\dag - J T\right)\otimes \ket{-}\bra{-}, \quad 
J\equiv t+\Delta,  \:f\equiv t-\Delta.
\end{equation*}
When $t=\Delta$ $(f=0$), we see that the generalized eigenvectors are constructed from those of $T$ and $T^\dag$. Specifically, $\ket{\chi_{1k}} = (-i J)^{-k+1}\ket{k}\ket{-}$ and $\ket{\chi_{2k}} = (i J)^{-k+1}\ket{N+k-1}\ket{+}$, with $k=1,\ldots, N$ in both cases, provide two length-$N$ Jordan chains at eigenvalue $\omega=0$.

Henceforth, we restrict to the case $t\neq \Delta$.
Thus, the problem reduces to diagonalizing an $N\times N$ matrix of the form $M=\left( a T+ b T^\dag\right)/2$, with $a,b\in \mathbb{R}\setminus\{0\}$. The reduced bulk Hamiltonian of Eq.\,\eref{eq:reduced} is $M(z,z^{-1}) = (az + b z^{-1})/2$ and the corresponding characteristic polynomial $P(z,\omega) = z\left( M(z,z^{-1})-\omega\right) = (a z^2  +b)/2-\omega z$. The roots are $z_\pm = (1/a)\left( \omega \pm \sqrt{\omega^2 -  ab}\right)$, which satisfy $z_-=c/z_+$. These roots only coalesce when $\omega=\omega_\pm \equiv \pm \sqrt{ab}$.

For the case $\omega\neq \omega_\pm$, the two bulk eigenstates are $\ket{z_\pm,1}$ which yields the boundary matrix
\begin{equation*}
B(\omega) = \frac{1}{2}\left[\matrix{
 -b & -b \cr z_+^{N-1}\left( b - 2\omega z_+\right) & (c/z_+)^{N-1} (b - 2\omega z_+^{-1})
}\right] ,\qquad c\equiv b/a.
\end{equation*}
It can be quickly checked that $B(-\omega)$ is similar to $B(\omega)$ and so the spectrum is necessarily symmetric about $\omega=0$.   The condition for a nontrivial kernel ($\det B(\omega) = 0$) reduces to the equation
\begin{equation*}
z_+^{2N-2}(b - 2\omega z_+) = c^N\left( a -2\omega z_+^{-1}\right).
\end{equation*}
The $2N$ roots (of which only $N$ are distinct) are given by $z_+=\pm \sqrt{c} e^{im \pi/(N+1)}$ with $m=1,\ldots N$. The corresponding $N$ distinct eigenvalues are $\omega_m = \sgn(a)\sqrt{ab}\cos(m\pi/(N+1))$. Note that $\omega_m\neq \omega_\pm$ and so we need not address the case of two coalescing roots. Taking the roots $z_+=\sqrt{c} e^{im\pi/(N+1)}$ yields the kernel vector $\bm{\alpha}=[1,-1]$. The (unnormalized) eigenvectors are then
\begin{equation*}
\ket{\psi_m} = \ket{z_m,1} - \ket{c/z_m,1} = \sum_{j=1}^N c^{j/2}\sin\left( \frac{m \pi j}{N+1}\right)\ket{j},\quad \left( M-\omega_m\mathds{1}_{N}\right)\ket{\psi_m} = 0.
\end{equation*}
With these solutions, we define 
\begin{equation*}
\ket{\phi_m^\pm} \equiv \sum_{j=1}^N\left(-\sigma\right)^{j/2}e^{\pm jr} \sin\left( \frac{m\pi j}{N+1}\right)\ket{j}\ket{\pm}, \quad \omega_m \equiv \sqrt{t^2-\Delta^2}\cos\left(\frac{m\pi}{N+1}\right) ,
\end{equation*}
where $r=1/2\ln(J/|f|)$. These satisfy
\begin{equation*}
G_O\ket{\phi_m^\pm} = \left\{\matrix{
\omega_m \ket{\omega_m}, &  \sgn(t-\Delta) = 1,
\cr
\pm \omega_m \ket{\omega_m}, & \sgn(t-\Delta)=-1.
}\right.
\end{equation*}
These eigenvectors can then be combined to form the bosonic eigenvectors $\ket{\psi_{m,\sigma}^\pm}$ in Eq.\,\eref{obcevecs} in the text.

\subsection{Twisted boundary conditions}
\subsubsection{The parameter regime $s=1$, $\varphi=\pi/2$, $t\neq \Delta$.}
\label{apptbc}
Instead of diagonalizing $G_T=G(1,\pi/2)$ directly, we will first perform a unitary rotation $G'_T \equiv U^\dag G_T U  = G_O' + V'$ where $U=\mathds{1}_N \otimes u$ with
\begin{equation*}
u = \frac{1}{\sqrt{2}}\left[\matrix{
1 & i \cr 1 & -i
}\right].
\end{equation*}
Physically, this unitary manifests at the many-body level as the basis transformation $(a_j,a_j^\dag)\mapsto (x_j,p_j)$. The rotated effective SPH has a very simple structure in this basis;
\begin{eqnarray}
\label{GOp}
&G_O' = T \otimes g_{1}' + T\otimes g_{-1} , 
\quad
&V' = \ket{N}\bra{1}\otimes v_1' + \ket{1}\bra{N}\otimes v_{-1}' ,
\\
&g_1' = -\frac{i}{2}\left[\matrix{
f & 0 \cr 0 & J
}\right] = \sigma_y g_{-1}'^\dag \sigma_y,
\quad
&v_1' = -\frac{i}{2}\left[\matrix{
0 & f \cr -J & 0
}\right] = \sigma_y v_{-1}'^\dag \sigma_y = v_{-1}' ,
\end{eqnarray}
where again $J=t+\Delta$ and $f=t-\Delta$.  The relevant matrix Laurent polynomial is given by
\begin{equation}
G'_T(z,z^{-1})= g_{1}'z +  g_{-1}z^{-1}= -\frac{i}{2}\left[\matrix{
 fz - J z^{-1}  & 0 \cr 0 & Jz - fz^{-1}
}\right] .
\end{equation}
The characteristic polynomial $P(\omega,z)\equiv z^2\det(G'_T(z,z^{-1})-\omega \mathds{1}_2)$ has four roots 
\begin{eqnarray}
&z_1 = \frac{1}{f}\left( i\omega - \sqrt{Jf - \omega^2}\right), \quad &z_2 = \frac{1}{J}\left( i \omega - \sqrt{Jf - \omega^2}\right),
\\
&z_3 = \frac{1}{f}\left( i\omega + \sqrt{Jf - \omega^2}\right), \quad &z_4 = \frac{1}{J}\left( i\omega + \sqrt{Jf - \omega^2}\right),
\end{eqnarray}
which are all distinct as long as $\omega\not\in\mathcal{S}\equiv \{\pm \sqrt{Jf},\pm(J+f)/2\}$. We will first assume that $\omega\not\in\mathcal{S}$. With this, we can easily find the bulk solutions
\begin{equation}\label{TBSs}
\ket{\psi_1} = \ket{z_1,1}\left[\matrix{
1 \cr 0 
}\right], 
\quad
\ket{\psi_2} = \ket{z_2,1}\left[\matrix{
0 \cr 1
}\right], 
\quad
\ket{\psi_3} = \ket{z_3,1} \left[\matrix{
1 \cr 0
}\right], 
\quad
\ket{\psi_4} = \ket{z_4,1} \left[\matrix{
0 \cr 1
}\right] ,
\end{equation}
from which we can construct the boundary matrix
\begin{equation*}
\fl B(\omega) = \frac{i}{2}\left[\matrix{
-J & -fz_1^Ne^{-2Nr'} & -J & -f(-z_1)^{-N}
\cr
Jz_1^N & -f & Je^{2Nr'}(-z_1)^{-N} & -f 
\cr
iz_1^{N-1}\left( 2z_1 \omega -iJ\right) & -fz_1e^{-2r'} & e^{2Nr'}(-z_1)^{-N}\left( 2i\omega-fz_1\right) & fz_1^{-1}
\cr
Jz_1 & z_1^Ne^{-2Nr'}(Jz_1^{-1}+2i \omega) & -Je^{2r'}z_1^{-1} & (-z_1)^{-N}(2i\omega-fz_1)
}\right] ,
\end{equation*}
where $r'=r$ for $t>\Delta$ and $r'=r+i\pi/2$ for $\Delta>t$. 
The condition for $\omega$ to be an eigenvalue is $\det B(\omega)=0$. From the expression for $z_1$, we can see that $\omega= i(fz_1-Jz_1^{-1})$.  Inserting this into $B(\omega)$ and taking the determinant introduces $4$ fictitious roots of $\det B(\omega) =0$, which we will identify after finding all of the roots
\begin{equation*}
 \left( z_1^{2N}+e^{2Nr'}  \right)^2\left( J+fz_1^2\right)^2 =0 .
\end{equation*}
If $z_1=\pm i\sqrt{J/f}$, then $\omega=\pm \sigma \sqrt{Jf}\in\mathcal{S}$, with $\sigma=\sgn(f)$, which must be considered separately.  The remaining roots are $z_1= \pm z_m\equiv \pm e^{r'} e^{ik_m}$, where $k_m=(m+1/2)\pi/N$ and $m=0,\ldots, 2N-1$. This gives the $2N$ potential eigenvalues $\omega_m \equiv \sigma\sqrt{Jf}\sin(k_m) = \sqrt{t^2-\Delta^2}\sin(k_m)$, with $m=0,\ldots,2N-1$. 

Now, we must split into separate cases: if $N$ is even, $\omega_m\neq \pm \sqrt{Jf}$ for all $m$ and so we have all $2N$ eigenvalues of $G'_T$, and hence for $G_T$.   If $N$ is odd, then when $m=(N-1)/2$, $\omega_m=\sigma\sqrt{Jf}$ and when $m=(3N-1)/2$, $\omega_m=-\sigma\sqrt{Jf}$. Since these are in $\mathcal{S}$, we must handle these separately. We do this after finding the eigenvectors for the remaining eigenvalues. 
The kernel vectors of $B(\omega_m)$ are
\begin{eqnarray*}
\bm{\alpha}_m  = [e^{-(N+2)r'},i(-1)^m,0,0]^T ,\qquad 
\bm{\beta}_m  = [0,0,e^{-(N+2)r'},i(-1)^{N-1-m}]^T ,
\end{eqnarray*}
with degeneracy arising due to the fact that each $\omega_m\neq \pm\sqrt{Jf}$ is doubly degenerate. The degenerate eigenvectors of $G'_T$ corresponding to eigenvalue $\omega_m$ are $\bm{\alpha}_m^T\ket{\Psi}$ and $\bm{\beta}_m^T\ket{\Psi}$, with 
$\ket{\Psi}\equiv[\ket{\psi_1},\ket{\psi_2},\ket{\psi_3},\ket{\psi_4}]^T.$ Rotating back via the unitary transformation $U$ gives the eigenvectors of $G$ corresponding to eigenvalue $\sigma \omega_m$ as
\begin{equation}\label{TBCevecs}
 \ket{\psi_{m,\sigma}} = \frac{1}{\sqrt{N}}\sum_{j=1}^N e^{ij k_m}\ket{j}\ket{\zeta_m(j)},\qquad \ket{\zeta_m(j)} = \sigma_1^m\left[\matrix{
\sinh\left[\left( j - \frac{N+2}{2}\right) r'\right]
\cr
\cosh\left[\left( j - \frac{N+2}{2}\right) r'\right]
}\right]. 
\end{equation}

For $N$ even, the above procedure exhausts all possibilities. For $N$ odd, we consider the case $\omega=\sigma\sqrt{J f}$ explicitly and note that the case $\omega=-\sigma\sqrt{J f}$ can be handled in an analogous way. In this case, the characteristic polynomial has two distinct roots $z_1= i e^{r'}$ and $z_2=-1/z_1$. The corresponding eigenvectors of $G'_T(z_j,z_j^{-1})$ are $\ket{u_1}=[1,0]^T$ and $\ket{u_2}=[0,1]^T$ giving two bulk solutions
\begin{equation*}
\ket{\psi_{1,1}}=\ket{z_1,1}\left[\matrix{
1 \cr 0
}\right],\qquad \ket{\psi_{2,1}} = \ket{z_2,1}\left[\matrix{
0 \cr 1
}\right].
\end{equation*}
The remaining two bulk solutions arise from the eigenvectors of $G'_{T,1}(z_j,z_j^{-1})$ where
\begin{equation*}
G'_{T,1}(z,z^{-1})=\left[\matrix{
G'(z,z^{-1}) & \partial_z G'(z,z^{-1})
\cr
0 & G'(z,z^{-1})
}\right].
\end{equation*}
These yield two more bulk solutions
\begin{equation*}
\ket{\psi_{1,2}}=\ket{z_1,2}\left[\matrix{
1 \cr 0
}\right],\qquad \ket{\psi_{2,2}} = \ket{z_2,2}\left[\matrix{
0 \cr 1
}\right].
\end{equation*}
The boundary matrix at $\omega=\sigma\sqrt{J f}$ is then
\begin{equation*}
 B(\sigma\sqrt{J f}) = \frac{i}{2}\left[\matrix{
-J & -f(-z_1)^{-N} & 0 & -Nf(-z)^{1-N}
\cr
J z_1^N & -f & NJz_1^{N-1} & 0
\cr
-J z_1^{N-1} & -\sigma f/z_1 & (N+1)f z_1^N & -f
\cr
J z_1 & -f (-z_1)^{1-N} & J  & \sigma(N+1)J (-z_1)^{-N}
}\right].
\end{equation*}
Then $\det B(\sigma\sqrt{ Jf}) \propto 1+(-1)^N = 0$ for $N$ odd. The kernel is one dimensional and is spanned by
\begin{equation*}
\bm{\alpha} = [e^{-(N+2)r'},i(-1)^{(N-1)/2},0,0]^T .
\end{equation*}
Hence, the eigenvector corresponding to $\sigma\sqrt{Jf}$ is $\ket{\psi_{(N-1)/2,\sigma}}$ where $\ket{\psi_{m,\sigma}}$ is exactly as in Eq.\,\eref{TBCevecs}. Similarly, the eigenvector corresponding to $-\sigma\sqrt{Jf}$ is $\ket{\psi_{(3N-1)/2,\sigma}}$.

\subsubsection{Dynamical phase boundaries.}
\label{GBTzero} 
In this section, we determine analytically the dynamical phase boundaries in boundary parameter space. An important assumption of this derivation is that certain phase boundaries are characterized by the emergence of zero modes and that the phase diagram is symmetric about $\varphi=\pi/2$. Thus, we will uncover the conditions on $s$ and $\varphi$ for $G(s,\varphi)$ to possesses zero as an eigenvalue. 

As in the preceding Appendix, we will rotate via the unitary $U$ and study the unitarily equivalent matrix $G'(s,\varphi)$. In contrast to the preceding section, however, we keep $\varphi$ arbitrary and restrict to the non-open case $s\in(0,1]$. Since the bulk ($G_O'$) is unchanged, and the roots of the characteristic polynomial $P(\omega=0,z)$ are distinct, we have the same four bulk solutions $\ket{\psi_j},$ $j=1,2,3,4$, given in Eqns. \eref{GOp}-\eref{TBSs}. On the other hand, the boundary modification is now given by
\begin{eqnarray*}
&V'(s,\varphi) = \ket{N}\bra{1}\otimes v_1'(s,\varphi) + \ket{1}\bra{N}\otimes v_{-1}'(s,\varphi) ,
\\
&v_1' = -\frac{is}{2}\left[\matrix{
f\cos(\varphi) & f\sin(\varphi) \cr -J\sin(\varphi) & J\cos(\varphi)
}\right] = \sigma_y v_{-1}'^\dag \sigma_y .
\end{eqnarray*}
Since the boundary condition is different, the boundary matrix becomes
\begin{equation*}
 B(\omega=0) = \frac{i}{2}\left[\matrix{
C_1(z_1) & C_2(z_1^{-1}) & C_1(-z_1) & C_2(-z_1^{-1}) 
}\right] ,
\end{equation*}
where 
\begin{eqnarray*}
C_1(z) &\equiv & [-J\left(1-sz^N\cos(\varphi)\right),sJz^N\sin(\varphi),fz\left(z^N-s\cos(\varphi)\right),Jzs\sin(\varphi)]^T ,
\cr
C_2(z) &\equiv & [-sf z^N\sin(\varphi),-f\left(1-sz^N\cos(\varphi)\right),-sfz\sin(\varphi),Jz\left(z^N-s\cos(\varphi)\right)]^T .
\end{eqnarray*} 
Demanding that the determinant vanishes, we obtain the conditions
\begin{equation}
\cos(\varphi^\pm) =  \frac{1}{2}\left\{\matrix{
(s+s^{-1})\text{sech}(Nr),& N \text{ even},
\cr
\pm 2\, \text{sech}(Nr),& N\text{ odd} .
}\right. 
\label{phi-}
\end{equation}
For $N$ even, this specifies one angle $\varphi^+=\varphi^-$ in the interval $[0,\pi]$, in fact, smaller than $\pi/2$. On the other hand, for $N$ odd, there are two distinct angles $\varphi^\pm$ symmetric about each side of $\pi/2$. Thus, both phase boundaries host zero modes for $N$ odd and just the left boundary for $N$ even.

When Eq. \eref{phi-} is satisfied, the kernel of $B(0)$ can be determined analytically. The cases $s\neq 1$ and $s= 1$ must be handled separately. We begin by taking $s\neq 1$. For $N$ even, $\ker B(0)$ is two-dimensional and spanned by the vectors
\begin{eqnarray*}
\bm{\alpha} &=& \frac{1}{s-s^{-1}}\left[\left(4-(s+s^{-1})^2\text{sech}^2(Nr)\right)^{1/2}e^{-(N+2)r},(s+s^{-1})\tanh(N r),0,s-s^{-1}\right]^T ,
	\\
\bm{\beta} &=& \frac{1}{s-s^{-1}}\left[(s+s^{-1})\tanh(N r),\left(4-(s+s^{-1})^2\text{sech}^2(Nr)\right)^{1/2}e^{(N+2)r},s^{-1}-s,0\right]^T.
\end{eqnarray*}
For $N$ odd and $\varphi=\varphi^\pm$, $\ker B(0)$ is one-dimensional and spanned by
\begin{equation*}
\bm{\alpha_\pm} = \left[\left(\frac{s\mp 1}{s\pm 1}\right)e^{-(N+2)r},\frac{s\mp 1}{s\pm 1},e^{-(N+2)r},1\right]^T .
\end{equation*}
For $s=1$, the analogous kernel vectors for $N$ even are
\begin{eqnarray*}
\bm{\alpha}  = \left[e^{-(N+2)r},1,0,0\right]^T, \quad 
\bm{\beta} = \left[0,0,e^{-(N+2)r},1\right]^T, 
\end{eqnarray*}
whereas for $N$ odd are
\begin{equation*}
\bm{\alpha_+} = \left[0,0,e^{-(N+2)r},1\right]^T, \quad \bm{\alpha_-} = \left[e^{-(N+2)r},-1,0,0\right]^T.
\end{equation*}

The important thing to note is that these calculations reveal that the dimension of the zero-mode subspace is one (two) for $N$ odd (even). The four-fold symmetry of the spectra of bosonic effective SPHs implies that the {\em algebraic multiplicity of the zero eigenvalue must always be even}. This confirms that for $N$ odd, there must be a Jordan chain of length two at zero, along the phase boundaries $(s>0)$. An additional symmetry of the even chain implies that all non-zero eigenvalues of $G(s,\varphi)$ are at least doubly degenerate, implying that the zero eigenvalue has algebraic multiplicity four. Thus, the even chain possesses two length-two Jordan chains at zero, along the left phase boundary. Alternatively, this can be concluded by checking that the dimension of kernel of the boundary matrix of $G^2$ at zero frequency is four.

\subsubsection{The parameter regime $s=1$, $\varphi\in(0,\pi)$, $t=\Delta$.}
\label{sub:jordan}
At $\varphi=\pi/2$, $G_T$ is non-diagonalizable when $t=\Delta$. The Jordan chains can be constructed by inspection and are given by
\begin{eqnarray}
\label{TBCgevecsNeven}
\ket{\chi_{1k}} &=&  \left(\frac{i}{t}\right)^{k} \ket{k}\ket{-}, \quad  k = 1,\ldots,N ,
\\
\ket{\chi_{2k}} &=& \left(\frac{i}{t}\right)^{k}\left\{\matrix{
i \ket{k+1}\ket{-}+(-1)^{k+1}\ket{N+1-k}\ket{+}, & 1\leq k<N , 
\cr
-\ket{1}\ket{+}, & k=N ,
}\right.
\end{eqnarray}
for $N$ even, and 
\begin{eqnarray}
\ket{\chi_{1k}} &=& \left(\frac{i}{t}\right)^{k} \left\{\matrix{
2\ket{1}\ket{-},& k=1,
\cr
 \ket{k}\ket{-}+i(-1)^k \ket{N+2-k}\ket{+}, & 2\leq k \leq N+1 ,
}\right.
\\
\ket{\chi_{2k}} &=& \left(\frac{i}{t}\right)^k \left(i\ket{k+1}\ket{-}+(-1)^{k+1}\ket{N-k+1}\ket{+}\right), \quad k = 1,\ldots,N-1 ,\label{TBCgevecsNodd}
\end{eqnarray}
for $N$ odd. Specifically, these satisfy $G_T\ket{\chi_{jk)}} = G_T\ket{\chi_{j(k-1)}}$, with $k\neq 1$ and $G_T\ket{\chi_{j1}}=0$ for $j=1,2$. It is interesting to note that for $N$ even there are two length-$N$ Jordan chains, whereas for $N$ odd there is a Jordan chain of length $N+1$ and one of length $N-1$. 

For $\varphi\neq \pi/2$ we define $G_T(\varphi) \equiv G(1,\varphi)$. Again, we simplify the problem by first diagonalizing $G'_T(\varphi) \equiv U^\dag G_T(\varphi)U$. In this case, $f=0$ and $J=2t$, and the corner modification takes the form
\begin{eqnarray*}
V'(\varphi) &=& U^\dag V(1,\varphi)U  = \ket{N}\bra{1}\otimes v_1'(\varphi) + \ket{1}\bra{N}\otimes v_{-1}'(\varphi),
\\
v_1'(\varphi) &\equiv& it\left[\matrix{
0 & 0 \cr \sin(\varphi) & -\cos(\varphi)
}\right],\quad v_{-1}'(\varphi) \equiv it\left[\matrix{
\cos(\varphi) & 0 \cr \sin(\varphi)  & 0
}\right].
\end{eqnarray*}
In particular, we note that $\det g_1'=\det g_{-1}'=0$ and so we expect emergent solutions to the bulk equation. The reduced bulk effective Hamiltonian is given by
\begin{equation*}
G'_T(\varphi,z,z^{-1}) = it\left[\matrix{
z^{-1} & 0 \cr 0 & -z
}\right].
\end{equation*}
The roots of the characteristic polynomial are $z_1=it/\omega$ and $z_2=-1/z_1$ wish coalesce only for $\omega=\pm t$. The eigenvectors are $\ket{u_1} = [1,0]^T$ and $\ket{u_2}=[0,1]^T$ which provide two bulk solutions
\begin{equation*}
\ket{\psi_1} = \ket{z_1,1}\left[\matrix{
1 \cr 0
}\right],\qquad \ket{\psi_2} = \ket{z_2,1}\left[\matrix{
0 \cr 1
}\right].
\end{equation*}
The remaining two bulk solutions come from the kernels of the matrices
\begin{equation*}
K^-(\omega) = \left[\matrix{
g_{-1}' & -\omega \mathds{1}_2 & g_1' & 0 
\cr
0 & g_{-1}' & -\omega \mathds{1}_2 & g_1'
\cr
0 & 0 & g_{-1}' & -\omega \mathds{1}_2
\cr
0 & 0 & 0 & g_{-1}'
}\right],
\qquad
K^+(\omega)\equiv
\left[\matrix{
g_1' & 0 & 0 & 0 
\cr
-\omega\mathds{1}_2 & g_1' & 0 & 0 
\cr
g_{-1}' & -\omega \mathds{1}_2 & g_1' & 0
\cr
0 & g_{-1}' & -\omega \mathds{1}_2 & g_1'
}\right],
\end{equation*}
which are spanned by $\ket{u_-}=[0,1,0,0,0,0,0,0]^T$ and $\ket{u_+} = [0,0,0,0,0,0,1,0]^T$ respectively. With these, the two additional bulk solutions
\begin{equation*}
\ket{\psi_-} = \ket{1} \left[\matrix{
0 \cr 1
}\right],\qquad \ket{\psi_+} = \ket{N}\left[\matrix{
1\cr 0
}\right] .
\end{equation*}
The corresponding boundary matrix is
\begin{equation*}
B(\omega) = it\left[\matrix{ 
z_1^N \cos(\varphi) -1 & 0 & 0  & \cos(\varphi)
\cr
z_1^N \sin(\varphi) & 0 & z_2 & \sin(\varphi)
\cr
0 & 0 & 0 & z_2
\cr
z_1 \sin(\varphi) & z_2(z_2^N-\cos(\varphi)) & -\cos(\varphi) & 0
}\right] ,
\end{equation*}
where we have used $\omega=it/z_1$. The condition for a vanishing determinant is
\begin{equation*}
( z_1^N \cos(\varphi)-1) (z_2^N -\cos(\varphi) ) = 0.
\end{equation*}
For $N$ even, the roots are doubly degenerate and given by $z_1=z_m\left(\cos(\varphi)\right)^{-1/N} e^{2m\pi i/N}$, with $m=1,\ldots,N$. For $N$ odd, the roots are $z_m=(\cos(\varphi))^{-1/N}e^{im\pi/N}$, with $m=1,\ldots 2N$. In both cases we 
let 
$\left(\cos(\varphi)\right)^{-1/N} \equiv e^{-i\pi/N}|\cos(\varphi)|^{-1/N}$, for $\varphi\in(\pi/2,\pi)$. The eigenvalues are then given by $\omega_m = it/z_m$. Equivalently, the spectrum $\sigma(G_T(\varphi))$ is related to the periodic and anti-periodic cases as
\begin{equation*}
\sigma(G_T(\varphi)) = |\cos(\varphi)|^{1/N}\left\{\matrix{
\sigma(G_P),& \varphi\in(0,\pi/2] ,
\cr
\sigma(G_A),& \varphi\in(\pi/2,\pi) ,
}\right.
\end{equation*}
with $G_P$ ($G_A$) the effective SPH of the chain under periodic (anti-periodic) BCs with $t=\Delta$.
Note that $|\omega_m|< t$ for all $m$ and $\varphi\in(0,\pi)$ and so we need not address the case $\omega_m=\pm t$. 

Now, for $N$ even the kernel of $B(\omega)$ is 2 dimensional and spanned by
\begin{eqnarray*}
\bm{\alpha}_m = [ e^{2m\pi i/N}(\cos(\varphi))^{1/N},0,\tan(\varphi),0]^T], \qquad 
\bm{\beta} = [0,1,0,0]^T .
\end{eqnarray*}
After rotating back to the $(a,a^\dag)$ basis, the (doubly degenerate) eigenvectors of $G_T(\varphi)$  for $N$ even, corresponding the eigenvalue $\omega_m$, are
\begin{eqnarray*}
\ket{\psi_{m,1}} &=& \mathcal{N}_{m,1}\left( z_m^{-1} \ket{z_m,}\ket{+} + i \tan(\varphi)\ket{1}\ket{-}\right),
\\
\ket{\psi_{m,2}} &=& \mathcal{N}_{m,2} \ket{-z_m^{-1},1}\ket{-} ,
\end{eqnarray*} 
with $\mathcal{N}_{m,\ell}$, $\ell=1,2$ normalization constants. For $N$ even the kernel of $B$ is one-dimensional and is spanned by $\bm{\alpha}_{m/2}$, for $m$ even, and $\bm{\beta}$, for $m$ odd. Hence, the eigenvector of $G_T(\varphi)$ for $N$ odd corresponding to the eigenvalue $\omega_m$ is given, up to a normalization constant, by
\begin{equation*}
\ket{\psi_m} = \mathcal{N}_m \left\{\matrix{
z_m^{-1} \ket{z_m,1}\ket{+} + i\tan(\varphi)\ket{1}\ket{-},& m \text{ even},
\cr
\ket{-z_m^{-1},1}\ket{-},& m\text{ odd} .
}\right. 
\end{equation*}

\vspace{5mm}
\noindent\textbf{References}
\vspace{.3cm}
\providecommand{\newblock}{}

\end{document}